\pdfoutput=1

\documentclass[de]{template/rrlab}
\usepackage{pdfpages} 

\RRLABtitle{Qualitätsmaße binärer Klassifikatoren im Bereich kriminalprognostischer Instrumente der vierten Generation
}

\RRLABauthor{Tobias D. Krafft}

\RRLABtype{Masterarbeit}

\RRLABinception{15. November 2016}

\RRLABsubmission{12. Mai 2017}

\RRLABfirstreviewer{Prof. Dr. Katharina A. Zweig}

\RRLABsecondreviewer{Prof. Dr. Nicholas A. Diakopoulos}

\RRLABsupervisor{}

\RRLABdegree{Master of Science (M.Sc.)}

\RRLABdefense{irgendwann}

\RRLABchair{irgendwer}

\RRLABdean{irgendwer}

\newcommand{\qed}{\nobreak \ifvmode \relax \else
      \ifdim\lastskip<1.5em \hskip-\lastskip
      \hskip1.5em plus0em minus0.5em \fi \nobreak
      \vrule height0.75em width0.5em depth0.25em\fi}

\graphicspath{%
 {./pictures/}
}
\newtheorem{Beobachtung}{Beobachtung}
\begin{document}

\RRLABtitlepage{
}{\currentdate\today}


\RRLABsecondpage


\thispagestyle{empty}
\vspace*{15cm}

Ich erkl\"are hiermit, die vorliegende Masterarbeit selbst\"andig verfasst zu
haben. Die verwendeten Quellen und Hilfsmittel sind im Text kenntlich
gemacht und im Literaturverzeichnis vollst\"andig aufgef\"uhrt.\\
\ \\
Kaiserslautern, den 12. Mai 2017

\vspace{3cm}

(\RRLABVARauthor )

\RRLABpreface{Abstract}{

This master’s thesis discusses an important issue regarding how algorithmic decision making (ADM) is used in crime forecasting. In America forecasting tools are widely used by judiciary systems for making decisions about risk offenders based on criminal justice for risk offenders. By making use of such tools, the judiciary relies on ADM in order to make error free judgement on offenders. For this purpose, one of the quality measures for machine learning techniques which is widly used, the $AUC$ (area under curve), is compared to and contrasted for results with the $PPV_k$ (positive predictive value). Keeping in view the criticality of judgement along with a high dependency on tools offering ADM, it is necessary to evaluate risk tools that aid in decision making based on algorithms.\\
In this methodology, such an evaluation is conducted by implementing a common machine learning approach called binary classifier, as it determines the binary outcome of the underlying juristic question. This thesis showed that the $PPV_k$ (positive predictive value) technique models the decision of judges much better than the $AUC$.  Therefore, this research  has investigated whether there exists a classifier for which the $PPV_k$ deviates from $AUC$ by a large proportion. It could be shown that the deviation can rise up to 0.75. In order to test this deviation on an already in used Classifier, data from the fourth generation risk assement tool COMPAS was used. The result were were quite alarming as the two measures derivate from each other by 0.48.\\
In this study, the risk assessment evaluation of the forecasting tools was successfully conducted, carefully reviewed and examined. Additionally, it is also discussed whether such systems used for the purpose of making decisions should be socially accepted or not.
}

\RRLABpreface{Kurzzusammenfassung}{

Die vorliegende Masterarbeit setzt sich kritisch mit der Integration von algorithmischen Entscheidungssystemen (ADM) in der Kriminalprognostik auseinander, und versucht die Begutachtungsstrategien der Justiz zu beleuchten, indem sie die AUC als bewährtes Qualitätsmaß aus dem maschinellen Lernen dem $PPV_k$ gegenüberstellt. \\
Zunächst wird ein Überblick über die standardisierten Prognoseinstrumente zur Individualvorhersage der deutschen sowie US-amerikanischen Kriminalprognostik gegeben.\\
Da das Justizsystem in den USA aktuell bei der Beurteilung von Straftätern stark auf ADMs baut, stellt sich die zentrale Frage, wie solche Instrumente bewertet werden und welche gesellschaftliche Akzeptanz sie erfahren.\\
Die hier Anwendung findenden Instrumente aus dem maschinellen Lernen münden meist in binären Klassifikatoren, da die juristische Fragestellung nur duale Urteile zulässt. Das vorherrschende Repertoire zur Bewertung solcher Klassifikatoren ist zwar differenziert, jedoch wird in diesem Umfeld vorwiegend die \glq Area under the curve\grq~($AUC$) zur Evaluation herangezogen.\\
Diese Arbeit überprüft deshalb, ob es Klassifikatoren gibt, für die dieses Qualitätsmaß deutlich vom \glq Positive Predictive Value\grq~($PPV_k$) abweicht, der als Evaluationsmöglichkeit den richterlichen Entscheidungsprozess realistischer abbildet. Die Ergebnisse sind besorgniserregend, da Abweichungen von bis zu $0.75$ nachweislich möglich sind.\\
Dieser Effekt konnte abschließend an dem vom Justizsystem in Wisconsin eingesetzten Prognoseinstrument der vierten Generation ‚COMPAS‘ auf realen Daten nachgewiesen werden. Hier sind Abweichungen von bis zu $0.48$ zu verzeichnen. 

}


\RRLABcontents



\RRLABpagenumbers

\chapter{Einleitung}
Der Bürger eines jeden demokratischen Rechtsstaates erwartet von der Justiz Gerechtigkeit und einen höchstmöglichen Schutz vor jeglicher Art von Angriff auf seine Person und Rechte.
Jedoch sollte ein Angeklagter im Falle der nachgewiesenen Schuld ein gerechtes Urteil im Hinblick auf seine Straftat und eine dementsprechend angemessene Strafe erwarten können. In früheren Zeiten oblag es allein dem Richter und etwaigen Beratern, Persönlichkeit und zukünftiges Legalverhalten eines Angeklagten einzuschätzen und beispielsweise zu entscheiden, ob dieser eine Gefahr für die Öffentlichkeit darstellt und verwahrt werden muss, oder ob die Strafe zur Bewährung ausgesetzt werden kann.\\
Diese über Jahrhunderte praktizierte, allein auf der subjektiven Einschätzung und Prognose des Richters basierende Urteilsfindung war natürlich extrem anfällig für Fehlurteile, auch hinsichtlich der Höhe des verhängten Strafmaßes. Die Wissenschaftswelt setzte sich daher mit der Problematik schon im Zuge der Aufklärung auseinander. So forderte der italienische Rechtsphilosoph Cesare Beccaria bereits in seinem 1764 erschienenen Werk "`\textit{Dei delitti e delle pene}"'\footnote{zu deutsch: Über Vergehen und deren Strafe}~\cite{Cesare1764} die Verhältnismäßigkeit der Sanktion zur begangenen Straftat und gilt insofern heute als Wegbereiter der Kriminologie~\cite{Roessner2000}, denn die durch ihn ausgelösten Diskussionen führten insgesamt zum Bemühen, die Rechtsprechung von subjektiver Willkür zu befreien und objektive Kriterien für die Urteilsfindung zu entwickeln. Ein erster Ansatz eines kriminalprognostischen Versuchs lässt sich bei dem Turiner Arzt Lombroso erkennen \cite[S.96]{Schwind2010}, der 1887 die These vertrat, dass Verbrecher an äußeren Merkmalen (stigmata) erkennbar seien\footnote{ Dieser sogenannte \glq Etikettierungsansatz\grq~setzte sich im Übrigen im angloamerikanischen Raum durch und wurde in Deutschland von Sack noch in den 1970-er Jahren vertreten, so Sack in seinem "`Handbuch der empirischen Sozialforschung"' ~\cite{Sack1978}} und dies mit umfangreichen empirisch-biometrischen Studien zu belegen versuchte. Anfang des 20. Jahrhunderts wurden schließlich erste Kriterienkataloge entwickelt, sogenannte Prognosetafeln, mittels derer man potenzielle Straftäter zu identifizieren hoffte~\cite[S.43f]{Nedopil2005} und die somit den Grundstein zur statistischen Kriminalprognose legten (Zur Entwicklung der kriminalprognostischen Instrumente, siehe Kapitel \ref{hist_krim}).

\section{Kriminalprognosen und ihre immanente Problematik}
Es empfiehlt sich zunächst, eine kurze Begriffsklärung vorzunehmen, da sich in der Literatur keine einheitliche Definition der \glq Kriminalprognose\grq~findet. Nach Nedopil zielt eine Kriminalprognose auf die Frage, ob ein Mensch oder eine bestimmte Gruppe von Menschen zukünftig gegen das Strafgesetz verstoßen werden, unabhängig davon, ob sie zuvor kriminell aufgefallen sind~\cite[S.17]{Nedopil2005}.\\
Im Strafrecht geht es jedoch primär um die Feststellung der individuellen Schuld, sodass nur Individualprognosen von Relevanz sind. Außerdem steht ein Angeklagter wegen einer eventuell begangenen Straftat vor Gericht, weswegen hier der praxisnäheren Definition von Kröber gefolgt wird: 
\begin{center}
\textit{"`Die gutachterliche Beurteilung der Kriminalprognose ist eine Risikobeurteilung zu der Frage, mit welcher Wahrscheinlichkeit eine bestimmte Person mit bestimmten Delikten straffällig werden wird."'}~\cite[S.88]{Kroeber2006}
\end{center}
Ähnlich wird auch die sogenannte Gefährlichkeitsprognose definiert als eine "`\textit{wissenschaftlich fundierte Wahrscheinlichkeitseinschätzungen darüber, (…) in welchem Maße eine bereits mit Straftaten in Erscheinung getretene Person in der Zukunft erneut rechtswidrige Taten begehen wird. Es handelt sich also um eine Verhaltensprognose für diese spezielle Person}"'~\cite[S.422]{Bliesener2014}.
Jede Kriminalprognose sieht sich grundsätzlich mit dem Problem konfrontiert, mit welcher Genauigkeit sich zukünftiges kriminelles Verhalten vorhersagen lässt, denn das menschliche Verhalten resultiert nun einmal nicht nur aus individuellen Persönlichkeitsmerkmalen, sondern wird ebenso durch verschiedenste situative Faktoren beeinflusst, die aufgrund ihrer Variabilität allenfalls vage abschätzbar sind~\cite[S.425f]{Bliesener2014}. Da in dieser Arbeit jedoch davon Abstand genommen wird, den Determinismus des menschlichen Verhaltens näher zu analysieren, wird es bei der Feststellung belassen, dass eine hundertprozentige Vorhersagbarkeit auch mit heutigen kriminalprognostischen Mitteln grundsätzlich nicht garantiert werden kann. \\
Von der Richtigkeit dieser These zeugt auch der Baxstrom-Fall aus dem Jahr 1966, welcher über die USA hinaus für mediales Aufsehen sorgte und die Zuverlässigkeit der Gewaltprognose ins Zentrum der Kritik rückte. \\
Bei diesem unbeabsichtigten Experiment mussten aus formal-juristischen Gründen der Gewalttäter Johnnie Baxstrom sowie 967 weitere, als gefährlich eingeschätzte Straftäter, im Bundesstaat New York freigelassen werden. Nach insgesamt vier Jahren in Freiheit waren aber lediglich 14,2\% der als gefährlich eingestuften Täter erneut straffällig geworden, darunter nur etwa 2,5\% wegen schwerer Gewaltstraftaten, also der Taten, weswegen ihr Risiko als besonders hoch eingeschätzt worden war. Als Ergebnis dieses ungewollten Experiments lässt sich daher feststellen, dass die Rückfallquote jener Freigelassenen äußerst gering war~\cite[S.17]{Obergfell-Fuchs}\cite{USSC1966}. \\
Daraus resultierende Studien stellten Mitte der 1970-er Jahre bei der klinischen Gewaltprognose (zu den Methoden der Kriminalprognostik, siehe Kapitel \ref{kap_hist}) erhebliche Mängel und Schwierigkeiten fest~\cite{Monahan1975,Monahan1981}, ohne die Methode allerdings grundsätzlich in Frage stellen zu wollen. Ziel war es vielmehr, Fehlerquellen zu ermitteln und durch rationale Handhabung sowie Einhaltung wissenschaftlicher Standards die aufgezeigten Mängel in der Gewaltprognostik zu verringern. Darüber hinaus wollte man die Justiz auf die Grenzen der Leistungsfähigkeit prognostischer Instrumente aufmerksam machen~\cite{Monahan1996}, denn die unbestrittenen Vorteile der nach wissenschaftlichen Standards entwickelten Gefahrenprognose für den juristischen Entscheidungsfindungsprozess gegenüber der früheren subjektiven Urteilsfindung eines Richters haben allgemein in der Praxis unrealistische Erwartungen an die tatsächlich aktuell mögliche Leistung von Prognosen geweckt, was einer dringenden Relativierung bedarf~\cite[S.475f]{Albrecht2004}.

\section{Der Einzug der algorithmischen Entscheidungsfindung (ADM) in die Kriminalprognostik}

In der heutigen digitalen Gesellschaft unterstützen und übernehmen algorithmische Prozesse, ADM ("`\textit{algorithmic decision making"'}~\cite{Zweig2016}) genannt, zunehmend die Entscheidungsfindung in den verschiedensten Bereichen, auch in der Justiz. Das \glq ADM-Manifest\grq~\cite{AlgorithmWatch2016} der 2016 gegründeten Bürgerinitiative ‚Algorithm Watch‘ differenziert folgende Aspekte als Bestandteil der algorithmischen Entscheidungsfindung: 
\begin{itemize}
	\item \textit{Prozesse zur Datenerfassung zu entwickeln,}
	\item	\textit{Daten zu erfassen,}
	\item \textit{Algorithmen zur Datenanalyse zu entwickeln, die die }
	
	\begin{itemize}
		\item \textit{Daten analysieren,}
\item	\textit{auf der Basis eines menschengemachten Deutungsmodells interpretieren,}
\item	\textit{automatisch handeln, indem die Handlung mittels eines menschengemachten Entscheidungsmodells aus dieser Interpretation abgeleitet wird.}

	\end{itemize}
\end{itemize}
Während in den USA aufgrund des anders strukturierten Justizsystems schon seit längerem ADM-gesteuerte Kriminalprognosen eingeführt sind, steht die deutsche Justiz in dieser Hinsicht noch relativ am Anfang dieser Entwicklung, denn deutsche Gerichte nutzen derzeit noch keine ADM-Risikoprognosen. Insofern ist es ein zentrales Anliegen der vorliegenden Arbeit, eine Diskussion über Chancen und Risiken beim Einsatz algorithmenbasierter Risikoprognosen anzuregen. Die Gesellschaft sollte sich möglichst zeitnah der zentralen Frage widmen, inwieweit sie es zulassen will, dass Algorithmen Entscheidungen treffen, die wichtig für das Leben des Einzelnen sein können, denn Deutschland besitzt jetzt noch die einmalige Möglichkeit, aus den Erfahrungen anderer Länder zu lernen und als Gesellschaft den Umgang mit Algorithmen mitbestimmen zu können. \\
Im Folgenden wird daher zunächst ein kurzer Überblick über die Entwicklung der Kriminalprognostik in den USA gegeben (siehe Abschnitt \ref{admusa}), gefolgt vom Abschnitt \ref{admde} der knapp die Situation im deutschen Justizwesen in Bezug auf kriminalprognostische Gutachten skizziert. 
\newpage
\subsection{Die Integration von ADMs in die kriminalprognostischen Instrumente im US-amerikanischen Justizwesen\label{admusa}}
Das Justizsystem der USA ist im Begriff zu kollabieren: Als Folge der etablierten Praxis, drakonische Strafen zur Abschreckung zu nutzen, sitzen derzeit in US-amerikanischen Gefängnissen knapp 2.15 Millionen Inhaftierte~\cite{statista2017} ein. Das bedeutet, 20\% aller weltweit Arrestierten befinden sich in US-amerikanischer Haft~\cite{Walmsley2014}, obwohl das Land lediglich fünf Prozent der gesamten Weltbevölkerung stellt~\cite{statista2017a}. Die hohe Zahl von Gefängnisinsassen beschert den USA infolgedessen explodierende Kosten im Strafvollzug, sodass die Suche nach einer effizienten Kostenreduzierung ein zentrales Anliegen der US-amerikanischen Justiz ist. \\
Da sich die Standards der Haftbedingungen in den USA jedoch laut Amnesty International bereits am unteren Limit befinden~\cite{AmnestyInternational2013}, lassen sich bei Unterbringung und Verpflegung keine weiteren Kosten einsparen~\cite{Heiser2013}, sodass die zunächst reißerisch anmutende Schlagzeile der Süddeutschen Zeitung vom 23. August 2016 "`Nudeln sind die neue Währung in US-Gefängnissen"'~\cite{Werner2016} bei näherer Betrachtung durchaus ihre Berechtigung\footnote{"`Ein Bericht von Amnesty International über Einzelhaft in den Bundesgefängnissen stellte fest, dass die Haftbedingungen im - bislang einzigen - Hochsicherheitsgefängnis, das dem US-Standard Super-Maximum Security entspricht, in Florence im Bundesstaat Colorado gegen die Standards einer humanen Behandlung von Gefangenen verstoßen."'~\cite{AmnestyInternational2015}} besitzt. Auch andere Konzepte zur Kosteneinsparung erwiesen sich als Fehlschlag. Abgesehen von skurrilen  anmutenden Vorschlägen wie aus dem Jahr 2009, dass Häftlinge finanziell selbst für ihren Gefängnisaufenthalt aufkommen könnten~\cite{FocusOnline2009}, zeigte sich längerfristig auch die Idee, Privatgefängnisse einzuführen, als unrentabel~\cite{NTV2016}.
Die Erkenntnis, zur Kosteneinsparung kostenintensive Haftstrafen auf Bewährung auszusetzen, ließ die Kriminalprognose zunehmend in den Fokus juristischer Überlegungen rücken. Darüber hinaus resultiert das vermehrte Interesse an geeigneten prognostischen Methoden zur Einschätzung des Rückfallrisikos auch aus den Diskussionen rund um den eingangs erwähnten Baxstrom-Fall aus dem Jahr 1966. Dieser Vorfall warf seinerzeit die Frage auf, ob man die Personen zu Unrecht in diese Anstalten eingewiesen habe und wie es zu der Fehleinschätzung kommen konnte. Seitdem wurde immer wieder eine genauere Einschätzung des Kriminalitätsrisikos gefordert, wenn es in die Urteilsfindung einbezogen wird. So sah sich die Bürgerrechtsvereinigung ACLU (\textit{\textbf{A}merican \textbf{C}ivil \textbf{L}iberties \textbf{U}nion}) im Jahr 2011 zur Empfehlung veranlasst, durch eine genaue Datenanalyse das Risiko zu kalkulieren, ob Straftäter tatsächlich rückfällig und zu einer Gefahr für die Gesellschaft werden würden~\cite[S.11]{Chettiar2011}. \\
Mit Hilfe der ARAI, der \glq \textbf{A}ctual \textbf{R}isk \textbf{A}ssessment \textbf{I}nstruments\grq \footnote{ zu deutsch: Bewertungsinstrumente für das tatsächliche Risiko} , meint die US-amerikanische Strafjustiz die Wahrscheinlichkeit einer potenziellen Rückfälligkeit bei Gewalttaten genauer  ermitteln und eine vorurteilslose und neutrale Rechtsprechung erreichen zu können, deren Urteile frei von subjektiven und eventuell verzerrenden Einflüssen gefällt werden. Daher  benutzt das US-Justizsystem in 9 US-Bundesstaaten seit Jahren Risikobewertungs-Tools in verschiedenen Bereichen der Rechtsprechung~\cite{Angwin2016}, wobei von den Behörden mancher US-Staaten, wie zum Beispiel Florida, hauptsächlich das von der US-Firma Northpointe Ende der 1990-er Jahre entwickelte COMPAS Assessment Tool (\textit{\textbf{C}orrectional \textbf{O}ffender \textbf{M}anagement \textbf{P}rofiling for \textbf{A}lternative \textbf{S}anctions}) zum Einsatz kommt~\cite{Northpointe2012}. Northpointe propagiert, mit Hilfe ihres Algorithmus ließe sich eine genaue Prognose der Rückfallwahrscheinlichkeit (predicted recidivism) eines Angeklagten erstellen~\cite{Brennan2009}, was gern geglaubt wird, zumal die Möglichkeit einer präzisen Bestimmung des zukünftigen Legalverhaltens eines Straftäters die Urteilsfindung wesentlich erleichtern würde.

\subsection{Der Umgang der deutschen Justiz mit ADM gestützten Risikoprognosen \label{admde}}

Anders als im US-amerikanischen Justizsystem gilt in der deutschen Rechtsprechung uneingeschränkt das \glq Individualisierungsgebot\grq~\cite{Maschke2008}, sodass hier individuelle Kriminalprognosen zwingend vorgeschrieben sind. Individuelle kriminalprognostische Gutachten durch entsprechende Sachverständige müssen bei strafrechtlichen Entscheidungen in diversen Fällen mit einbezogen werden, nicht nur bei schweren Delikten oder bei der Beurteilung der (vorzeitigen) Entlassung eines verurteilten Täters aus einer freiheitsentziehenden Maßregel (§ 67d StGB). Gesetzlich konkret eingefordert und explizit im Gesetzestext vermerkt, werden individuelle Kriminalprognosen bei Entscheidungsfindungsprozessen, die bei Fehlurteilen eventuell mit dem Schutz der Öffentlichkeit vor Gewalttaten kollidieren. Etwaige Fehlurteile könnten hier zur Gefährdung des Lebens Unschuldiger führen, was  gesellschaftlich hochbrisante Folgen haben könnte. Dementsprechend heißt es im § 57 StGB zur Aussetzung des Strafrestes bei einer befristeten Freiheitsstrafe ausdrücklich: 
\begin{center}
\textit{"`Bei der Entscheidung sind insbesondere die Persönlichkeit der verurteilten Person, ihr Vorleben, die Umstände ihrer Tat, das Gewicht des bei einem Rückfall bedrohten Rechtsguts, das Verhalten der verurteilten Person im Vollzug, ihre Lebensverhältnisse und die Wirkungen zu berücksichtigen, die von der Aussetzung für sie zu erwarten sind."'\footnote{§ 57 StGB}}
\end{center}

Aufgrund oft mangelhafter Qualität und des besonderen Stellenwerts kriminalprognostischer Gutachten in der deutschen Justiz erarbeitete 2006 eine interdisziplinäre Expertengruppe von Richtern, Bundesanwälten, forensischen Psychiatern und Psychologen einen Katalog  zu den "`Mindestanforderungen für Prognosegutachten"'~\cite{Boetticher2007}.
Es soll dem Sachverständigen das Erstellen der Gutachten erleichtern und legt dabei zentrale Kriterien fest: Das Gutachten hat sich "`methodischer Mittel, die dem aktuellen wissenschaftlichen Kenntnisstand"' entspringen zu bedienen, es muss "`transparent"' und "`nachvollziehbar"'~\cite{Boetticher2007} sein. 
Obwohl die  \glq Mindestanforderungen\grq~ ausdrücklich feststellen, dass im Gutachten "`eine Wahrscheinlichkeitsaussage über das künftige Legalverhalten des Verurteilten"'~\cite{Boetticher2007}zu treffen ist, auf deren Basis das Gericht die Rechtsfrage zu beantworten hat, erwartet das deutsche Straf-  und Prozessrecht von den prognostischen Gutachten ein derart hohes Maß an Bestimmtheit, dass "`von einem fast naiven Vertrauen in die Leistungsfähigkeit von Prognosen"'~\cite[S.97]{Albrecht2003} gesprochen werden kann und angesichts der vielfach nachgewiesenen Fehleranfälligkeit "`ganz offensichtlich unrealistische Erwartungen an die Leistungsfähigkeit der Gefährlichkeitsprognose"'~\cite[S.122]{Albrecht2003} hat.\\
Die zunehmenden Erwartungen in die Prognosefähigkeit wird durch die Änderung des Strafgesetzbuchs § 67d "`Dauer der Unterbringung"' am 31.Januar 1998 deutlich. Hier änderte die Justiz die "`Anforderungen an die Sicherheit der Prognose (...) bei der Entlassung aus dem Maßregelvollzug"'~\cite[S.121]{Albrecht2003} wie in Tabelle \ref{67d} gezeigt. 

\begin{table}[]
\centering

\begin{tabular}{|c|c|}
\hline
	16. März 1994 – 31. Januar 1998& 31. Januar 1998 – heute\\ \hline
\begin{tabular}[c]{@{}c@{}}sobald \textbf{verantwortet werden } \\  \textbf{kann zu erproben}, \\ ob der Untergebrachte außerhalb\\  des Maßregelvollzugs \\ keine rechtswidrige Taten \\ mehr begehen wird\end{tabular} & \begin{tabular}[c]{@{}c@{}}\textbf{wenn zu erwarten ist},\\  dass der Untergebrachte außerhalb\\  des Maßregelvollzugs\\  keine rechtswidrigen Taten\\  mehr begehen wird\end{tabular} \\ \hline
\end{tabular}
\caption{Teile der Veränderung des § 67d StGB \glq Dauer der Unterbringung\grq~am 31.Januar 1998}
\label{67d}
\end{table}

Auch wenn die deutschen Gerichte noch keine ADM-Risikoprognosen~\cite[S.7]{Lischka2017} nutzen, muss im Hinblick auf die Aufmerksamkeitsökonomie sowohl der Richter als auch Sachverständigen bedacht werden, dass die benötigten Gutachten zwar noch von menschlichen Experten verfasst und prognostiziert werden. Sollte jedoch eine Firma aus dem amerikanischen Raum die deutsche Justiz mit Argumenten der Effizienz und angeblicher Objektivität überzeugen können, ihre ADM-gestützten Prognose-Instrumente einzusetzen, kann sich diese Tatsache schnell ändern. Insofern könnten die im folgenden Abschnitt umrissenen Fragestellungen in nicht allzu ferner Zukunft auch in Deutschland von hoher Brisanz sein. 

\section{Ziele der Arbeit}
Wenn sich eine Gesellschaft in einem derart relevanten Bereich wie der Justiz in ihrer Urteilsfindung auf algorithmenbasierte Risikoprognosen verlässt, so muss deren Evaluation ein zentrales Anliegen sein. 
Nach einer Überführung der Kriminalprognose in das mathematische Modell eines binären Klassifikators werden in dieser Arbeit dessen generelle Überprüfbarkeit statistisch ausgelotet und klassische Evaluationsstrategien vorgestellt (Kapitel \ref{def}). \\
Die vorliegende Arbeit verfolgt daraufhin das Ziel, den aktuellen Stand um die Beurteilung von ADM-Prozessen im kriminalprognostischen Bereich zu erfassen, wobei zuvor ein historischer Abriss über die Entwicklung eben dieser Instrumente gegeben wird (Kapitel \ref{kap_hist}). Das Qualitätsmaß, welches sich herauskristallisiert, ist die sogenannte  "`Area under the Reciever Operating Characteristic"' ($AUC$). Diese ist ein im maschinellen Lernen häufig verwendetes Maß, das ohne ersichtlichen Grund immer weitere Anwendung findet.\\
Da sich herausstellt, dass der "`Positive Predictive Value"' ($PPV_k$), als weitere Möglichkeit, einen solchen Klassifikator zu bewerten, dem richterlichen Entscheidungsprozess deutlich gerechter wird, nimmt sich die Arbeit dann der Frage an, wie weit diese beiden Werte für einen Klassifikator voneinander abweichen können (Kapitel \ref{kap_bez_auc_ppv}). \\
Abschließend wird die hierbei aufgezeigte Diskrepanz zwischen den beiden Qualitätsmaßen an einem realen Datensatz aus den USA für das sogenannte COMPAS-Tool, ein kriminalprognostisches Instrument der vierten Generation, nachgewiesen (Kapitel \ref{ergebnisse}), mögliche gesellschaftliche Folgen angerissen sowie Handlungsempfehlungen verfasst (Kapitel \ref{fazit}).

\chapter{Binäre Klassifikatoren und ihre Beurteilung\label{def}}
Für die anstehende Überprüfung der kriminalprognostischen Instrumente ist es notwendig, einige Fachbegriffe sowie deren Herkunft zu klären. Viele Fragestellungen, die mit Tools aus der Kriminalprognose beantwortet werden, führen zu dualen Entscheidungen. Zwar bauen die Algorithmen des „maschinellen Lernens“ mehr oder minder komplexe Entscheidungsstrukturen auf, aber die Frage, die den Richter beschäftigt und für welche er das Instrument zu Rate zieht, sind binärer Natur, wie zum Beispiel: \textit{Wird der Straftäter X rückfällig oder nicht, begeht Person Y ein weiteres Gewaltverbrechen --- ja oder nein?} Somit münden die hier Anwendung findenden ADMs in Entscheidungsoperatoren, welche zwischen zwei Klassen entscheiden können. Folglich werden sie im Zuge dieser Arbeit als binäre Klassifikatoren bezeichnet.\\
Das vorliegende Kapitel beabsichtigt zunächst, die hier zugrunde gelegte Definition binärer Klassifikatoren wiederzugeben. Im Anschluss wird erläutert, wie diese evaluiert werden, also wie entschieden werden kann, wie gut sie ihrer Aufgabe gerecht werden.

\section{Definition \glq binärer Klassifikator\grq \label{def_klas}}

Bei dem Begriff 'binärer Klassifikator' greift diese Arbeit auf eine Umsetzung zurück, die sich an folgender Definition orientiert: 
\begin{center}
\textit{Ein Klassifikator ist eine Zuordnungsvorschrift $x \rightarrow f(x)$ mit $x \in \mathbb{R}^N$, dem Informations- bzw. Merkmalsvektor und $f(x) \in \mathbb{Z}$ dem Klassifikationsergebnis~\cite{Hengen2004}.}\\
\end{center}
Ein Klassifikator ordnet also Objekte oder Datenpunkte anhand bestimmter Merkmale verschiedenen Eigenschaftsgruppen zu, den sogenannten Klassen. Aufteilungen in genau zwei Klassen werden somit durch einen binären Klassifikator umgesetzt, der für ein gegebenes Element entscheidet, ob es zu Klasse 1 oder Klasse 2 gehört.\\
Eine in dieser Arbeit betrachtete Umsetzung $C(E,f,S_f,\tau)$ ermittelt für eine gegebene Menge E mit $n :=|E|$ Elementen, ob ein Element zur ersten Klasse $(class~1)$ oder zur zweiten Klasse $(class~2)$ gehört. Sie besteht aus einer reellwertigen Bewertungsfunktion $f: E\rightarrow \mathbb R_0^+$, auch Scoringfunktion genannt, um Elemente einem Wert zuzuordnen, welche die Zugehörigkeit zu Klasse 1 widerspiegeln soll, einer bijektiven Sortierfunktion $S_f: E\rightarrow \{1,2,…,n\}$ und einem Parameter $\tau$. \\
Nachdem alle Elemente durch die Scoringfunktion bewertet wurden, werden die Elemente in einer Liste durch die Sortierfunktion $S_f$ absteigend sortiert, sodass also für $e_i,e_j \in E$, für die $f(e_i) < f(e_j)$ gilt, auch $S(e_i)<  S(e_j)$ gilt. Anhand des mitgelieferten $\tau$ entscheidet nun der Klassifikator, welche Elemente eine Bewertung größer und kleiner $\tau$ erhalten haben und unterteilt die Liste in genau zwei Bereiche. Elemente ($\ge\tau$) erhalten das Label \glq $(class~1)$ \grq~und ($<\tau$) \glq $(class~2)$\grq . \\
Die Vorhersage eines binären Klassifikators ist zwar im Einzelfall eine binäre Entscheidung, um jedoch Aussagen über die Qualität oder Güte treffen zu können, bedarf es einer statistischen Auswertung. Murphy und Winkler  haben hierzu in ihrem Artikel "`\textit{A general framework for forecast verification}"' \cite{Murphy1987} gezeigt, dass sich die Wahrscheinlichkeitsverteilungen in Bezug auf Prognose und Beobachtung als sinnvoll erweisen. Meist wird hierzu eine Menge von Elementen, für welche die Klassenzugehörigkeit a priori gegeben ist (Ground-Truth), durch den binären Klassifikator klassifiziert und anhand verschiedener Indizes Aussagen über die Qualität der Vorhersage getroffen.\\
 Um das Thema der Rückfälligkeitsvorhersage aufzugreifen, würde man  einen Algorithmus, der eine Vorhersage über die Zuordnung zu der Klasse "`hohes Rückfallrisiko"' bzw. "`niedriges Rückfallrisiko"' trifft, mit einer Ground-Truth überprüfen. Da dies in diesem Fall ein historischer Datensatz wäre, von dem bekannt ist, welche Personen rückfällig geworden sind und welche nicht, könnte man bewerten, wie gut die getroffenen Prognosen zutrafen. 
\section{Einführung in die statistischen Gütekriterien}

\begin{figure}[htb]
 \centering
 \includegraphics[width=0.5\textwidth,angle=0]{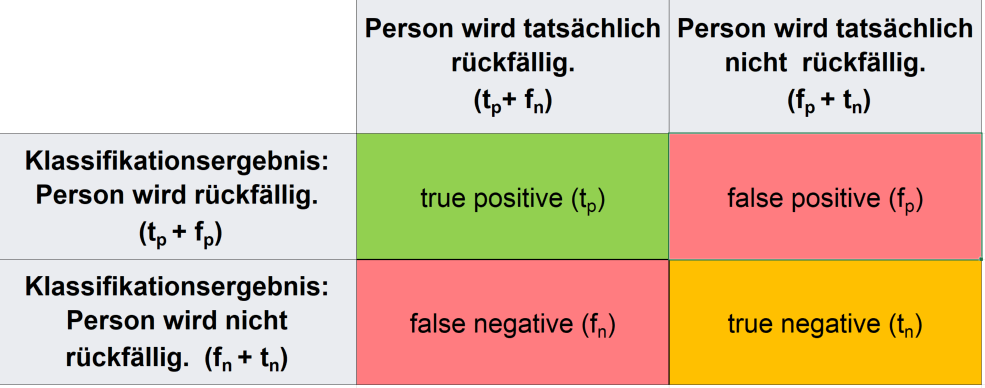}
 \caption[konfusionsmatrix a]{Konfusionsmatrix des vorgestellten binären Klassifikators}
\label{fig:konfusionsmatrix}
\end{figure}

An einem binären Klassifikator zur Vorhersage einer individuellen Rückfälligkeitswahrscheinlichkeit, welcher in die zwei Klassen "`hohes Rückfallrisiko"' $(class~1)$ und "`niedriges Rückfallrisiko"' $(class~2)$ aufteilt, ließen sich unter Zuhilfenahme einer Ground-Truth-Menge folgende vier verschiedenen Fälle identifizieren. Dem medizinischen Milieu entspringend (vgl. \cite[S.10]{Potapov2012}) haben sich folgende Fachbegriffe etabliert:
\begin{itemize}
	\item \textbf{true positive} ($t_p$): Die Person wird als potenziell rückfällig eingeschätzt und ist rückfällig geworden. 
	\item \textbf{false positive} ($f_p$): Die Person wird als potenziell rückfällig eingeschätzt und ist nicht rückfällig geworden. 
	\item	\textbf{false negative} ($f_n$): Die Person wird als nicht rückfällig eingeschätzt und ist rückfällig geworden. 
	\item	\textbf{true negative} ($t_n$): Die Person wird als nicht rückfällig eingeschätzt und ist nicht rückfällig geworden. 
\end{itemize}
Die entstehenden relativen Häufigkeiten werden meist, wie in Abbildung \ref{fig:konfusionsmatrix} gezeigt, in einer Wahrheitsmatrix, auch Konfusionsmatrix genannt~\cite{Bradley1997}, abgebildet. 

Diese relativen Häufigkeiten bilden zwar schon erste Kennzahlen für die Güte eines binären Klassifikators, jedoch werden sie meist noch in Wahrscheinlichkeiten umgerechnet. Um nun erweiterte Fragestellungen zu beantworten, bedarf es komplexerer statistischer Kennziffern. Diese können zum Beispiel intuitiv über die Größe der gesamten Stichprobe ($n=|E|$, Grundgesamtheit) normalisiert werden, woraus sich unter anderem die folgenden Kennziffern ergeben~\cite{Bradley1997}: 
\begin{itemize}
	\item \textbf{Accuracy} (Genauigkeit),  als der Anteil der korrekt kategorisierten Elemente an der Grundgesamtheit $(\frac{t_p+t_n}{n})$
	\item \textbf{Error rate} (Fehlerrate), als der Anteil der falsch kategorisierten Elemente an der Grundgesamtheit $(\frac{f_p+f_n}{n})$
	\item \textbf{Prevalence} (Prävalenz), als der Anteil der als positiv kategorisierten Elemente an der Grundgesamtheit $(\frac{t_p+f_p}{n})$
\end{itemize}
Da vor allem bei unbalancierten Klassen, also $|class~1|<<|class~2|$ oder $|class~1|>>|class~2|$, die Grundgesamtheit als Normalisierungsfaktor nicht immer sinnvoll ist, werden die verschiedenen Anteile jeweils auch durch die Anzahl der positiven/negativen Testergebnisse sowie Klassengrößen der Ground-Truth geteilt, um so, den Anwendungsfragen entsprechend, Aussagen über diese speziellen Anteile treffen zu können, wie im Folgenden anhand einiger klassischer Maße aufgezeigt wird. Ein sehr bekanntes Begriffspaar aus diesen Reihen, die \glq \textit{Sensitivität}\grq~und \glq \textit{Spezifität}\grq,~werden im folgenden Abschnitt näher beleuchtet.

\section{Sensitivität und Spezifität\label{sens_spez}}
Da sich die historischen Ursprünge der \glq \textit{Sensitivität}\grq ~sowie der \glq \textit{Spezifität}\grq~nicht mehr eindeutig bestimmen lassen, wird in dieser Arbeit die folgende konsensuale Definition verwendet, wobei die verwendete Notation sowohl in der Biologie als auch der Informatik im Bereich des maschinellen Lernens geläufig ist~\cite{Mohabatkar2013}.
Bei der Sensitivität wird die sogenannte \glq \textbf{R}ichtig-\textbf{P}ositiv-\textbf{
R}ate\grq~(RPR) gemessen, also der Anteil korrekt positiv kategorisierter Elemente ($t_p$) an der Menge aller tatsächlichen ($class~1$) Elemente ($t_p+f_n$)~\cite{Altman1994,Bradley1997}, die sich dem Beispiel entsprechend, als die folgende bedingte Wahrscheinlichkeit darstellen lässt: 
$$
Sensitivität = P(\textrm{Klassifikator~sagt~rückfällig}|\textrm{Person~wird~rückfällig})=\frac{t_p}{t_p+f_n}
$$
Die Spezifität dagegen wird auch als  \glq \textbf{F}alsch-\textbf{N}egativ-\textbf{R}ate \grq~(FNR) bezeichnet. Da hier die fälschlicherweise positiv kategorisierten Elemente ($f_p$) gegen alle ($class~2$) Elemente ($t_n+f_p$)~\cite{Altman1994, Bradley1997} abgewägt werden, lässt sich die Spezifität erneut dem Beispiel entsprechend als folgende bedingte Wahrscheinlichkeit darstellen: 
$$
Spezifität= P(\textrm{Klassifikator~sagt~rückfällig}|\textrm{Person~wird~nicht~rückfällig})=\frac{f_p}{t_n+f_p}
$$
Die Auswertung der Sensitivität und Spezifität gilt vor allem in der Medizin als wichtiges Maß zur Bewertung von Laborergebnissen. Honest und Kahn zeigten 2002, dass 72 \% der Autoren in den 90 analysierten Primärstudien aus den Jahren 1994-2000 auf eine Auswertung der Sensitivität und Spezifität als Gütekriterium ihrer Arbeit zurückgegriffen haben~\cite{Honest2002}.
Die Verbindung zur Informatik zieht in diesem Zusammenhang das Forschungsgebiet \glq \textit{Information Retrieval}\grq~(IR), das sich mit der Informationsgewinnung aus großen Datensätzen (vgl. \cite{Rijsbergen1979, Singhal2001}) beschäftigt. Hier hat sich für den Begriff der Sensitivität der im folgenden Abschnitt betrachtete Recall etabliert.

\section{Precision und Recall\label{pre_rec}}

Bei der Analyse und Bewertung von Suchanfragen in einer großen Menge Dokumente hat sich im IR als grundlegendes Gütekriterium das Begriffspaar \glq \textit{Precision}\grq\footnote{Genauigkeit oder positiver Vorhersagewert}~und \glq\textit{Recall}\grq\footnote{Trefferquote}~herauskristallisiert. 
Der Anteil aller relevanten Dokumente im Suchergebnis ($t_p$) in Bezug zu allen selektierten Dokumenten einer Suchanfrage ($t_p+f_p$) wird als \textit{Precision} bezeichnet:
$$
Precision=\frac{t_p}{t_p+f_p}~\cite[S.268]{Manning1999}
$$
Setzt man den Anteil aller relevanten Dokumente im Suchergebnis ($t_p$) in Relation mit allen für die Suchanfrage interessanten Ergebnissen ($t_p+f_n$), erhält man den sogenannten Recall, der sich also durch folgende Formel berechnen lässt:
$$
Recall = \frac {t_p}{t_p+f_n}~\cite[S.269]{Manning1999} 
$$

\section{F1-Score\label{def_f1score}}
Da vor allem im \glq Natural Language Processing\grq~(NLP) und IR großer Wert auf ein eindimensionales Evaluationskriterium gelegt wird~\cite{Goutte2005}, hat sich das harmonische Mittel von \glq \textit{Precision}\grq~und \glq \textit{Recall}\grq~unter dem Begriff \glq F1-score\grq~als weitere Evaluationsmöglichkeit etabliert: 
$$F_1=\frac{2\cdot Precision\cdot Recall}{Precision+ Recall}~\cite[S.340]{Aggarwal2015}~\cite[S.269]{Manning1999}$$
Dieses kombinierte Maß bietet einen gleichwertigen Einfluss von \glq \textit{Precision}\grq~und \glq Recall\grq~auf den Evaluationsprozess. Dies ist wichtig, denn eine gute Suche sollte alle relevanten Dokumente finden (hoher \glq \textit{Recall}\grq) und im gefundenen Datensatz eine hohe Dichte an wichtigen Dokumenten aufweisen (hohe \glq \textit{Precision}\grq).
Sollte eine andere Gewichtung gewünscht sein, so sei an dieser Stelle auf den Ursprung des F1-score verwiesen, denn bereits 1979 führte van Rijsbergen das sogenannte "`Effektivitätsmaß E"'\footnote{"`\textit{E measure}"'~(vgl. \cite[S.269]{Manning1999}} mit   
$$
E = 1- F \textrm{~und~}F=\frac {1}{\alpha \cdot \frac{1}{Precision}+(1-\alpha)\cdot \frac{1}{Recall}}, \alpha \in [0,1]
$$ 
 ein, um eine freie Skalierung des Einflusses der beiden Maßzahlen zu schaffen. Sollte $\alpha = 0$ gewählt werden, verhält sich F wie der Recall und für $\alpha = 1$ entspricht es der Precision. Für $\alpha =0.5$ erhält man einen ausgewogenen Einfluss von Precision und Recall und die oben angeführte Formel für den $F1-score$.

\section{Area under the Receiver Operating Characteristic ($AUC$)\label{def_auc}}
Eine weitere Evaluierungsmöglichkeit für binäre Klassifikatoren resultiert aus der \glq Receiver Operating Characteristic\grq~ (in der Psychologie und Statistik  teilweise auch \glq Relative Operating Characteristic\grq~genannt, vgl. ~\cite{Beck1986}).\\ Hierbei werden für verschiedene Schwellenwerte $\tau$ eines binären Klassifikators und seiner reellwertigen Bewertungsfunktion $f$ die Sensitivität und Falsch-Positiv-Rate ($FPR = 1-Spezifität$) gegeneinander abgetragen, wobei die FPR als Abszisse und die Sensitivität als Ordinate gewählt wird~\cite[S.340f]{Aggarwal2015}\cite[S.2]{Bradley1997}\cite{Peterson1954}.
Die Betrachtung der sogenannten "`Area under the Receiver Operating Characteristic"' (im Folgenden AUC genannt) ist in der Medizin und anderen Wissenschaften weit verbreitet. Vor allem für Metaanalysen wiesen 2002 Honest und Kahn eine Nutzungsquote von 73~\%~\cite{Honest2002} dieser grafischen Auswertung der Sensitivität und Spezifität nach.\\

\begin{minipage}[t]{0.5\linewidth}
\includegraphics[width=\linewidth]{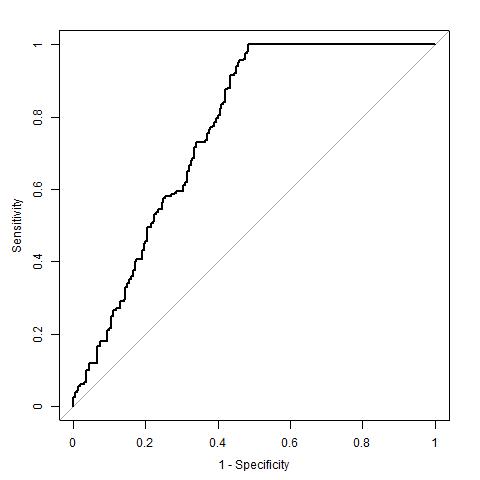}
\captionof{figure}{Darstellung der AUC\\für einen binären Klassifikator}
\label{ROCexample}
\end{minipage}
\begin{minipage}[t]{0.5\linewidth}
\includegraphics[width=\linewidth]{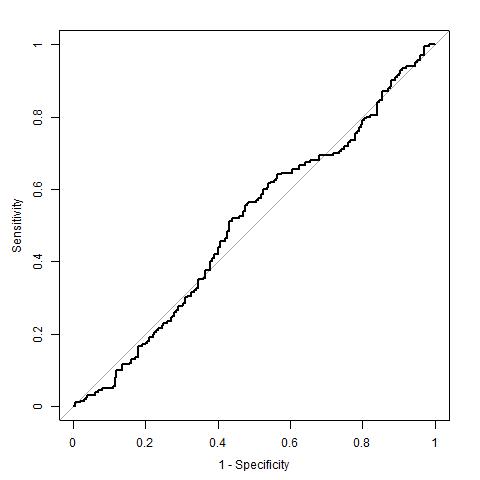}
\captionof{figure}{\label{ROCrandom} Darstellung der AUC eines zufallsnahen binären Klassifikator}
\end{minipage}
Hierbei wird die Fläche unter der ROC Kurve herangezogen, um so Aussagen über die Trennschärfe der Klassifizierung treffen zu können, also wie zuverlässig kann der Klassifikator zwischen Elementen der beiden Klassen unterscheiden.\\
Wie in Abbildung \ref{ROCexample} zu erkennen ist, wird häufig die Achsendiagonale eingezeichnet, da diese als Sinnbild für eine zufällige Klassifizierung als Maßstab genutzt wird, um die Qualität der abgetragenen Ergebnisse korrekt zu interpretieren. Sollte sich eine ROC-Kurve an dieser Diagonalen (vgl. Abbbildung \ref{ROCrandom}) orientieren, der Klassifikator also für verschiedene $\tau$ eine gleich große Trefferquote (Sensitivität) sowie  \textbf{F}alsch-\textbf{P}ositiv-\textbf{R}ate (FPR) aufweisen, kann er nicht besser separieren als eine zufällige Zuordnung, welche die Elemente mit gleicher Wahrscheinlichkeit einer der beiden Klassen zuweist. Dort wären Trefferquote und FPR ebenfalls gleich hoch.\\ Wenn die Kurve jedoch deutlich unter dieser Diagonalen liegt, lässt sich durch Umbenennen der beiden Klassen die Klassifizierung anpassen, weshalb die Diagonale die schlechteste Klassifizierung darstellt.\\
Bei der Betrachtung von binären Klassifikatoren lässt sich die AUC zusätzlich als die Wahrscheinlichkeit interpretieren, in einer zufälligen Stichprobe, bestehend aus je einem Element beider Klassen, dem Element der ersten Klasse eine höhere Wahrscheinlichkeit zuzuordnen zu dieser zu gehören~\cite[S.2]{Hanley1982}. 
Dieses Maß bildet also die Wahrscheinlichkeit ab, gegeben ein Element der Zielklasse ($class~1$) und ein Element der anderen Klasse ($class~2$), durch die Scoringfunktion $f$, des Klassifikators, dem ($class~1$) Element ein höheres Gewicht zuzuweisen als dem ($class~2$) Element. \\
Im Rahmen von Prognosen versteht man unter der Basisrate "`den theoretischen Anteil derjenigen Personen innerhalb der relevanten Population, für den das zu prognostizierende Ereignis eintreffen wird"'~\cite[S.426]{Bliesener2014}.
Als parameterloses Maß hat die AUC den Vorteil, nicht nur für eine spezielle Basisrate, also fixierte Klassengröße, eine Aussage über die Güte der Klassifizierung treffen zu können.

\section{Positive Predictive Value  und der $PPV_k$ \label{def_ppv}}

Ein besonderes Augenmerk wird in dieser Arbeit auf den sogenannten "`\textit{positive predictive value}"' (kurz: $PPV$) gelegt, welcher die korrekt positiv ($t_p$) klassifizierten Elemente mit der Anzahl aller positiv kategorisierten Elemente ($t_p+f_p$) normalisiert~(vgl. \cite[S.340]{Aggarwal2015}):
$$
PPV=\frac{t_p}{t_p+f_p} ~~\cite{Mohabatkar2013, Beigi2011}
$$
Dieses Maß findet hauptsächlich im medizinischen Umfeld Anwendung und wird dort in der Evaluation neuer Therapien, Medikamenten und Verfahren genutzt. Als bekanntes Beispiel sei hier nur die Brustkrebsvorsorge (im Speziellen der Mammographie) angeführt, welche maßgeblich über dieses Qualitätskriterium evaluiert wird~\cite{Kerlikowske1993, Burrell1996}.\\
Der PPV wird in den unterschiedlichsten informatiknahen Bereichen, wie zum Beispiel als Precision im Information Retrieval (vgl. Kapitel \ref{pre_rec}), angewendet und stellt ein Gütekriterium dar, das den Anteil an korrekten Ergebnissen zu allen zurückgelieferten ins Verhältnis setzt, hierfür jedoch die Basisrate, also die jeweiligen Klassengröße benötigt.  
Um einen binären Klassifikator mit diesem Qualitätsmaß zu bewerten, muss das $\tau$ passend gewählt sein. 
Wenn der Klassifikator $\tau = |class~1| = k$ wählt, wird über den $PPV_k$ überprüft, welcher Anteil an $(class~1)$ Elementen in den ersten k Elementen, der aus dem Klassifikator resultierenden Liste liegen. So kann die in der Ground-Truth vorliegende Basisrate an ($class~1$) Elementen genutzt werden, um das benötigte Verhältnis zu errechnen. In der Netzwerkanalyse zum Beispiel wird versucht, das \glq link prediction problem\grq~für Klassifikatoren zu evaluieren, bei denen das $\tau$ als Größe der Klasse 1 gewählt wurde~\cite{Liben-Nowell2003, Horvat2012}.\\
Auf den generellen Transfer dieser Methodik in den Bereich des maschinellen Lernens wird an dieser Stelle nicht näher eingegangen, jedoch bietet der $PPV_k$ eine Möglichkeit, den Fokus auf die tatsächliche Anzahl korrekt klassifizierter Objekte im vorderen Bereich der Sortierung zu legen und dabei fast vollständig auf eine Betrachtung der genauen Sortierung der Elemente zu verzichten. Einzig und allein die Eigenschaft "`ist unter den ersten k"' ist hier relevant, weshalb er sich von seiner Bewertungsnatur besser an dem Entscheidungsprozess eines Richters orientiert. Dies wird in Kapitel \ref{kap_bez_auc_ppv} genauer betrachtet.

\chapter{Entwicklung der standardisierten Prognoseinstrumente zur Rückfälligkeitsanalyse und des \grq State of the Art\grq~ihrer Beurteilung\label{kap_hist}}
Das vorliegende Kapitel beabsichtigt zunächst, die drei unterschiedlichen Methoden vorzustellen, derer sich ein Prognostiker grundsätzlich bedienen kann. Auf ein Abwägen der jeweiligen Vor- und Nachteile wird hier verzichtet, da der lange geführte \glq Methodenstreit\grq~heutzutage kaum noch von Relevanz ist.  
Anschließend soll ein kurzer Abriss einen knappen Überblick über die historische Entwicklung der standardisierten Prognoseinstrumente zur Rückfälligkeitsanalyse geben.
Eine abschließende Betrachtung der als \glq State of the Art\grq~geltenden $AUC$ in Bezug auf die Beurteilung von ADM-gestützten Prognoseinstrumenten rundet die historische Betrachtung in diesem Feld ab. 

\section{Methoden der Kriminalprognose \label{hist_krim}}
Seit der Veröffentlichung von "`Clinical versus statistical prediction"'~\cite{Meehl1954} von Paul Everett  Meehl im Jahre 1954 differenziert die einschlägige Fachliteratur üblicherweise drei Strategien kriminalprognostischer Herangehensweisen, die jedoch in der beschriebenen idealtypischen Ausprägung in der Realität nicht zu finden sind.
Da sie sich in der Praxis stark überschneiden~\cite{Dahle2006}, ist die folgende folgende Unterscheidung "`allenfalls ein Ordnungsversuch"'~\cite{Kroeber1999}. \newpage
\textbf{Die intuitive Methode}\\\\
Diese Methode ist zwar das wissenschaftlich am wenigsten fundierte, in der Praxis aber sicherlich am häufigsten benutzte Prognoseverfahren, da wohl die meisten Gerichtsurteile nach der ‚intuitiven Methode‘ des jeweiligen Richters gefällt werden und die Qualität des Urteils von der richterlichen Erfahrung und Fachkompetenz abhängt. Das prognostische Urteil basiert auf keiner expliziten Theorie, sondern auf gefühlsmäßiger Erfassung und globaler Eindrucksbildung vom Beschuldigten. Somit ist es weder überprüfbar noch transparent, sodass man im Prinzip von keiner Methode sprechen kann. Volckart bezeichnet diese Art des Vorgehens daher auch treffend als Gegenteil einer Methode~\cite{Volckart1997}.\\\\
\textbf{Die statistische Methode}\\\\
Die Entwicklung der statistischen Methode war die logische Konsequenz der intuitiven, sodass die Prognose nun auf einer vollständig regelgeleiteten Beurteilung einer Person basiert. Durch empirisch-statistische Vergleiche Rückfälliger und nicht Rückfälliger wurden einige Risikofaktoren ermittelt, die Durchschnittszusammenhänge zwischen Risikoprofil und späterer Legalbewährung abbilden und in Prognosetafeln zusammengefasst wurden. Im konkret zu beurteilenden Einzelfall wird der Beschuldigte dann einer Risikogruppe von mehr oder weniger Gefährlichen zugeordnet, sodass die Prognose quasi auf der Rückfallquote dieser Gruppe basiert, die aus der Konstitutionsstichprobe bekannt ist (vgl.~\cite[S.17]{Doebele2013}.\\
Die entscheidende Schwäche dieser Methode liegt im statischen Persönlichkeitskonzept, welches mögliche Wandlungen des Charakters in der Prognose nicht berücksichtigt. Ein weiterer Vorwurf bezieht sich auf die fehlende theoretische Basis~\cite[S.21]{Rettenberger2013}.\\\\
\textbf{Die klinische Methode}\\\\
Ziel der klinischen Prognose ist es, "`ein Erklärungskonzept für die Handlungen des 
Betreffenden und deren jeweilige Voraussetzungen und Bedingungen zu gewinnen, das auf der Grundlage theoretisch fundierter und empirisch abgesicherter Zusammenhänge fußt"'~\cite{Dahle1997}. \\
Der grundlegende Unterschied zur statistischen Methode besteht in der vom Einzelfall abhängigen Gewichtung der Faktoren. Das Urteil fußt insofern auf keinem festen Algorithmus. \\\\
Heutzutage favorisiert man bei prognostischen Stellungnahmen eine Verbindung der statistischen und klinischen Prognose. Man versucht, die jeweiligen Vorteile möglichst aufzugreifen und die beiden Methoden ohne die spezifischen Nachteile zu kombinieren. Nedopil~\cite[S.196]{Nedopil2005} rät, bei einem Prognosegutachten auf keine der drei Prognosemethoden zu verzichten. 
\section{Vorstellung der vier Generationen standardisierter Prognoseinstrumente zur Rückfälligkeitsanalyse}
 2008 zählte Guy~\cite{Guy2008}, dass in den vorhergehenden 50 Jahren 457 Verfahren zum \glq risk assessment\grq~(Risikobeurteilung) entwickelt wurden. Angesichts der Fülle von Prognoseinstrumenten über den langen Zeitraum wird in der Fachliteratur häufig eine Einordnung der Verfahren in verschiedene \glq Generationen\grq~vorgenommen, die allerdings von Autor zu Autor differiert~\cite[S.20-26]{Doebele2013}\cite[S.21f]{Rettenberger2013}.\\
Für die ersten drei Generationen wird der ausführlichen Darstellung von A.-L. Döbele~\cite[S.20-26]{Doebele2013}   gefolgt, für die von Döbele nicht behandelte vierte Generation den Ausführungen von Rettenberger et al.~\cite[S.21f]{Rettenberger2013}.\\\\
\textbf{Instrumente der ersten Generation}\\\\
Wie eingangs bereits erwähnt, wurden die ersten Prognoseinstrumente im Bestreben, die Kriminalprognose auf eine wissenschaftlich überprüfbare Basis zu stellen, Anfang des 20. Jahrhunderts in den USA entwickelt. Im Jahr 1928 veröffentlichte Ernest W. Burgess~\cite{Burgess1928} eine der ersten Prognosetafeln, die jahrelang das am häufigsten gebrauchte Prognoseinstrument der USA war~\cite{Schneider1967}.\\
Aus Akten von 3.000 entlassenen Straftätern extrahierte er 21  rückfallrelevante Faktoren. Im Gegensatz zu heutigen Prognoseverfahren verteilte Burgess ‚Gutpunkte‘ für jede positive Abweichung von der durchschnittlichen Rückfallerwartung, sodass seine Tafel nicht der Vorhersage der Rückfälligkeit dient, sondern die Wahrscheinlichkeit angibt, nicht rückfällig zu werden. Diese Art der Prognostik setze sich jedoch nicht durch.\\
1936 wurde die erste Prognosetafel in Deutschland von Robert Schiedt~\cite{Schiedt1936} entwickelt. Er extrahierte 15 rückfallrelevante Faktoren aus einer Gruppe von 500 Häftlingen. Für jedes vorliegende Merkmal wurde jeweils ein \glq Schlechtpunkt\grq~ohne weitere Gewichtung vergeben. Je höher die Punktzahl, desto größer war das Rückfallrisiko, was Schiedt durch Prozentpunkte anzeigte.  Die Prognosetafel von Fritz Meyer~\cite{Meyer1959} wurde 1959 entwickelt und in Deutschland noch bis 1990 verwendet~\cite[S.24]{Doebele2013}.\\
Die prozentuale Berechnung der Rückfallwahrscheinlichkeit durch die Anzahl der \glq Schlechtpunkte\grq~ist auch heute noch ein gebräuchliches Verfahren~\cite[S.25]{Doebele2013}, während die Liste der hier noch ausschließlich statischen und unveränderbaren Merkmale von der folgenden Generation erweitert wurde. \\\\
\textbf{Instrumente der zweiten Generation}\\\\
Bei diesen Instrumenten wurden die Merkmale um anamnestische Daten erweitert und personen- sowie tatbezogene Merkmale einbezogen. Die empirisch ermittelten Rückfallprädiktoren wurden in eine Prognoseformel übersetzt, die mittels der Akten des Straftäters eine schnelle und einfacher Handhabung des Instruments erlaubt, etwaige Wandlungen der Täterpersönlichkeit hingegen nicht erfasst.\\
Der von dem Kanadier Karl Hanson und dem Briten David Thornton 1999 entwickelte \glq Static 99\grq~gilt als Instrument der zweiten Generation und wurde speziell zur empirischen Ermittlung der Rückfallwahrscheinlichkeit von Sexualstraftätern entwickelt. Es zählt in dem Bereich zu dem am häufigsten verwendeten Instrument zur Einschätzung des Rückfallrisikos (vgl.~\cite[S.49 f]{Doebele2013}. \\
Die Gründe hierfür liegen wahrscheinlich an der Benutzerfreundlichkeit und den geringen Anforderungen, die es an den Anwender stellt, da die erforderlichen Faktoren ohne psychologisches Hintergrundwissen und ohne persönliches Gespräch mit dem Täter zu ermitteln sind. Dies führt allerdings zum Hauptkritikpunkt, nämlich dass ausschließlich statische, unveränderliche Fakten ermittelt werden, die etwaige Veränderungen des Risikopotenzials nicht erfassen. Dittmann fasst daher zutreffend zusammen, dass der Straftäter bei Anwendung der Instrumente dieser Generation "`zum Gefangenen seiner Biographie"'~\cite{Dittmann2003} wird.\\\\
\textbf{Instrumente der dritten Generation}\\\\
Wissenschaftliche Entwicklungen der 90er Jahre des letzten Jahrhunderts führte die Kriminalprognostik zur dritten Generation von Prognoseinstrumenten, indem zunehmend dynamische Merkmale herangezogen wurden. Die Datenbasis wurde so um grundsätzlich veränderbare Faktoren, wie zum Beispiel "`persönliche Einstellung der Straftäter, Persönlichkeitszüge, seine soziale Bindungen oder die Art seiner Freizeitgestaltung"'~\cite{Doebele2013} erweitert. Da hierbei meist bereits aufwändige Interviews mit einem ausgeklügelten Fragekatalog geführt werden müssen, ist die Tendenz in Richtung des klinischen Beurteilungskonzepts innerhalb der Instrumente erkennbar~\cite{Dahle2006}. \\
Das "`Historical-Clinical-Risk Managment-20"' (HCR-20) wurde 1995 von dem Kanadier \textit{Webster et al.}~\cite{Webster1997} auf der Grundlage eines nur 75 Personen umfassenden Datensatz erarbeitet, entwickelte sich aber nichtsdestotrotz bis heute zu einem der "`international bekanntesten Prognoseinstrumente"'~\cite{Doebele2013}. Während dieses Tool zwar auf 10 historische, nicht veränderbare Eigenschaften des Beurteilten zurückgreift, sind die restlichen 10 sogenannte Risikovariablen, die sich einer klinischen Beurteilung ähnlich aus dem persönlichen Umfeld und Verhalten der Person ergeben. Das Interesse der deutschen Justiz an solchen aktuarischen Prognoseinstrumenten wird deutlich an der zeitnahen Übersetzung ins Deutsche im Jahre 1998~\cite{Mueller-Isberner1998}.\\\\
\textbf{Instrumente der vierten Generation}\\\\
Instrumente der vierten Generation gehören zur letzten Stufe und stellen somit den aktuellsten Stand der Kriminalprognostik dar. Sie sehen ein breites Band an Einsatzmöglichkeiten für verschiedenste  Prognosebereiche vor. Zum einen fließen immer mehr variable Aspekte in den Beurteilungsprozess ein, zum anderen beschränken sich die Tools nicht mehr nur auf  Risikoprognosen von Verhaltensweisen, sondern bieten Empfehlungen für Therapiepläne und Therapieplatzvergabe an und können sogar Aussagen darüber machen, ob ein Straftäter vor Gericht erscheinen wird oder nicht (vgl. z.B. \cite{Northpointe2012}).\\
Eine Firma, die aktuell ein weitverbreitetes Vorhersageinstrument der vierten Generation entwickelt hat, ist die Firma Northpointe Inc. aus den USA. Sie hat 1998 begonnen, das COMPAS Assessment Tool (Correctional Offender Management Profiling for Alternative Sanctions) zu entwickeln, welches als eins der besten Systeme seiner Art gilt~\cite[S.8]{Andrews2006}.\\
Wie bei allen Instrumenten der vierten Generation stellt die geringe Transparenz der Algorithmen und die daraus resultierende fehlende Einsicht in den Bewertungsprozess auch beim  COMPAS Assessment Tool ein großes Problem dar. 
Mögliche Überprüfungsstrategien wären sogenannte \glq Blackbox\grq -Analysen, die eigentlich aus der Softwareentwicklung stammen~(vgl. \cite{Beizer1995}). Hierbei wird ohne Kenntnisse über die innere Funktionsweise  der Algorithmen versucht, die tatsächlichen Ergebnisse mit denen zu überprüfen, die erwartet werden würden. \\
Eine 2016 von ProPublica, einer durch Spenden finanzierte US-Rechercheorganisation, vorgelegte Studie wies für das COMPAS Assessment Tool eine dramatische Ungleichbehandlung von Schwarzen und Weißen nach~\cite{Angwin2016}.  Die Antwort aus dem Hause Northpointe~\cite{Dieterich2016} begründet die festgestellten Ergebnisse mit der Nutzung eines anderen Fairnesskriteriums und zeigt so den sehr weiten Modellierungsrahmen auf, dem solche Instrumente unterliegen. \\
Deshalb ist es wichtig, den abschließenden Bewertungsprozess dieser Algorithmen zu beleuchten, um einschätzen zu können, ob und wann ein solches ADM zur Beurteilung von Menschen, und dazu in einem für die Betroffenen essenziellen Bereich wie in der Justiz, eingesetzt werden sollte.  \\
Im folgenden Abschnitt wird daher der historische Werdegang der $AUC$ betrachtet, die im Allgemeinen als Bewertungsmaßstab solcher Klassifikatoren gilt.

\section{ State of the Art --- Wo und wie wird die $AUC$ verwendet\label{soa_auc}}

Eine der ersten Anwendungen der $AUC$ als Evaluationsmöglichkeit von binären Entscheidungen geht auf die Signalentdeckungstheorie von David M. Green und John A. Swets~\cite{Swets1964} zurück. Eine Ursache für den Beginn dieser Forschungen war der fatale Flugzeugangriff auf Pearl Harbor im Jahre 1941, bei dem sowohl auf maschineller als auch menschlicher Ebene grobe Fehlinterpretationen über das Vorhandensein von feindlichem Funkverkehr zu einer der größten amerikanischen Militärkatastrophen des 20. Jahrhunderts geführt haben ~\cite{Wohlstetter1962, McDermott2016}.
Um diese Problemstellung wissenschaftlich zu fassen, hat Swets die Signalentdeckungstheorie auf Basis folgender Definition formuliert:
\begin{center}
\textit{"`A diagnostic system looks for a perticular \glq signal\grq, however defined, and attempts to ignore or reject other events, which are called \glq noise\grq"'}\footnote{ Ein Diagnosesystem sucht nach einem spezifischen \glq Signal\grq, das irgendwie definiert ist und versucht andere Ereignisse zu ignorieren oder auszusortieren. Diese werden als \glq Störgeräusch\grq~bezeichnet.}\cite{Swets1988}.
\end{center}

In ihrem Buch \glq Signal Detection Theory and Psychophysics\grq~\cite{Swets1966} aus dem Jahr 1966 entwickelten Green und Swets unterschiedliche Experimente rund um die Detektion von Signalen in höchst verrauschten (Audio-)Daten. Zusätzlich beschäftigten sie sich mit Evaluationsmöglichkeiten dieser Problemstellung für Mensch und Maschine.
Eines dieser Experimente stellt Probanden vor die Frage, in welchem von zwei betrachteten Zeiträumen ein Stimulus auftaucht. Da hierbei eine binäre Entscheidung forciert wird, sprechen die Autoren von einer \glq two-alternative forced choice\grq~(2AFC)~\cite[S.44ff]{Swets1966}, welche sie mit der vorgestellten ROC $AUC$ evaluieren.\\
Die graphische Auswertung, wie in Kapitel \ref{def_auc} beschrieben, geht jedoch bereits auf einen Techreport aus dem Jahre 1954 \cite[S.176]{Peterson1954} zurück.

Die psychologisch-statistische Nutzungshistorie der $AUC$ wurde 1973 von Swets \cite{Swets1973} umfassend ausgearbeitet, weshalb an dieser Stelle nur kurz das Themengebiet des \glq \textit{operant conditioning}\grq \footnote{ Instrumentelle und operante Konditionierung}~\cite{Nevin1969} genannt wird, bei welchem durch positives Feedback eine von zwei möglichen Handlungsoptionen erlernt werden soll. Die Anzahl an korrekten Entscheidungen ist hier das übliche Maß für den Lernstatus des Probanden. Da diese Entscheidungsart binärer Natur ist (korrektes Verhalten $\Leftrightarrow$ falsches Verhalten), mündet die Evaluation in dieselben statistischen Kriterien und zieht die $AUC$ zu Rate.

Den zuweilen sehr erfolgreichen Transfer dieses Gütekriteriums in verschiedene Forschungsbereiche wies Swets in seinem Science-Artikel "`Measuring the Accuracy of Diagnostic Systems"'\cite{Swets1988} im Juni 1988 nach. Vereinzelte Nutzung konnte im Materialtest und in der Forschung um polygraphe Lügendetektoren festgestellt werden, auch in der Wettervorhersage und dem in Kapitel \ref{pre_rec} vorgestellten Information Retrieval hatte sich die $AUC$ als bevorzugtes Gütekriterium durchgesetzt. In den 90ern wurde von Beck und Shultz aufgezeigt, dass sich die Nutzung der $AUC$ ebenso in der Medizin etabliert hat und dort einen hohen Stellenwert bei Wirksamkeitsnachweisen und Überprüfung der Effektivität verschiedenster Medikamente einnimmt~\cite{Beck1986}.\\ Dennoch existieren in diesem Forschungsfeld auch kritische Ansichten, die sich unter anderem mit der Validität von Prognose und deren Bewertung beschäftigen~\cite{Leushuis2009}. Ansätze bezüglich der Anwendbarkeit der $AUC$ für unbalancierte Klassengrößen wurden zwar mit dem Argument, dass die AUC, selbst wenn die Klassengröße nicht bekannt ist, ein gutes Bewertungskriterium eines Entscheiders~\cite{Provost1998, Bradley1997} darstellt, vermeintlich beantwortet, jedoch zeigen die im Folgenden aufgezeigten Ergebnisse (vgl. Kapitel \ref{propub_analyse}) auf, dass (aktuellen) binäre Klassifikatoren vor allem in stark unbalancierten Klassen trotz hoher AUC eine schlechte Klassifikation ermöglicht wird. Hier fällt es dem Klassifikator letztendlich sehr schwer, die Scoring-Funktion $(f: E\rightarrow \mathbb R_0^+)$ sowie auch das $\tau$ so zu wählen, dass die Sensitivität ausreichend hoch ist, die Falsch-Positiv-Rate jedoch verträglich gering ausfällt.\\ 
Bei der Nutzung der AUC als Gütekriterium gibt es je nach Feld unterschiedliche Schwellwerte, ab denen eine Klassifizierung als angemessen anerkannt wird. In der Medizin folgt man so folgender Interpretation der $AUC$: 
\begin{center}
\textit{"`A model is considered to have poor performance if the AUC lies between 0.50 and 0.70. An AUC between 0.70 and 0.80 represents fair performance, and an AUC between 0.80 and 0.90 represents good performance}"'\footnote{ Einem Modell wird eine schlechte Performance zugesprochen, wenn die AUC zwischen 0.50 und 0.70 liegt. Eine AUC zwischen 0.70 und 0.80 repräsentiert eine faire Performance, eine $AUC$ zwischen 0.80 und 0.90 repräsentiert eine gute Performance}~\cite{Leushuis2009}.
\end{center}

Ab einer AUC von 0.80 wird von einer guten Performance, d. h. einem guten Klassifikationsergebnis gesprochen, was im Hinblick auf das Anwendungsgebiet 'Mensch' in der Humanmedizin als wichtiger Eckpunkt zu erfassen ist, denn in Bezug auf Rückfälligkeitsvorhersagen werden Algorithmen bereits mit einer deutlich niedrigeren AUC akzeptiert: 
 \begin{center}
 \textit{"`The lower half of acceptable AUC value ranges (0.65 to 0.75) reported in other criminal justice risk-classification studies."'}\footnote{ Die untere Hälfte des akzeptierten AUC-Wertebereichs (0.65 bis 0.75), wie es in anderen Strafjustiz-Risikoklassifikatoren-Studien berichtet wird.}  ~\cite[S.22]{Lansing2012}
 \end{center}

\begin{center}
\textit{"`In most of the studies, the area under the curve ($AUC$) ranged from .70 to .80
[...], which reflects a satisfactory to good prognostic accuracy"'}\footnote{ In den meisten Studien bewegt sich die $AUC$ im Bereich 0.70 bis 0.80, was eine zufriedenstellende bis gute Genauigkeit der Prognose widerspiegelt.}~\cite{Endrass2008}

\end{center}
Bereits 1977 wurden die ersten Rückfälligkeitsvorhersage-Statistiken mit einer $AUC$ bewertet \cite{Fergusson1977} und sie wird bis heute als wichtigste Kennzahl zur Bewertung dieser Klassifikatoren herangezogen, was in Kapitel \ref{compas} näher betrachtet wird. Es ist somit wichtig, den Grund für die Etablierung dieses Werkzeugs in durch maschinelles Lernen erarbeitete Analysetools zu verstehen.\\
Es kann nicht nachvollzogen werden, warum unterschiedliche Schwellenwerte in unterschiedlichen Disziplinen Anwendung finden und warum gerade bei kriminalprognostischen Instrumenten ein deutlich niedrigerer Wert als Garant für eine gute Klassifikation angesehen wird.\\
Im Jahre 1997 hat Andrew E. Bradle mit seinem Artikel "`The use of the area under the ROC curve in the evaluation
of machine learning algorithms"'~\cite{Bradley1997} in den verschiedenen Feldern des maschinellen Lernens die Möglichkeiten und Grenzen der $AUC$ statistisch austariert. Er kommt zu dem Schluss, dass gerade in Bereichen, in denen häufig verschiedenartige Klassifikatoren verglichen werden müssen, bekannte statistische Kennziffern keine valide Vergleichbarkeit schaffen. Er führt zum Beispiel an, dass das häufig angegebene Tupel aus Sensitivität und Spezifität für die Evaluation bei maschinellem Lernen keine Indikation bietet, wie sich ein veränderter Schwellenwert $\tau$ auf die Klassifizierung auswirkt~\cite[S.2]{Bradley1997} und  somit als Optimierungskriterien für das maschinelle Lernen nicht nutzbar sind. 
Im angesprochenen Artikel bestätigt er am Beispiel von sechs Anwendungsgebieten aus dem maschinellem Lernen\footnote{"`Multiscale Classifier, Perceptron, Multi-layer Perceptron, k-Nearest Neighbours, and a Quadratic Discriminant Function"'~\cite{Bradley1997}} folgende Eigenschaften für die $AUC$:
\begin{enumerate}
	\item Sie bietet die Möglichkeit, die Sensitivität in Bezug auf eine ANOVA (vgl. \cite[S.391]{Walpole1993}) zu steigern.
	\item Sie ist unabhängig vom gewählten $\tau$.
	\item Sie zeigt sich invariant zu a priori festgelegten Klassengrößen.
	\item Sie eignet sich als Indikator, wie gut die zwei Klassen getrennt werden konnten.
	\item Sie lässt zufällige Verteilungen oder Zuordnungen zu nur einer der beiden Klassen gut erkennen. 
\end{enumerate}
Auch wenn einige dieser Punkte bereits 1995 von Marnie E. Rice and Grant T. Harris thematisiert wurden \cite[S.738]{Rice1995}, stellt Bradley im Hinblick auf lernende Algorithmen fest:
\begin{center}
\textit{"`The $AUC$ [...] appears to be one of the best ways to evaluate a classifier's performance on a data set when a "'single number"' evaluation is required"'\footnote{ Die $AUC$ scheint einer der besten Möglichkeiten zu sein, die Performance einer Klassifizierung zu evaluieren, wenn nur eine einzelne Maßzahl gefordert ist.}\cite{Bradley1997} }
\end{center}
Dieser Fokus auf eine Kennzahl mag zwar für die Evaluation von Algorithmen des maschinellen Lernens benötigt werden um themen- und anwendungsübergreifende Vergleichbarkeit zu schaffen, jedoch wiesen Rice und Harris gerade im Spezialfall der Vorhersage von Gewalttaten auf den Umstand hin, dass viele Studien mit sehr wenigen Datenpunkten versuchten, die ROC-Kurve zu approximieren\footnote{ \textit{"`ROC methods cannot make up for data that have not been gathered. Most studies of the prediction of violence yield only one or two ROC data points. Reliable calculation of ROC effect size is impossible with so few points without making untenable parametric assumptions."'\cite[S.738]{Rice1995} }}, wodurch dem Maßstab die statistische Grundlage völlig entzogen wäre.\\ Zusätzlich steht dieses spezielle Forschungsfeld, wie bereits in der Einleitung erläutert, vor dem Problem der einseitigen Evaluation, denn schon in richterlichen Entscheidungen konnte kaum ein Straftäter, welcher für gefährlich gehalten wurde und dementsprechend im Gefängnis verblieb, seine potenzielle Unschuld beweisen.\\ Es ist also bedenklich, dass sich der $AUC$ als meist genutztes Bewertungskriterium für Risiko-Klassifikatoren~\cite{Lansing2012} herauskristallisiert und in der Betrachtung der Rückfälligkeit von Straftätern eine solch zentrale Rolle einnimmt: 
\textit{
\begin{center}
"`The best measure for determining how accurately a score predicts an event like recidivism is a statistic called the area under the receiver operating characteristic ([ROC] $AUC$)."' \footnote{Das beste Maß zur Feststellung, wie gut ein Ereignis durch eine Maßzahl vorhergesagt werden kann, wie zum Beispiel die Rückfälligkeit, ist die $AUC$.}\cite{Barnoski2007} 
\end{center}
}
Selbst wenn ein Klassifikator eine $AUC$ von 1.00 erreicht, könnte er jedem Rückfälligen eine Rückfallswahrscheinlichkeit von 10 \% und jedem nicht Rückfälligen eine von 9\% prognostizieren. Obwohl laut $AUC$ diese Klassifizierung eine perfekte Trennschärfe aufweist, ließe sich einerseits sehr schwer zwischen den beiden Klassen separieren, andererseits ist eine solch niedrige Rückfallswahrscheinlichkeit in keiner Art hilfreich, wenn die Basisrate höher liegt als die Prognose. \\
Diese Diskrepanz würde sich zwar auch bei einer Nutzung des $PPV_k$ widerspiegeln, jedoch bildet dieser, wie im folgenden Kapitel erläutert wird, den generellen Entscheidungsprozess eines Richters deutlich besser ab. Insofern ist die anschließende Betrachtung der Abweichung dieser beiden Maße von großer Relevanz, zumal aktuell Firmen lediglich eine hohe $AUC$~\cite{Northpointe2012} angeben, um ihre vermeintlich guten Klassifikatoren zu bewerben.

\chapter{Abweichung zwischen $AUC$ und $PPV_k$\label{kap_bez_auc_ppv}}
\begin{figure}[h!]
\centering
\includegraphics[width=\linewidth]{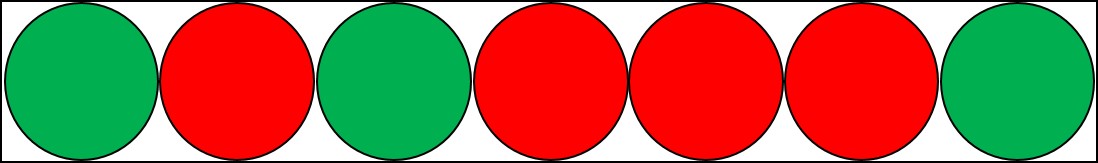}
\caption{\label{class1}Mögliche Sortierung der Elemente anhand der Scoringfunktion $(f: E\rightarrow \mathbb R_0^+)$ und der bijektiven Sortierfunktion $S_f: E\rightarrow \{1,2,…,n\}$ eines $(class~1)$-Klassifikators}
\end{figure}

Ausgehend von der aktuell hohen Nutzungsquote der AUC als Bewertungsmaßstab in der ADM-gestützten Kriminalprognose überprüft dieses Kapitel die Eignung der AUC bei binären Klassifikatoren für Rückfälligkeitsvorhersage und beabsichtigt mathematisch zu bestimmen, ob und wie weit sie vom $PPV_k$ abweichen kann, der deutlich näher am richterlichen Entscheidungsprozess evaluiert.\\
Der juristische Prozess, bei welchem einem Richter ein Delinquent vorgeführt wird, bildet keine \glq two-alternative forced choice\grq~(2AFC) (siehe Kapitel \ref{soa_auc}) ab, da in den seltensten Fällen ein Richter zwischen zwei Straffälligen entscheiden muss, von denen er zu 100\% weiß, dass genau einer rückfällig wird. Somit kann die Gültigkeit dieser Evaluationsmöglichkeit bereits in seinem Ursprung für dieses Anwendungsfeld angezweifelt werden.\\ Deutlich näher an der Fragestellung zeigt sich der $PPV_k$ (vgl. Kapitel \ref{def_ppv}), da dieser überprüft, wie viele rückfällige Delinquenten der Klassifikator in die ersten k seiner Sortierung bringen konnte.
In Kapitel \ref{soa_auc} wurde der sehr hohe Stellenwert der AUC für die Evaluation von binären Klassifikatoren bei der Rückfälligkeitsvorhersage beschrieben und hinterfragt. Diese soll nun mit dem $PPV_k$ (vgl. Kapitel \ref{def_ppv}) auf Korrelation überprüft werden, indem mathematisch ausgelotet wird, welchen Wert ein Klassifikator bei einem Maß minimal und maximal erreichen kann, wenn das andere fixiert bleibt.\\
Hierzu wird zunächst das Symmetrieverhalten der beiden Qualitätsmaße im Bezug auf ein Austauschen der betrachteten Klassen beleuchtet und im Anschluss ausführlich besprochen, wie sich die jeweiligen Extrema bestimmen lassen. \\
Zur Veranschaulichung wird im Folgenden ein Beispiel mit $K_1=3$ grünen $(class~1)$ und $K_2=4$ roten $(class~2)$ Elementen verwendet. Es kann zum Beispiel angenommen werden, dass die drei grünen Kugeln Straftäter sind, welche innerhalb eines bestimmten Zeitraums rückfällig geworden sind, während die vier roten jene sind, deren Legalverhalten rechtskonform verlief. Die Sortierung ergibt sich anhand der Zugehörigkeitswahrscheinlichkeit zu $(class~1)$, welche über die Scoringfunktion f des  $(class~1)$-Klassifikators ermittelt wird (vgl. Abbildung \ref{class1}). 
Wird das k als die im Datensatz vorliegende Klassengröße von $|class~1|=3$ gewählt, ergibt sich in Abbildung \ref{class1} für den $PPV_k=PPV_3=\frac{2}{3}\approx0.66$, da der Klassifikator in die ersten drei Elemente genau zwei $(class~1)$ Elemente sortieren konnte. Die AUC würde wie folgt evaluiert: Von allen 12 Paaren aus roten und grünen Kugeln sind genau 7 Paare korrekt sortiert, die grüne Kugel ist also vor der roten, weshalb eine AUC von $\frac{7}{12}\approx0.58$ vorliegt.
\newpage
\section {Symmetriebetrachtung \label{chap_Symmetrie}}

\begin{figure}[h!]
\centering
\includegraphics[width=\linewidth]{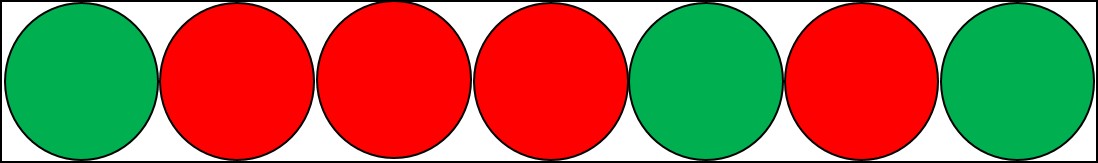}
\caption{\label{abb_Symmetrie}"'Umgekehrte"' Klassifizierung, diese entspricht einem $(class~2)$-Klassifikator}
\end{figure}

Da eine sehr schlechte binäre Klassifizierung durch einfaches \glq Umlabeln\grq~der beiden Klassen\footnote{ $(class~1)$ in $(class~2)$ umbenennen und vice versa} in eine gute bis sehr gute umgewandelt werden kann, ist es für die folgende Korrelationsbetrachtung wichtig, die beiden Qualitätsmaße bezüglich ihres Symmetrieverhaltens und des Austauschens der Klassennamen zu betrachten.
Die Ergebnisse der Scoringfunktion des Klassifikators werden hierfür invers interpretiert, denn eine $70\%$ige Wahrscheinlichkeit zu $(class~1)$ zu gehören, bedeutet bei einem binären Klassifikator gleichzeitig eine $30\%$ige ($100\%-70\%$) Wahrscheinlichkeit zu $(class~2)$ zu gehören. Ob hierfür die Scoringfunktion $(f'=1-f)$ oder die Sortierfunktion angepasst wird, ist irrelevant. Es ergibt sich also für vorheriges Beispiel (siehe Abbildung \ref{class1}) eine invers sortierte Liste, wie in Abbildung \ref{abb_Symmetrie} zu erkennen ist. 

\subsection{Symmetriebetrachtung der $AUC$ \label{sym_auc}}
\begin{figure}[h!]
\centering
\includegraphics[width=\linewidth]{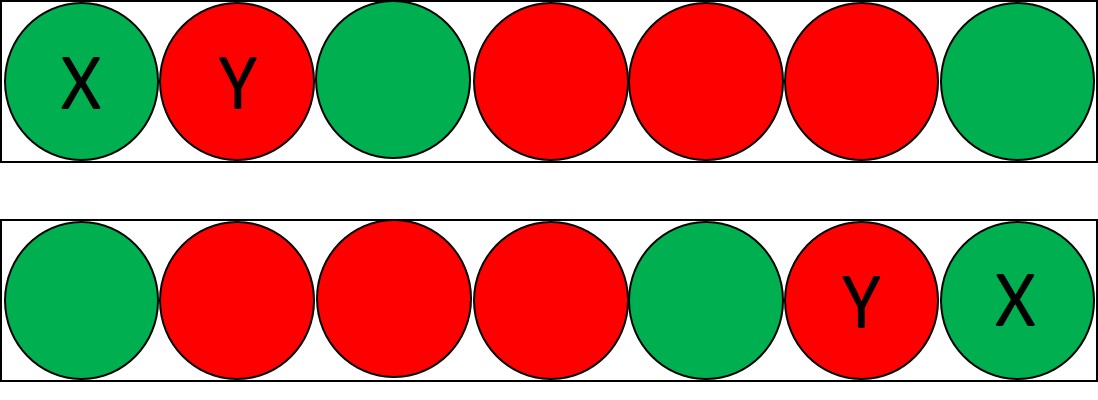}
\caption{\label{AUC_bsp}AUC: Betrachtung eines Paares bei einer Sortierung eines $(class~1)$-Klassifikator (oben) und einer Sortierung welche aus einem $(class~2)$-Klassifikator entspringt (unten)}
\end{figure}

Nach Definition (siehe Kapitel \ref{def_auc}) normalisiert die $AUC$ die Anzahl korrekt sortierter Paare [zwischen $(class~1)$ und $(class~2)$ Elementen] über die Summe aller Vergleiche. Im Folgenden wird von $AUC_1$ gesprochen, wenn sie sich auf einen $(class~1)$-Klassifikator bezieht, also zwei Elemente dann korrekt sortiert sind, wenn die Scoringfunktion f dem $(class~1)$ Element eine höhere Zugehörigkeitswahrscheinlichkeit zu $(class~1)$ zu gehören, gegeben hat als dem $(class~2)$ Element. Die entsprechende Sortierfunktion würde dieses Element folglich weiter nach vorne sortieren. Analoges gilt für die $AUC_2$ bei einer Bewertung eines $(class~2)$-Klassifikators, welcher nach oben beschriebener Konstruktion entstanden ist. Generell wird also eine vom Klassifikator stammende Sortierung bewertet.
Des Weiteren wird mit $\#AUC_1$/$\#AUC_2$ die absolute Anzahl an korrekt sortierten Paaren des jeweiligen Klassifikators bezeichnet; $s_f^1$ / $s_f^2$ sind die jeweiligen Sortierfunktionen.\\
Der $AUC_1$ für Abbildung \ref{class1} setzt sich im genannten Beispiel wie folgt zusammen:
$$
AUC_1=\frac{\#AUC_1}{|class~1|\cdot |class~2|} = \frac{7}{3\cdot 4}=\frac{7}{12}
$$
Bei der Berechnung von $\#AUC_2$ fallen bei der Betrachtung eines Paares folgende Eigenschaften auf (vgl. Abbildung \ref{AUC_bsp}): 

\begin{itemize}
	\item Die Sortierung aller Paare wird umgekehrt, was nach Konstruktion aus der umgekehrten Sortierfunktion oder angepassten Scoringfunktion resultiert.
	\item Der $\#AUC_2$ erfasst exakt alle Paare, die vor der Anpassung vom $\#AUC_1$ erfasst werden. Falls damals $s_f^1(x) > s_f^1(y)$ galt, so gilt nach der Anpassung $s_f^2(y) > s_f^2(x)$, womit stets $\#AUC_1=\#AUC_2$ gilt.
\end{itemize}

Sowohl für den $AUC_1$ als auch für den $AUC_2$ sind die Größen der betrachteten zwei Klassen identisch. Somit wird, dem Kommutativgesetz der Multiplikation folgend, für die Anzahl aller Möglichkeiten von Paaren derselbe Wert herausgezogen\footnote{ $|class~1|\cdot |class~2|=|class~2|\cdot |class~1|$}. Da sowohl Zähler als auch Nenner identisch sind, verändert sich der AUC für das beschriebene Symmetrieverhalten nicht und es gilt stets: 
$$
AUC_1=\frac{\#AUC_1}{|class~1|\cdot |class~2|}=\frac{\#AUC_2}{|class~2|\cdot |class~1|}=AUC_2
$$

\subsection{Symmetriebetrachtung des $PPV_k$\label{sym_ppv}}
\begin{figure}[h!]
\centering
\includegraphics[width=\linewidth]{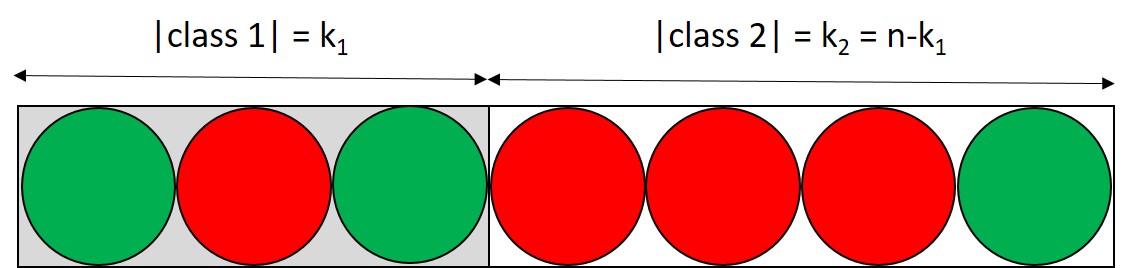}
\caption{\label{PPV_1}Visualisierung einer durch einen $(class~1)$-Klassifikators sortierten Ground-Truth Menge zur Veranschaulichung der Berechnung des \textbf{$PPV_{k_1}$}}
\end{figure}

Wie in Kapitel \ref{def_ppv} beschrieben, evaluiert der $PPV_{k}$ einen binären $(class~1)$-Klassifikator über die Anzahl der $(class~1)$ Elemente in den ersten $k$ Elementen. Wobei $k := |class~1|=k_1$ (siehe Abbildung \ref{PPV_1}) gewählt wird, weshalb im Folgenden von $PPV_{k_1}$ gesprochen wird und die 1 im Index lediglich der Unterscheidung zum $PPV_{k_2}$ für den konstruierten $(class~2)$ Klassifikator dient.
In Abbildung \ref{PPV_1} wird die Anzahl der $(class~1)$ Elemente im grau hinterlegten Bereich herangezogen (im Weiteren mit $\#PPV_{k_1}$ bezeichnet), um den $PPV_{k_1}$ zu bestimmen, und es ergibt sich folgender Wert:

$$
PPV_{k_1,1}=\frac{\#PPV_{k_1}}{|class~1|}=\frac{\#PPV_{k_1}}{k_1}=\frac{2}{3}
$$
 
Der $PPV_{k_2}$ orientiert sich nach Definition an der Größe der zweiten Klasse $(class~2)$, welche $k_2=|class~2|=n-k_1$ groß ist. Im Gegensatz zur zuvor betrachteten $AUC$ verändert sich der $PPV_k$ deutlich bei der Anwendung der in Kapitel \ref{chap_Symmetrie} vorgestellten Symmetrie-Betrachtung. Der Zähler, also $\#PPV_{k_2}$, ist zwar abhängig von $\#PPV_{k_1}$, aber nicht gleich. Da durch die reverse Sortierung des $(class~2)$-Klassifikators für die Berechnung der $\#PPV_{k_2}$ Elemente exakt die $k_2$ letzten Elemente der Sortierung des $(class~1)$-Klassifikators genutzt werden, lässt sich dieser Wert bereits auf der Sortierung des $(class~1)$-Klassifikators berechnen.\\

Offensichtlich gilt $\#PPV_{k_1}=PPV_{k_1}\cdot k_1$ (I), da es die Anzahl aller $(class~1)$ Elemente in den ersten $k_1$ Elementen aufsummiert. Durch das Subtrahieren von $k_1$ erhalten wir in II die Anzahl aller $(class~2)$ in den ersten $k_1$ Elementen. Da exakt diese Anzahl an Elementen offensichtlich nicht in den letzten $k_2$ Elementen enthalten sein können, ergibt sich im Zähler (III) durch das Subtrahieren von allen $(class~2)$ Elementen die Anzahl an $(class~2)$ Elementen in den letzten $k_2$ Elementen, was nach Definition $\#PPV_{k_2}$ entspricht.

\begin{figure}[h!]
$$
PPV_{k_2}
=\frac{\overbrace{\#PPV_{k_2}}^{III}}{k_2}
=\frac{\overbrace{k_2-(\overbrace{k_1-\overbrace{(\#PPV_{k_1}}^{I})}^{II})}^{III}}{k_2}
=\frac{\overbrace{k_2-(\overbrace{k_1-\overbrace{(PPV_{k_1}\cdot k_1}^{I})}^{II})}^{III}}{k_2}
$$
\end{figure}

Da der $\#PPV_{k_2}$ nur noch durch die Größe der zweiten Klasse normalisiert werden muss ($k_2$), um den $PPV_{k_2}$ zu erhalten, ist gezeigt, dass der $PPV_{k_1}$ in den $PPV_{k_2}$ für festes $k_1,k_2$  überführbar aber nicht gleicht ist, bzw. lediglich für den Fall $k_1=k_2$.

\begin{figure}[h!]
\centering
\includegraphics[width=\linewidth]{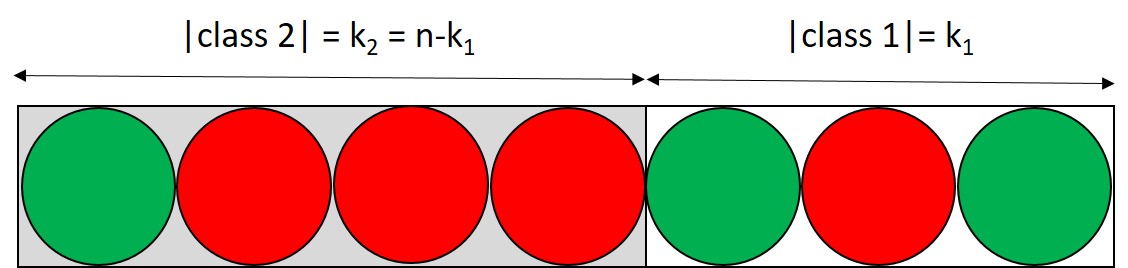}
\caption{\label{PPV_3}$PPV_{k_2}$}
\end{figure}

Für das Beispiel in Abbildung \ref{PPV_3} ergibt sich also: 
$$PPV_{k_2}
=\frac{k_2-(k_1-(PPV_{k_1}\cdot k_1))}{k_2}
=\frac{4-(3-(\frac{2}{3}\cdot 3))}{4}
=\frac{3}{4}
$$
Durch einfache Termumformung lässt sich die oben aufgestellt Formel in $$PPV_{k_2}=1- \frac{k_1}{k_2}\left( 1- PPV_{k_1}\right)=\frac{k_1}{k_2}\cdot PPV_{k_1}+1-\frac{k_1}{k_2}$$ überführen, und es ist ersichtlich, dass sich der $PPV_k$ für feste Klassengrößen bei einem Perspektivenwechsel, wie eingangs erläutert (vgl. Kapitel \ref{chap_Symmetrie}), direkt berechnen lässt.

\newpage
\section{Extremwertbetrachtung für die $AUC$ und den $PPV_k$}
Der $PPV_k$ beruht auf einer festen Aufteilung der sortierten Liste in zwei Teilbereiche~\textendash~die AUC dagegen auf der Anzahl korrekt sortierter Paare. Um festzustellen, wie weit die AUC mit dem $PPV_k$ korreliert, ist es wichtig festzustellen, wie hoch oder niedrig die AUC werden kann, wenn man einen $(class~1)$-Klassifikator konstruiert, der einen  bestimmten $PPV_k$-Wert erreicht. Selbiges gilt natürlich auch für den $PPV_k$ bei fixierter AUC. 
Im Folgenden bezeichne ich die ersten $k_1$ Elemente der aus dem $(class~1)$-Klassifikator entspringenden Sortierung als Abschnitt A und die letzten $k_2=n-k_1$ Elemente als Abschnitt B (s. Abb. \ref{abb_auc_teilbarkeit}). \\
Die $\#AUC(A)$ beinhaltet dann sämtliche Vergleiche zwischen $(class~1)$ und $(class~2)$ Elementen im Bereich A, für die die Scoring-Funktion des zugrunde liegenden Klassifikators dem $(class~1)$ Element einen höheren Wert zugewiesen hat als dem $(class~2)$ Element.  Analoges gilt für die $\#AUC(B)$ im Bereich B. Die $\#AUC(zwischen~A~und~B)$ dagegen sammelt sämtliche paarweisen Vergleiche zwischen $(class~1)$ und $(class~2)$ Elementen  aus unterschiedlichen Bereichen, die das oben beschriebene Kriterium erfüllen.

\subsection{Betrachtung der Abhängigkeit der $AUC$ vom $PPV_k$\label{3_1_auc_betrachtung}}

\begin{figure}[h!]
\centering
\includegraphics[width=\linewidth]{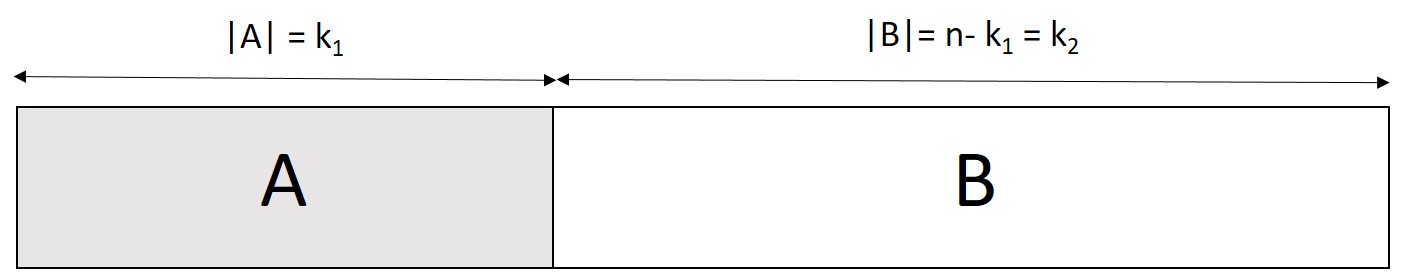}
\caption{\label{abb_auc_teilbarkeit}Visualisierung zur Aufteilbarkeit bei der $AUC$ Berechnung}
\end{figure}

Um die maximal sowie minimal mögliche AUC zu bestimmen, werden zunächst notwendige Beobachtungen notiert. Da im späteren Verlauf die Fallunterscheidung bezüglich der Klassengrößen, also ob die betrachtete $(class~1)$ größer oder kleiner als $(class~2)$ ist, zu deutlich komplexeren Termen führen würde, wird in Beobachtung \ref{groesen} gezeigt, wie sich die Fälle ineinander überführen lassen. Zudem ist es wichtig festzustellen, dass die absolute Anzahl korrekt sortierter Paare für eine Aufteilung, wie in Abbildung \ref{abb_auc_teilbarkeit}, in drei Terme zerfällt: 
\begin{enumerate}
\item  Die Anzahl korrekt sortierter Paare innerhalb von A.
\item  Die Anzahl korrekt sortierter Paare innerhalb von B.
\item  Die Anzahl korrekt sortierter Paare zwischen A und B.
\end{enumerate}
Deshalb ist diese Eigenschaft in  Beobachtung \ref{teilbarkeit_auc} festgehalten. Weiterhin kann man feststellen, dass sich der $PPV_k$ nicht verändert, wenn die Elemente innerhalb von Abschnitt A oder B in ihrer Reihenfolge permutiert werden. Daher können dort jeweilig den AUC maximierende/minimierende Sortierungen vorgenommen werden, ohne den $PPV_k$ eines Klassifikators zu beeinflussen, da hier lediglich dieselbe Anzahl von Elementen beider Klassen in Abschnitten A und B enthalten sein müssen. Diese Sortierungen sind in Beobachtung \ref{max_auc_area} sowie \ref{min_auc_area} erläutert.
In den folgenden Abschnitten wird dann jeweils betrachtet, wie man für den festen $PPV_k$ eines Klassifikators die AUC minimiert oder maximiert.


\begin{Beobachtung}\textbf{: Für das Klassenverhältnis kann bei der $AUC$/$PPV_k$ Betrachtung stets $k_1 \le k_2$ angenommen werden.}\label{groesen}\\
	
Bei der Betrachtung der Klassenverhältnisse treten folgende zwei Fälle auf:
\begin{enumerate}
	\item $k_1 >k_2$
	\item $k_1 \le k_2$
\end{enumerate}

Wenn nun der 1. Fall eintritt, kann der Klassifikator durch einen nach Kapitel \ref{chap_Symmetrie}  konstruierten $(class~2)$ Klassifikator ersetzt werden, bei dem durch die Vertauschung der Klassen und deren Größen Fall 2 ($k_1 \le k_2$) eintritt. Es wurde bereits gezeigt, dass sich beim Vertauschen der betrachteten Klassen:
\begin{itemize}
	\item der $AUC$  nicht ändert (vgl. Kapitel \ref{sym_auc}),
	\item der $PPV_{k_2}$ durch die Formel $PPV_{k_2}=1- \frac{k_1}{k_2}\left( 1- PPV_{k_1}\right)$ aus dem $PPV_{k_1}$ berechnen lässt (vgl. Kapitel \ref{sym_ppv}),
\end{itemize}
weshalb der $AUC$ einfach gleich bleibt und der $PPV_k$ entsprechend umgerechnet werden kann. 
Unter Beachtung der unter Umständen notwendigen Umstrukturierung bezüglich der Klassen und des $PPV_k$ kann also bei der Betrachtung der $AUC$ oder des $PPV_k$ stets $k_1 \le k_2$ angenommen werden.
\end{Beobachtung}

\begin{Beobachtung} \textbf{: Die Berechnung der $AUC$ lässt sich in Abschnitte (siehe Abbildung \ref{abb_auc_teilbarkeit}) untergliedern.\label{teilbarkeit_auc}}\\

Unterteilt man die gegebene Liste in beliebige, aber konsekutive Abschnitte A und B, wie in Abbildung \ref{abb_auc_teilbarkeit} skizziert, so gilt für den $AUC$ : 
$$
AUC= \frac{\#AUC(A)+\#AUC(B)+\#AUC(zwischen~A~und~B)}{|class~1|*|class~2|}
$$

\textbf{Beweis:}\\
Der alternativen Definition der $AUC$ in Kapitel \ref{def_auc} folgend, können zur Berechnung der $AUC$ sämtliche Vergleiche der Scorings aller $(class~1)$ und $(class~2)$ Elemente herangezogen werden. Da die oben angesprochene Unterteilung, die in der Mengenlehre eine Partition widerspiegelt, die gesamte Liste in 2 disjunkte, nicht leere Teillisten unterteilt,  befinden sich alle Vergleichspartner für alle $(class~1)$ Elemente in Abschnitt A entweder in Abschnitt A und werden von der $\#AUC(A)$ erfasst oder in Abschnitt B und werden von der $AUC(zwischen~A~und~B)$ erfasst. Selbiges gilt vice versa für alle $(class~1)$ in B, womit mathematisch alle Vergleiche genau einmal erfasst werden und alle nötigen Vergleiche analog zur Definition des $AUC$ korrekt behandelt werden.\\
\end{Beobachtung}
\begin{figure}[h!]
\centering
\includegraphics[width=\linewidth]{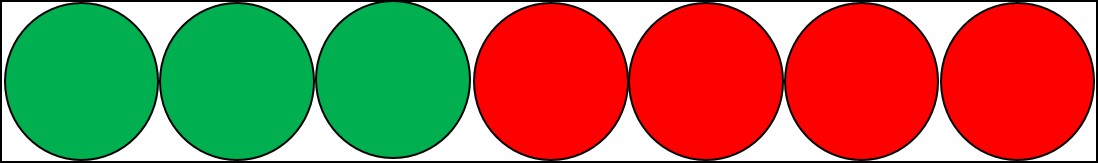}
\caption{\label{abb_max_auc1}Sortierung von 3 ($class~1$) Elementen (grün) und 4 ($class~2$) Elementen (rot), die den $AUC$ maximiert.}
\end{figure}

\begin{Beobachtung} \textbf{: Maximale $AUC$}\label{max_auc_area}\\
Die $AUC$ ist für eine Sortierung maximal, wenn alle $(class~1)$ Elemente eine höhere Wahrscheinlichkeit erhalten haben zu $(class~1)$ zu gehören als alle $(class~2)$ Elemente. Somit befinden sich alle (!) $(class~1)$ Elemente vor den $(class~2)$ Elementen und die Sortierung des Klassifikators kann nur wie in Abbildung \ref{abb_max_auc1} aussehen:
\begin{itemize}
	\item $\#AUC_1 =|class~1|*|class~2|=k_1*(n-k_1)=k_1*k_2$
	\item $AUC_1 =\frac{\#AUC_1}{|class~1|*|class~2|}=\frac{|class~1|*|class~2|}{|class~1|*|class~2|}=1$
\end{itemize}

\end{Beobachtung}

\begin{figure}[h!]
\centering
\includegraphics[width=\linewidth]{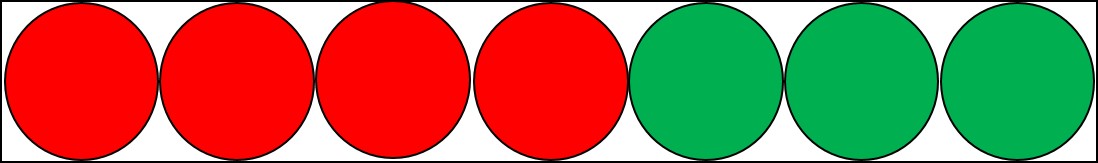}
\caption{\label{abb_min_auc1}Klassifikation von 3 ($class~1$) Elementen (grün) und 4 ($class~2$) Elementen (rot), die den $AUC$ minimiert.}
\end{figure}

\begin{Beobachtung} \textbf{: Minimale $AUC$}\label{min_auc_area}\\
Die $AUC$ ist für eine Sortierung minimal, in der alle $(class~1)$ Elemente eine niedrigere Wahrscheinlichkeit erhalten haben zu $(class~1)$ zu gehören als alle $(class~2)$ Elemente. Somit befinden sich alle (!) $(class~1)$ Elemente hinter den $(class~2)$ Elementen und die Sortierung des Klassifikators kann nur wie in Abbildung \ref{abb_min_auc1} aussehen:
\begin{itemize}
	\item $\#AUC_1 =0$
	\item $AUC_1 =\frac{\#AUC_1}{|class~1|*|class~2|}=\frac{0}{|class~1|*|class~2|}=0$
\end{itemize}
\end{Beobachtung}

\subsubsection{Maximale $AUC$ für festen $PPV_k$\label{max_auc_1}}
Auf Grundlage der gezeigten Beobachtungen lässt sich die aus dem binären Klassifikator entspringende Sortierung in zwei Abschnitte A und B mit $|A|=k_1$ und $|B|=k_2$, wie in Abbildung \ref{abb_auc_teilbarkeit} visualisiert, untergliedern. Da der $PPV_k$ zwar die Anzahl der $(class~1)$ Elemente in den beiden Abschnitten fixiert, jedoch eine veränderte Anordnung innerhalb dieser keinen Einfluss auf ihn hat, können unter Zuhilfenahme der Teilbarkeitsbeobachtung \ref{teilbarkeit_auc} diese beiden Bereiche und somit jeweils die internen \#AUC ($\#AUC(A)_{max}$ und $\#AUC(B)_{max}$) für eine hohe AUC optimiert werden: 
$$
AUC_{max}
= \frac{\#AUC(A)_{max}+\#AUC(B)_{max}+\#AUC(\textrm{zwischen~A~und~B})}{|class~1|\cdot |class~2|}
$$
Nach der Definition des $PPV_k$ (vgl. Kapitel \ref{def_ppv}) befinden sich, wenn dieser $a$ annimmt, genau $a\cdot k_1$ $(class~1)$ Elemente in den ersten $k_1$ Elementen, woraus sich die in Abbildung \ref{abb_auc_teilbarkeit} aufgeführten Werte ableiten lassen.    

\begin{figure}[h!]
\centering
\includegraphics[width=\linewidth]{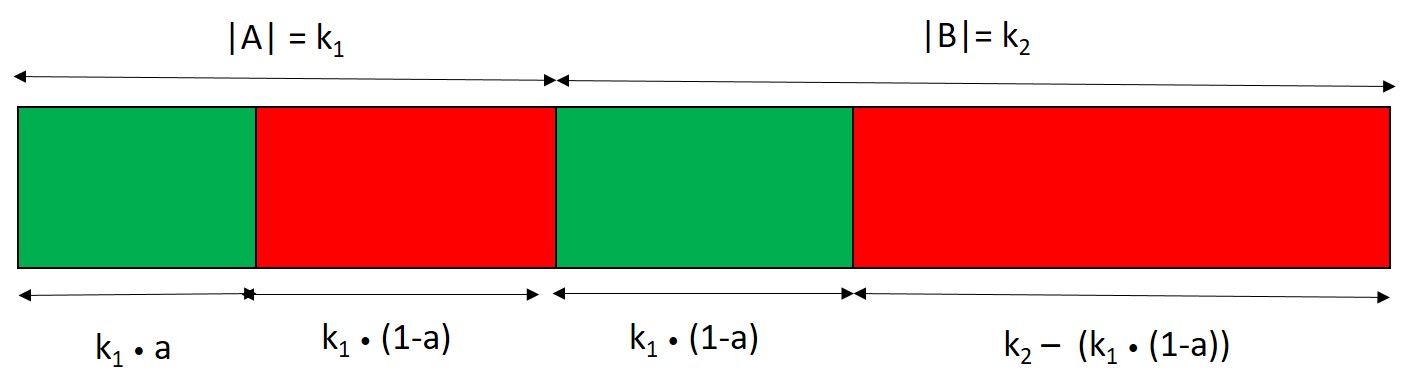}
\caption{\label{abb_max_auc_fest_ppv}Sortierung mit maximaler AUC für gegebenes k und $PPV_k=a$. Grüne Abschnitte symbolisieren die Anzahl der $(class~1)$-Elemente, rote die der $(class~2)$-Elemente.}
\end{figure}

Da innerhalb der Bereiche A und B zur AUC-Maximierung eine optimale Sortierung (vgl. Beobachtung \ref{max_auc_area}) vorliegen sollte, ergeben sich unter Berücksichtigung der Anzahl von $(class~1)$ und $(class~2)$ in den jeweiligen Bereichen (vgl. Abbildung \ref{abb_max_auc_fest_ppv}) folgende Werte: 

\begin{align*}
\#AUC(A)_{max} =&~|(class~1)~\textrm{Elemente~in~A}|\cdot |(class~2)~\textrm{Elemente~in~ A}|\\
=&~(k_1\cdot a)\cdot (k_1\cdot (1-a))\\
\#AUC(B)_{max} =&~|(class~1)~\textrm{Elemente~in~B}|\cdot |(class~2)~\textrm{Elemente~in~ B}|\\
=&~( k_1\cdot(1-a))\cdot (k_2-(k_1\cdot (1-a) ))
\end{align*}

Für die $\#AUC(\textrm{zwischen~A~und~B})$ sind nach Konstruktion lediglich alle $(class~1)$ Elemente aus Bereich A zu allen $(class~2)$ in Bereich B korrekt sortiert: 

\begin{align*}
\# AUC(\textrm{zwischen~A~und~B})=&~|(class~1)~\textrm{Elemente~in~A}|\cdot |(class~2)~\textrm{Elemente~in~ B}|\\
=&~(k_1\cdot a)\cdot (k_2-( k_1\cdot (1-a)))
\end{align*}

Setzt man nun diese Werte in die oben genannte Formel ein, lässt sich diese durch Termumformung wie folgt reduzieren:
\begin{align*}
AUC_{max}=&~\frac{\#AUC(A)+\#AUC(B)+\#AUC(\textrm{zwischen~A~und~B})}{|class~1|\cdot |class~2|}\\
=&~\frac{(k_1\cdot a)\cdot (k_1\cdot (1-a))}{k_1\cdot k_2}\\
+&~\frac{((1-a)\cdot k_1)\cdot (k_2-((1-a)\cdot k_1))}{k_1\cdot k_2}\\
+&~\frac{(k_1\cdot a)\cdot (k_2-((1-a)\cdot k_1))}{k_1\cdot k_2}\\\\
=&~\frac{(a\cdot (k_1\cdot (1-a))}{k_2}\\
+&~\frac{((1-a)\cdot (k_2-((1-a)\cdot k_1))}{k_2}\\
+&~\frac{a\cdot (k_2-((1-a)\cdot k_1))}{k_2}\\
=&~\frac{[ak_1-a^2k_1]+[k_2-k_1+ak_1-ak_2+ak_1-a^2k_1]+[ak_2-ak_1+a^2k_1]}{k_2}\\
=&~\frac{-a^2k_1+2ak_1-k_1+k_2}{k_2}\\
=&~\frac{k_2-(a^2k_1-2ak_1+k_1)}{k_2}\\
=&~\frac{k_2-((a-1)^2\cdot k_1)}{k_2}\\
=&~1-\frac{((a-1)^2\cdot k_1)}{k_2}\\
\end{align*}

\begin{figure}[h!]
\begin{minipage}[t]{0.45\textwidth}
\includegraphics[width=\linewidth]{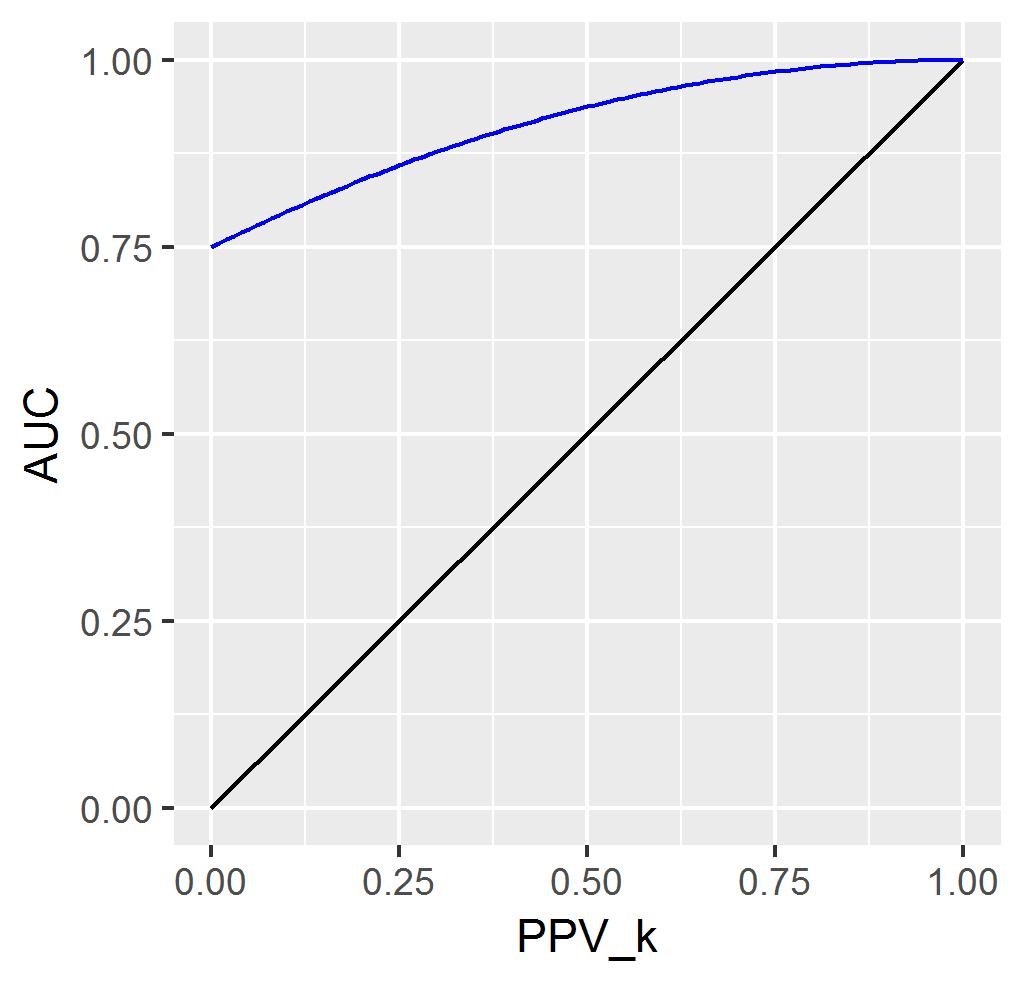}
\caption*{ $k_1: k_2$  ist $1:4$}

\end{minipage}
\hspace{\fill}
\begin{minipage}[t]{0.45\textwidth}
\includegraphics[width=\linewidth]{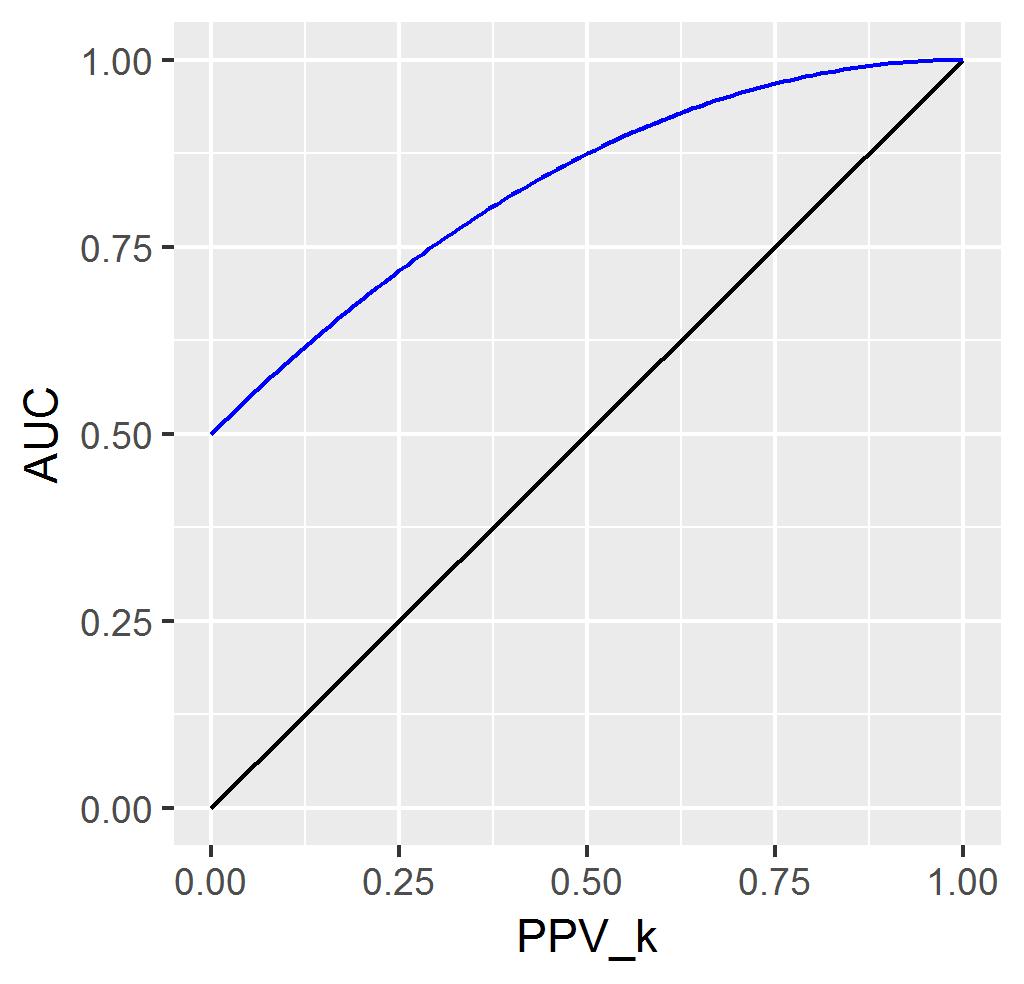}
\caption*{$k_1: k_2$  ist $2:4$}

\end{minipage}

\vspace*{0.5cm}
\begin{minipage}[t]{0.45\textwidth}
\includegraphics[width=\linewidth]{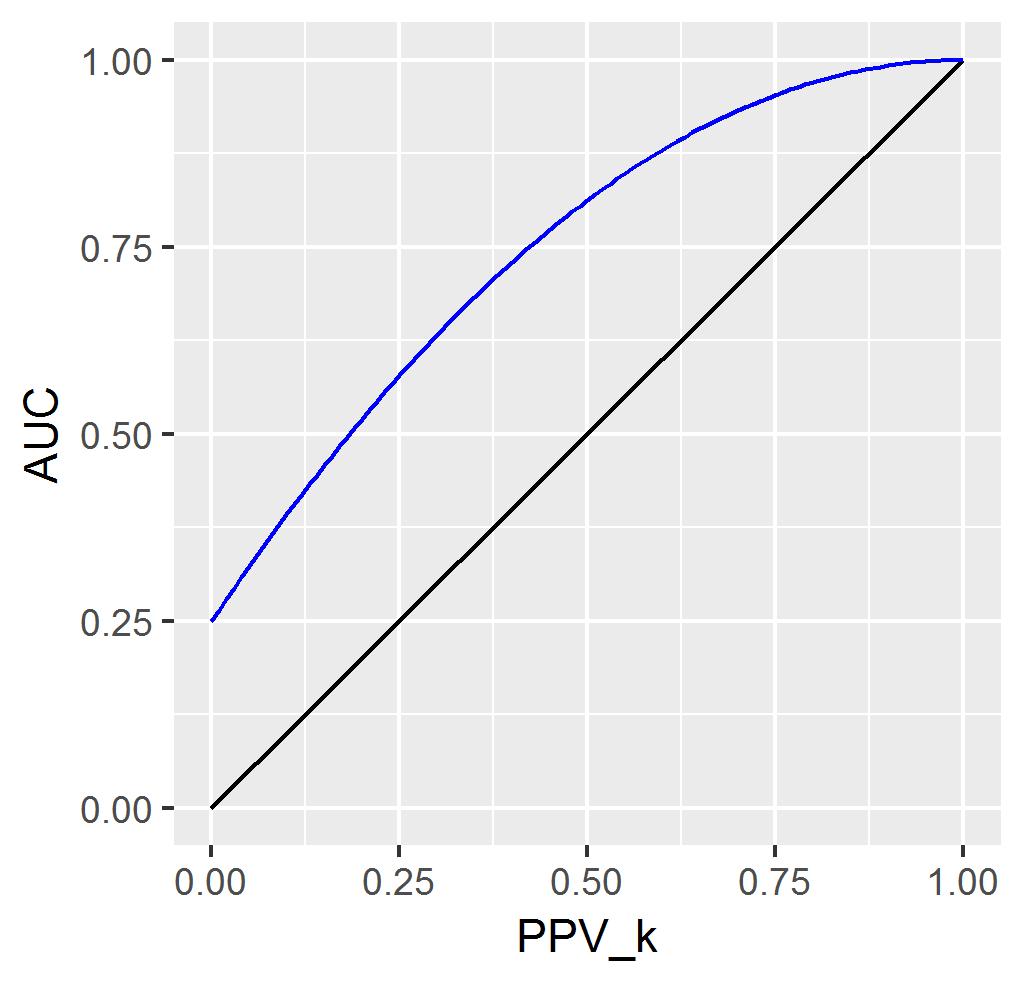}
\caption*{$k_1: k_2$  ist $3:4$}

\end{minipage}
\hspace{\fill}
\begin{minipage}[t]{0.45\textwidth}
\includegraphics[width=\linewidth]{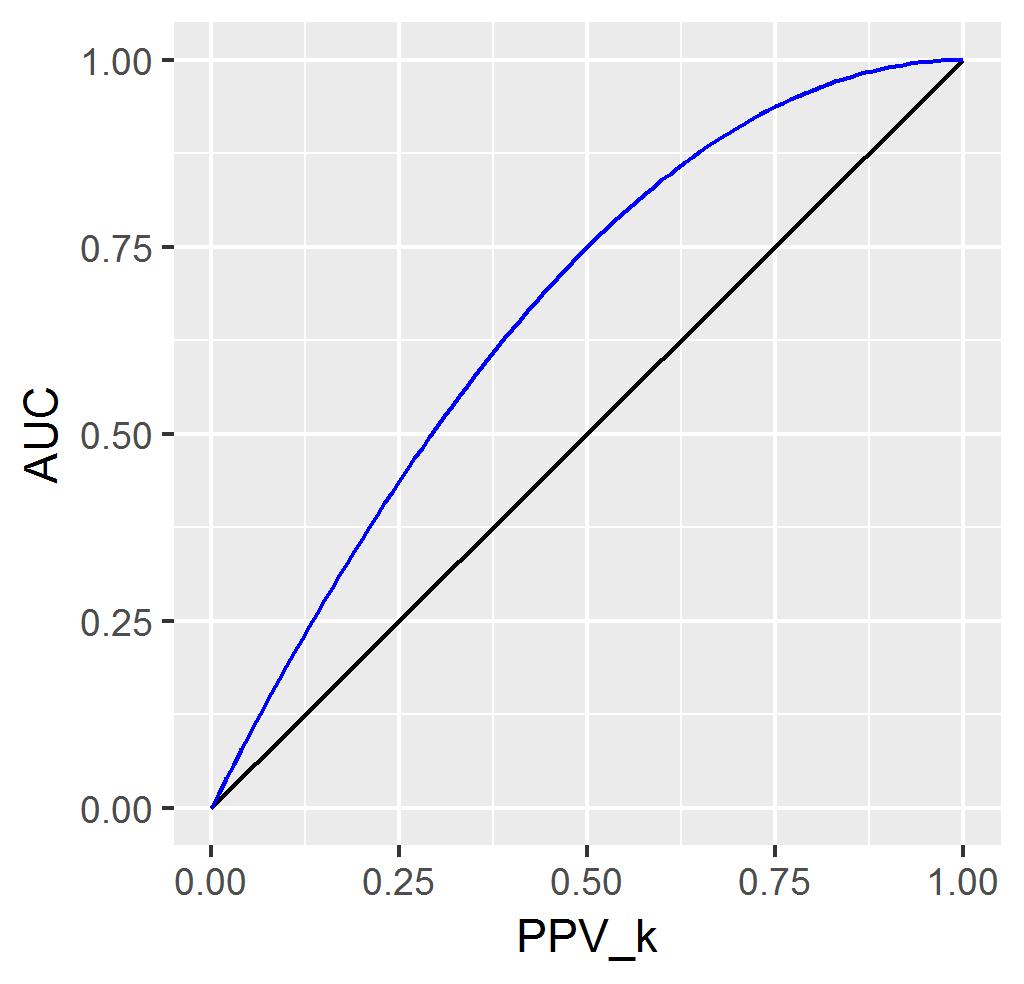}
\caption*{$k_1: k_2$  ist $1:1$}

\end{minipage}
\caption{Maximale $AUC$ (blau) für fixierten $PPV_k$ bei verschiedenen Klassengrößenverhältnissen ($k_1: k_2$)}
\label{abb_max_auc}
\end{figure}

Da $k_1 \le k_2$ angenommen werden kann, liegt die $AUC_{max}$ im Intervall  $[0,1]$. In Abbildung \ref{abb_max_auc} wird diese für verschiedene Klassenverhältnisse in blau abgetragen und die Abweichung von der Identitätslinie (schwarz) ist deutlich zu erkennen. Eine maximal mögliche AUC ist somit gefunden. Da jedoch auch nach unten eine starke Abweichung möglich sein kann, wird im folgenden Abschnitt überprüft, welche Untergrenze für die AUC durch einen Klassifikator erreicht werden kann, der dennoch den festgesetzten $PPV_k$ aufweist.	

\subsubsection{Minimale $AUC$ für festen $PPV_k$ \label{min_auc_kap}}
Analog zur Berechnung der maximalen $AUC$ in Abschnitt \ref{max_auc_1} wird auch hier eine Aufteilung in die Abschnitte A und B vorgenommen.
Die $AUC_{min}$ setzt sich also über die identische Formel zusammen (vgl. Beobachtung \ref{teilbarkeit_auc}), wobei versucht wird, den Einfluss der Teilbereiche auf die AUC jeweils zu minimieren:
$$
AUC_{min}
= \frac{\#AUC_{min}(A)+\#AUC_{min}(B)+\#AUC(\textrm{zwischen~A~und~B})}{|class~1|\cdot |class~2|}
$$
\begin{figure}[h!]
\centering
\includegraphics[width=\linewidth]{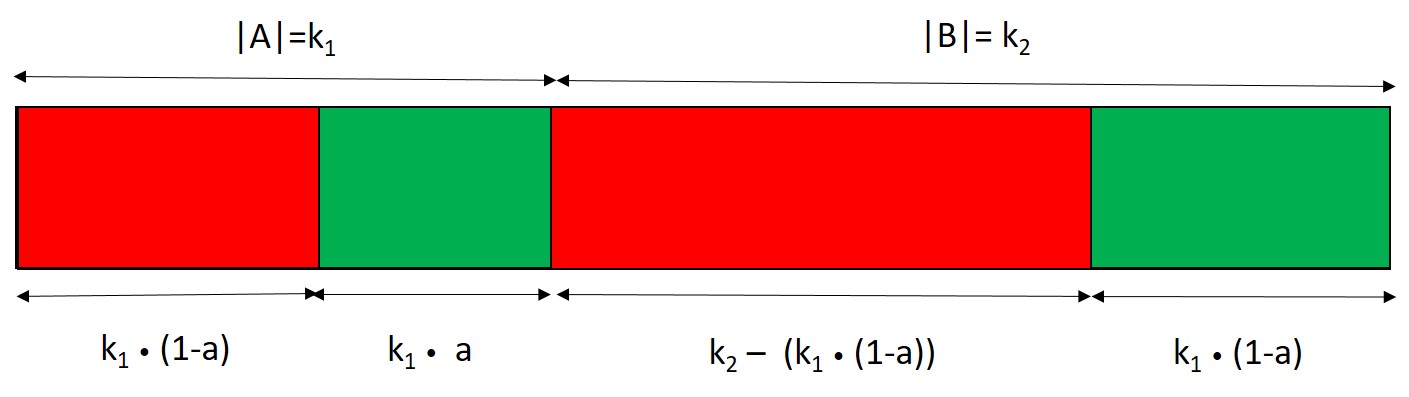}
\caption{\label{3_3_min_auc}Sortierung mit minimaler AUC für gegebenes k und $PPV_k=a$. Grüne Abschnitte symbolisieren die Anzahl der $(class~1)$-Elemente, rote die der $(class~2)$-Elemente.}
\end{figure}
Für die $\#AUC(\textrm{zwischen~A~und~B})$ bleiben nach Konstruktion alle $(class~1)$ Elemente aus Bereich A zu allen $(class~2)$ in Bereich B korrekt sortiert, um den $PPV_k$ nicht zu verändern. Lediglich die Bereiche A und B sehen intern anders aus. Um die $AUC_{min}$ zu erhalten, muss für diese vom Klassifikator jeweils eine minimierende Sortierung, wie in Beobachtung \ref{min_auc_area} aufgezeigt, angestrebt werden. Die Konstruktion in Abbildung \ref{3_3_min_auc} resultiert also aus dem Versuch innerhalb der Bereiche zu minimieren, weshalb dort die Anzahl an korrekt sortierten Paaren ($\#AUC(A)_{min}$ und $\#AUC(B)_{min}$) auf 0 sinkt und lediglich die $$
\# AUC(zwischen~A~und~B)= (k_1\cdot a)\cdot (k_2-((1-a)\cdot k_1)) $$ als Term in der Formel der $AUC_{min}$ übrig bleibt: 
\begin{align*}
AUC_{min}=&~\frac{\#AUC(A)_{min}+\#AUC(B)_{min}+\#AUC(\textrm{zwischen~A~und~B})}{|class~1|\cdot |class~2|}\\
=&~ \frac{0+0+[(k_1\cdot a)\cdot (k_2-((1-a)\cdot k_1))]}{k_1\cdot k_2}\\
=&~\frac{a\cdot (k_2-((1-a)\cdot k_1))}{k_2}\\
=&~\frac{ak_2-ak_1+a^2k_1}{k_2}\\
=&~a+\frac{(a-1)\cdot ak_1}{k_2}\\
=&~a\cdot (1+\frac{(a-1)\cdot k_1}{k_2}) \\
=&~a\cdot (1-\frac{(1-a)\cdot k_1}{k_2})
\end{align*}

\begin{figure}[h!]
\begin{minipage}[t]{0.45\textwidth}
\includegraphics[width=\linewidth]{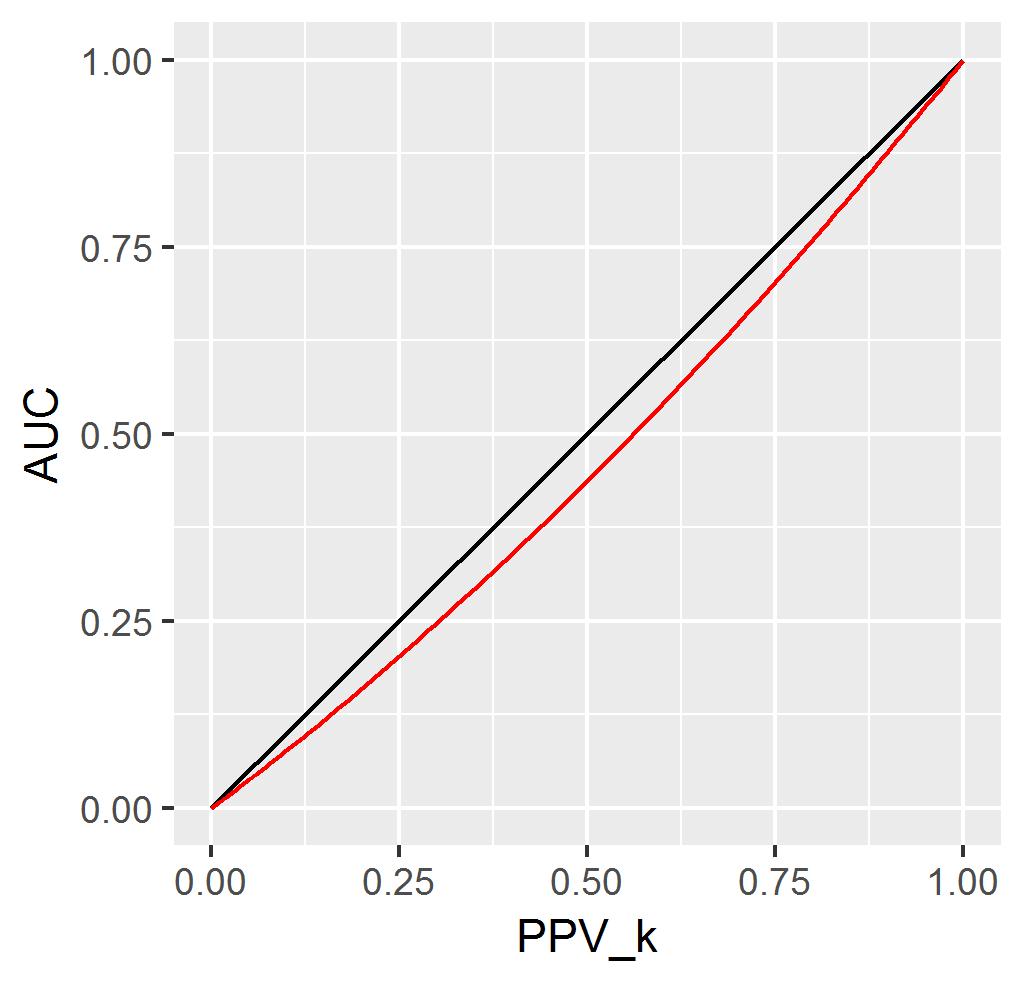}
\caption*{ $k_1: k_2$  ist $1:4$}

\end{minipage}
\hspace{\fill}
\begin{minipage}[t]{0.45\textwidth}
\includegraphics[width=\linewidth]{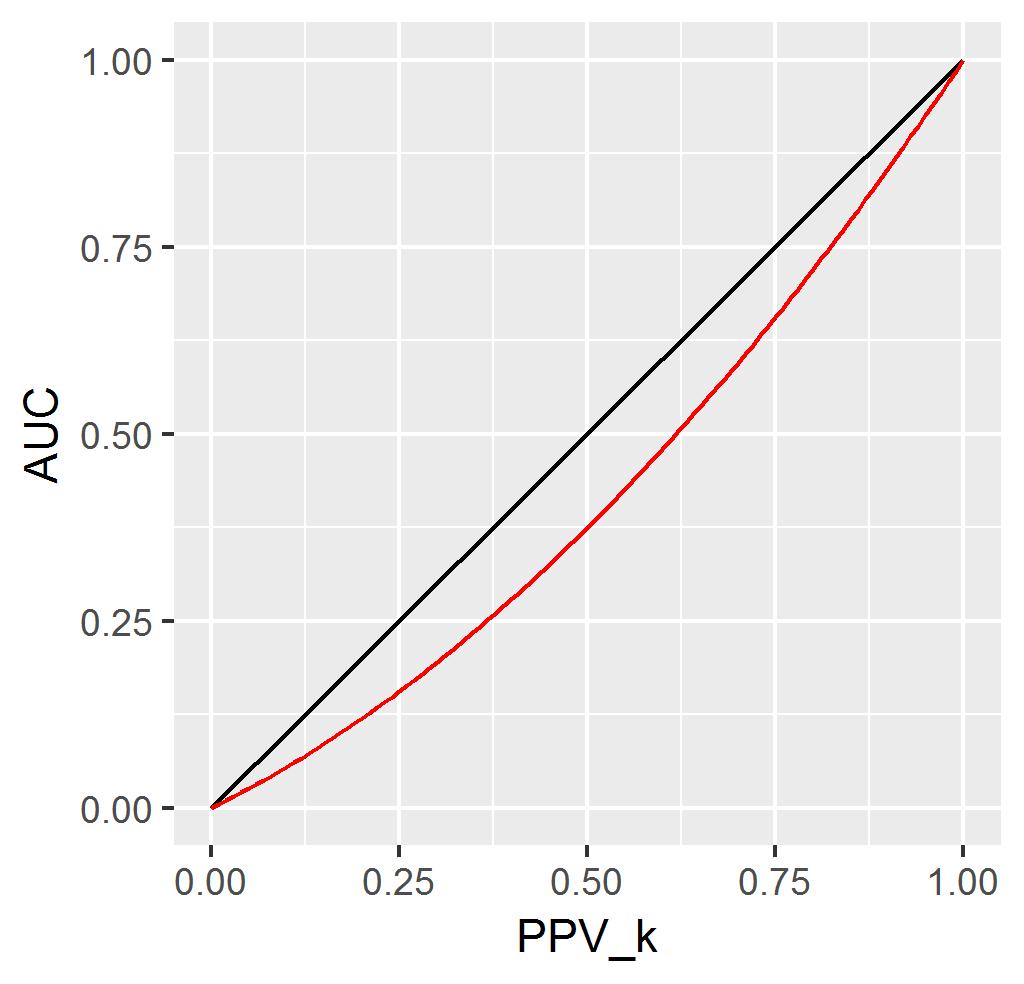}
\caption*{$k_1: k_2$  ist $2:4$}

\end{minipage}

\vspace*{0.5cm}
\begin{minipage}[t]{0.45\textwidth}
\includegraphics[width=\linewidth]{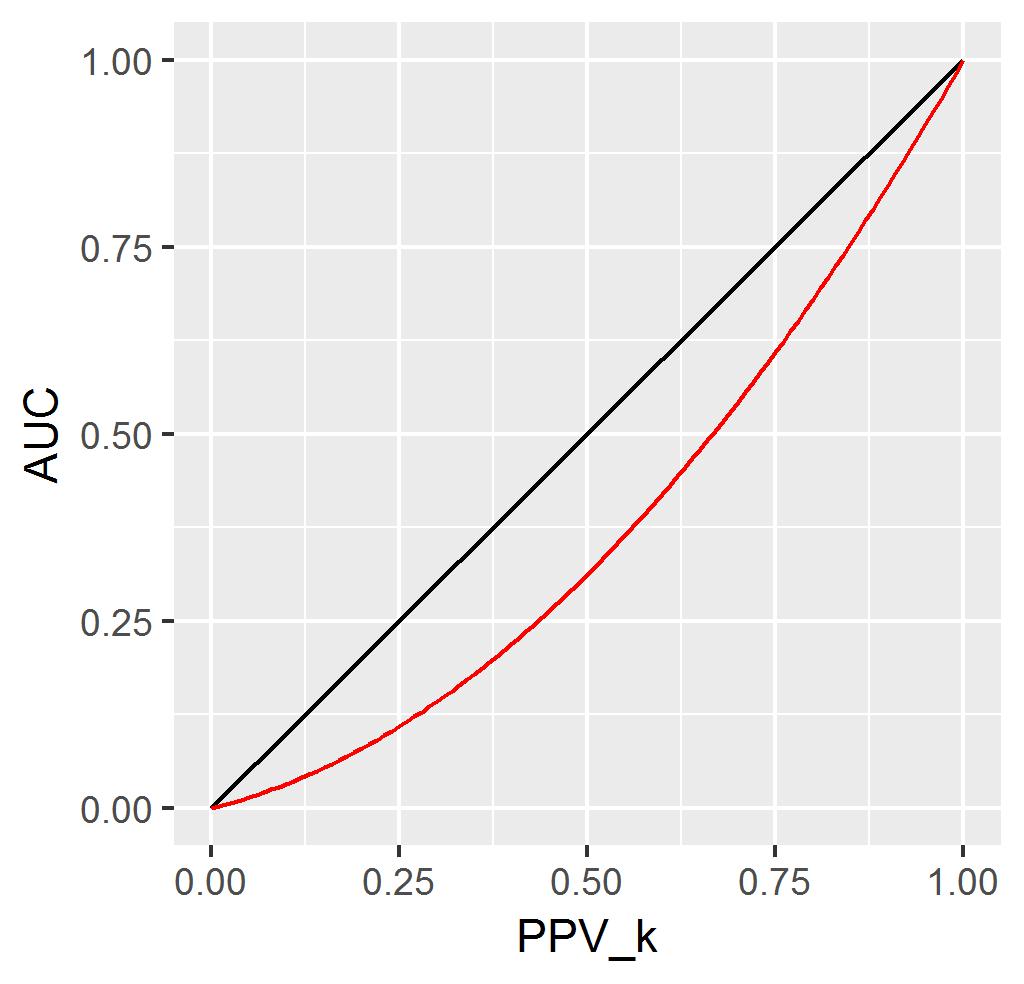}
\caption*{$k_1: k_2$  ist $3:4$}

\end{minipage}
\hspace{\fill}
\begin{minipage}[t]{0.45\textwidth}
\includegraphics[width=\linewidth]{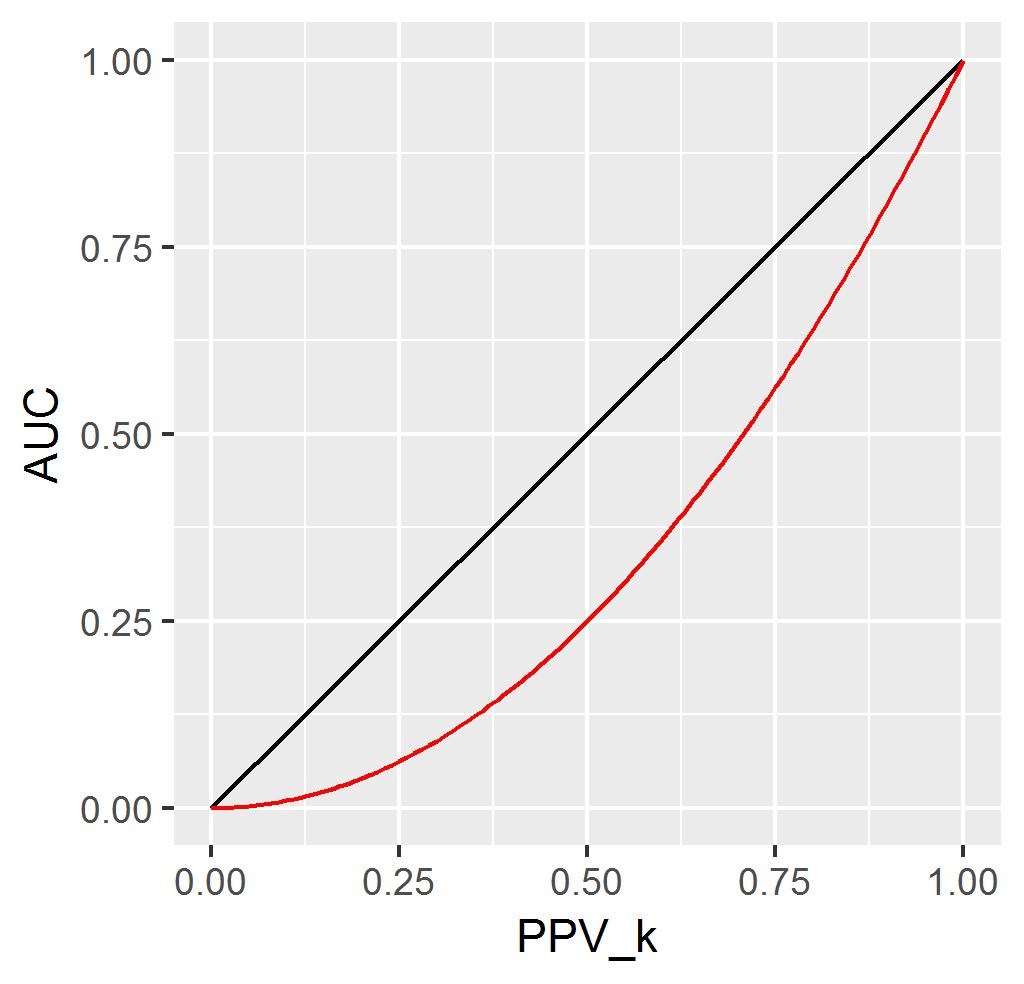}
\caption*{$k_1: k_2$  ist $1:1$}

\end{minipage}
\caption{Minimaler $AUC$ (rot) für fixierten $PPV_k$ bei verschiedenen Klassengrößenverhältnissen ($k_1: k_2$).}
\label{abb_min_auc}
\end{figure}

Da auch hier $k_1 \le k_2$ angenommen werden kann, liegt die $AUC_{min}$, wie in Abbildung \ref{abb_min_auc} zu erkennen ist, meist deutlich unterhalb der Identitätslinie und zeigt somit auf, dass es bei gleichbleibendem $PPV_k$, Klassifikatoren gibt, die im Hinblick auf die AUC deutlich schlechter arbeiten. \\
Mathematisch zu erkennen ist, dass der $PPV_k$ stets eine obere Grenze der $AUC_{min}$ darstellt, jedoch nicht, wie für eine perfekte Korrelation benötigt, auch als eine untere Schranke fungiert. Daher soll im folgenden Abschnitt das Zusammenspiel von maximal und minimal möglicher AUC genauer betrachtet werden.

\subsubsection{Vergleich maximaler und minimaler $AUC$ für festen $PPV_k$}
\begin{figure}[h!]
\begin{minipage}[t]{0.45\textwidth}
\includegraphics[width=\linewidth]{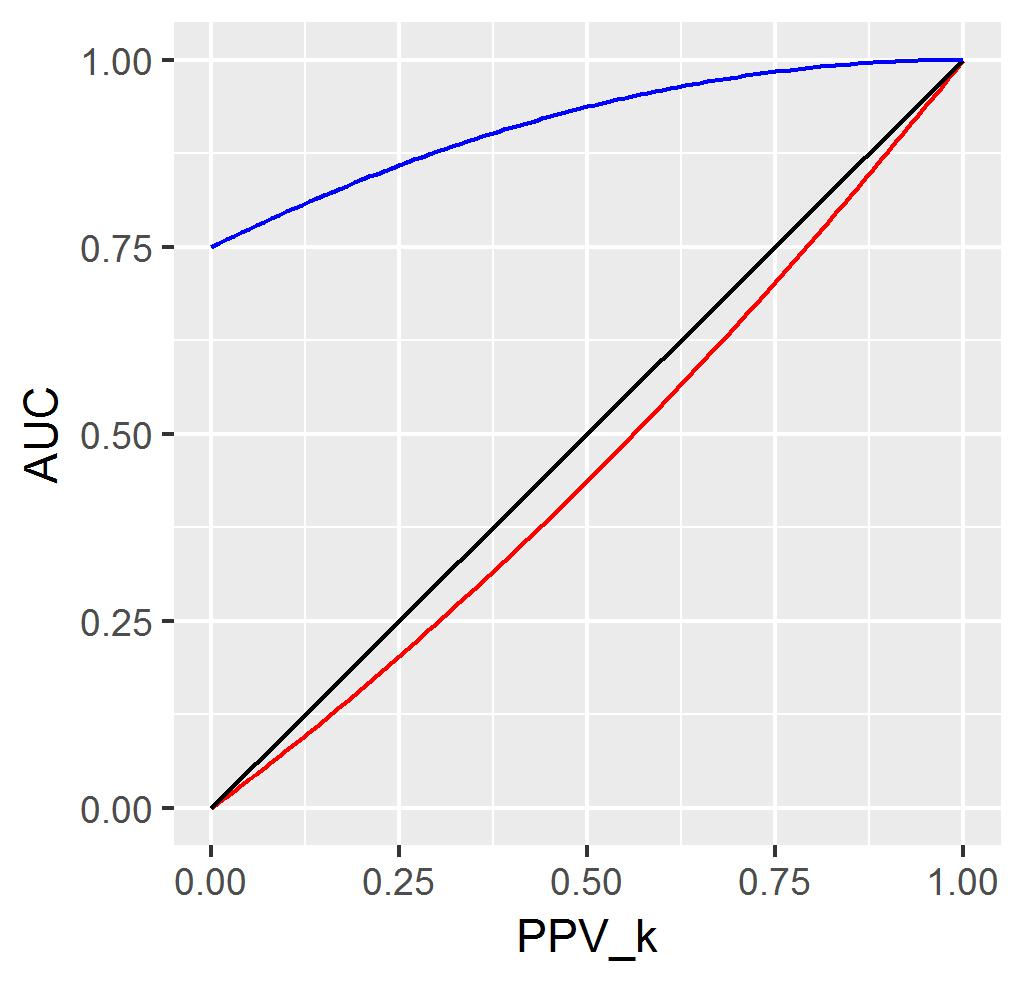}
\caption*{ $k_1: k_2$  ist $1:4$}

\end{minipage}
\hspace{\fill}
\begin{minipage}[t]{0.45\textwidth}
\includegraphics[width=\linewidth]{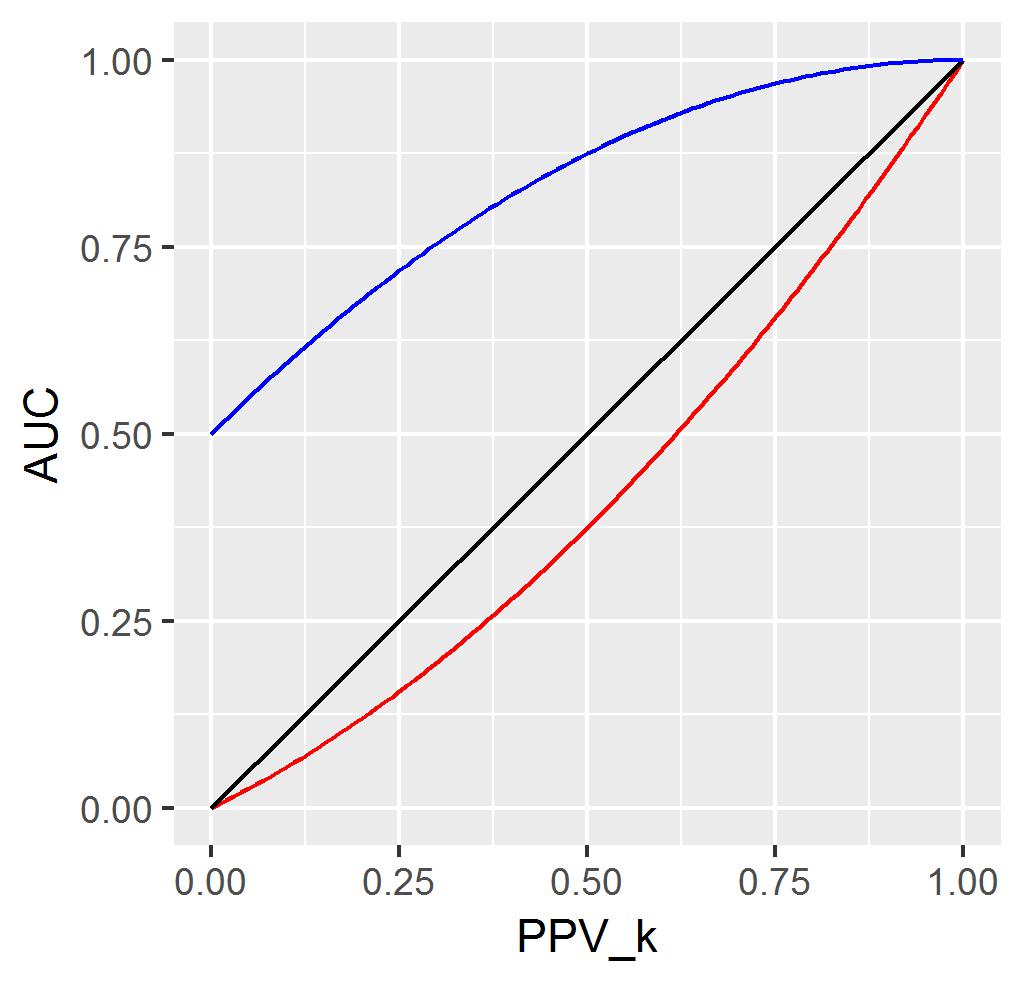}
\caption*{$k_1: k_2$  ist $2:4$}

\end{minipage}

\vspace*{0.5cm}
\begin{minipage}[t]{0.45\textwidth}
\includegraphics[width=\linewidth]{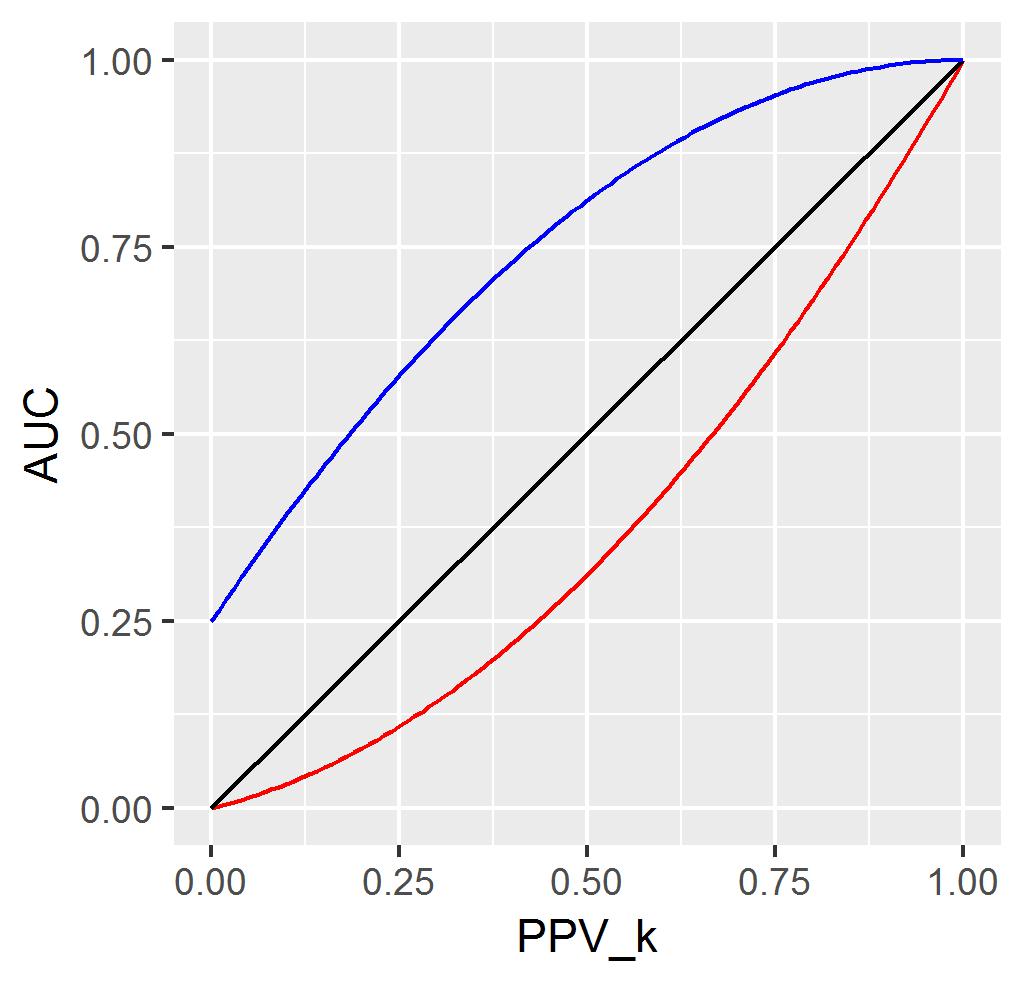}
\caption*{$k_1: k_2$  ist $3:4$}

\end{minipage}
\hspace{\fill}
\begin{minipage}[t]{0.45\textwidth}
\includegraphics[width=\linewidth]{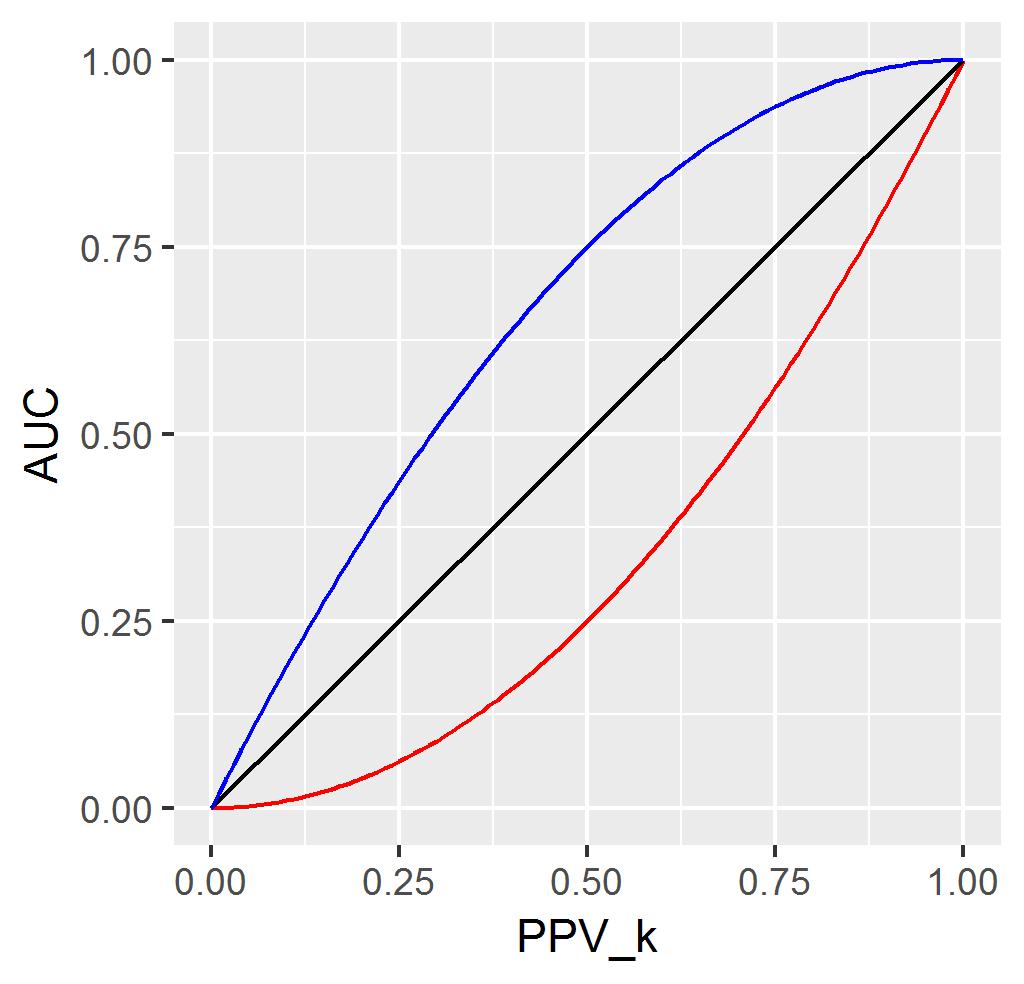}
\caption*{$k_1: k_2$  ist $1:1$}

\end{minipage}
\caption{$AUC_{max}$ (rot) und $AUC_{min}$ (blau) für fixierten $PPV_k$ bei verschiedenen Klassengrößenverhältnissen}
\label{abb_auc}
\end{figure}

Bei der Betrachtung der maximal (blau) sowie minimal (rot) möglichen AUC für festen $PPV_k$ in Abbildung \ref{abb_auc} wird ihr großer  Spielraum deutlich, denn es gibt Klassifikatoren, die außerhalb mancher Sonderfälle, wie zum Beispiel für $PPV_k=1$, Unterschiede zwischen AUC und $PPV_k$ von bis zu 0.75 aufweisen. 

Auch wenn sich mit zunehmendem Ungleichgewicht des Klassenverhältnisses ($k_1:k_2$) die Abweichung der $AUC_{min}$ von der Identitätslinie reduziert (vgl. Abbildung \ref{abb_min_auc}), überwiegt die Diskrepanz der $AUC_{max}$ deutlich. Es kann festgestellt werden, dass eine vollständige Korrelation ausgeschlossen werden kann, denn es existieren zumindest die zwei konstruierten Abweichungen, die für verschiedene Klassenverhältnisse berechnet und visualisiert wurden.\\

Zu beobachten ist auch, dass im Spezialfall $k_1=k_2$ der $$AUC_{max} =1-PPV_k^2+2PPV_k-1 = 2PPV_k-PPV_k^2$$ ist und die Differenz zwischen $PPV_k$ und der $AUC_{max}$ somit genau bei $PPV_k-PPV_k^2$ liegt. Die $AUC_{min}$ wird in diesem Fall $$AUC_{min}= PPV_k^2$$ und die Differenz zwischen $AUC_{max}$ und $AUC_{min}$ bei gleich großen Klassen liegt bei: $$2PPV_k-2PPV_k^2$$, was auch das sehr symmetrisch wirkende Schaubild in Abbildung \ref{abb_auc} ($k_1: k_2$  ist $1:1$) belegt.\\	
Interessant ist es zusätzlich das Verhalten der AUC für festen $PPV_k$ und $\lim\limits_{\frac{k_1}{k_2} \to 0}$ zu betrachten, denn hier nähert sich die $AUC_{min}$ zwar dem $PPV_k$ an, jedoch strebt die $AUC_{max}$ gegen 1. Daraus kann man schließen, dass für unbalancierte Klassen durch den $PPV_k$ zwar eine gute untere Schranke für die AUC gegeben ist, jedoch ein deutliches Verbesserungspotenzial existieren könnte. Deshalb soll im folgenden Kapitel überprüft werden, inwieweit bei fixierter AUC ein Spielraum für den $PPV_k$ existiert. 

\subsection{Betrachtung der Abhängigkeit des $PPV_k$ von der $AUC$\label{3_5_ppvk_betrachtung}}

\begin{figure}[h!]
\centering
\includegraphics[width=\linewidth]{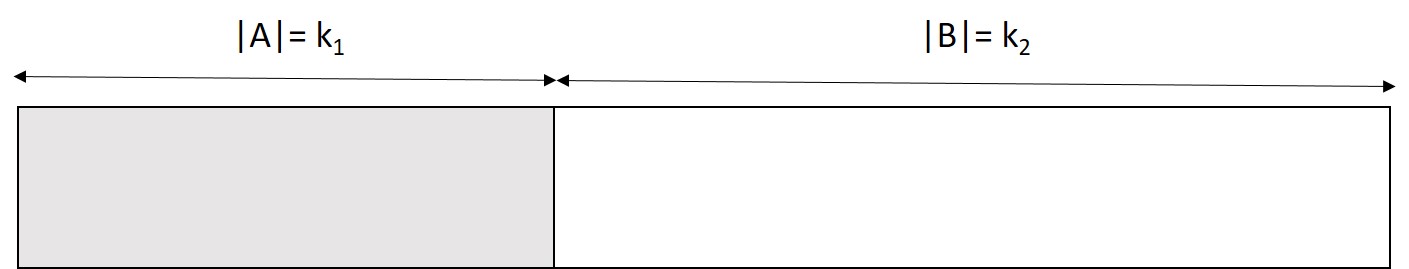}
\caption{\label{abb_ppv_aufteilung}Visualisierung der Aufteilung für die Bestimmung der Extrema für den $PPV_{k}$ bei fixiertem $AUC$.}
\end{figure}

Wie bereits ausführlich im Kapitel \ref{soa_auc} aufgezeigt, wird die $AUC$ bei fast allen binären Klassifikatoren aus dem maschinellem Lernen als bevorzugtes Evaluationskriterium genutzt und dient auch in der Rückfälligkeitsvorhersage als das beste Maß zur Bestimmung, wie präzise ein Wert die Rückfälligkeit vorhersagen kann\footnote{ "`The best measure for determining how accurately a score predicts an event like recidivism"'~\cite[S.3]{Barnoski2007}}. Wenn nun aber viele dieser Klassifikatoren mit der $AUC$ evaluiert werden~\cite{Lansing2012} und, wie zuvor gezeigt, das Entscheidungsmodell eines Richters nicht durch die $AUC$ abgebildet wird, ist es wichtig zu verstehen, welchen Spielraum der $PPV_k$ bei fixierter $AUC$ hat.\\In diesem Fall wird $AUC = b$ angenommen und wie zuvor mit $\#AUC$ die Anzahl an korrekt sortierten Paare notiert. Zur Bestimmung der Maxima/Minima des $PPV_{k}$, die ein Klassifikator annehmen kann, wenn er zusätzlich eine AUC von b erreichen soll, entstehen, ähnlich der $AUC$ Betrachtung zwei Fälle. Unter Zuhilfenahme der Beobachtung \ref{groesen} lassen sich diese beiden  auf $k_1\le k_2$ reduzieren.
Des Weiteren wird mit $\#PPV_k$  die Anzahl aller $(class~1)$ Elemente der Ground-Truth aufsummiert, die der Klassifikator in einer auf seiner Sortierfunktion $s_f$ beruhenden Sortierung in die ersten k befördern konnte.\\
Der im Folgenden skizzierte Lösungsansatz dieser Fragestellung ist noch nicht optimal, da jedoch die erarbeiteten Optimierungen in Zusammenarbeit mit Prof.~Dr.~Katharina~A.~Zweig entstanden sind, wird an dieser Stelle nur auf das angehängte Paper (Appendix A) verwiesen, um die Eigenständigkeit der Arbeit nicht zu gefährden. Lediglich zur Visualisierung der Ergebnisse in Abschnitt \ref{verg_ppv} wurde die geschlossene Formel herangezogen, da sich diese leichter verarbeiten lässt und die Ergebnisse ohnehin identisch waren.\\
Da zur Bestimmung des $PPV_{k}$ die ersten $k$ Elemente der Liste betrachtet werden, ist es sinnvoll, die Liste zunächst wieder in die zwei Abschnitte A und B, wie in Abbildung \ref{abb_ppv_aufteilung} gezeigt, aufzuteilen. 
Bei der Betrachtung einer fixierten AUC fällt auf, das es kompliziert werden würde, die durch die AUC vorgegebene Anzahl an korrekt sortierten Elementen (vgl. Definition AUC in Kapitel \ref{def_auc}) so zu nutzen, dass sich der $PPV_k$ maximiert oder minimiert, weshalb im Folgenden ein quasi iteratives Verfahren genutzt wird. Hierbei wird ausgenutzt, dass in Abschnitt \ref{3_1_auc_betrachtung} bereits Formeln vorgestellt wurden, welche für fixierten $PPV_k$ minimale und maximale AUC-Werte berechnen können. Mit Hilfe dieser ist es möglich die minimalen und maximalen Schwellenwerte zu finden, um so im Vergleich mit der gegebenen AUC den entsprechenden $PPV_k$ zu ermitteln.

\newpage
\subsubsection{Maximaler $PPV_k$ für feste $AUC$}

\begin{figure}[h!]
\centering
\includegraphics[width=\linewidth]{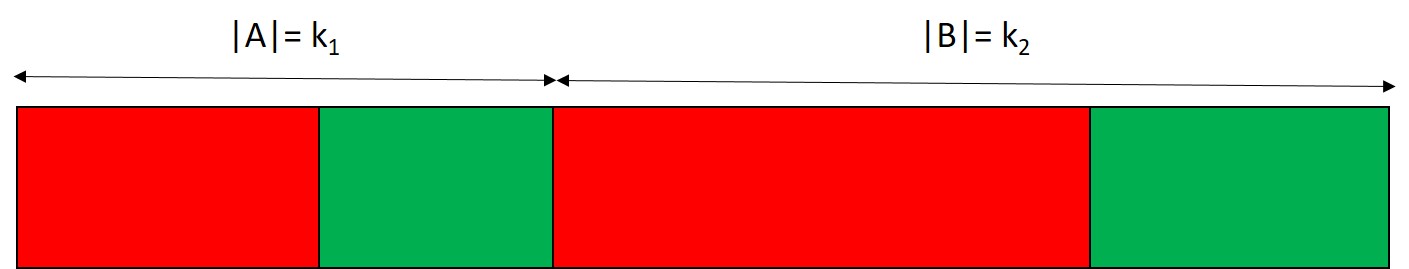}
\caption{\label{abb_ppv_min_auc}Minimale $AUC$ in den Bereichen A und B}
\end{figure}

Da nach der Untergliederung in die zwei Abschnitte A und B die jeweilige interne Sortierung für die Berechnung des $PPV_{k_1}$ nicht relevant ist (vgl. Definition $PPV_{k_1}$ in Kapitel \ref{def_ppv}), kann die in Abbildung \ref{abb_ppv_min_auc} skizzierte Sortierung innerhalb der Bereiche angenommen werden. Hier wird die AUC minimierende Sortierung, wie sie in Beobachtung \ref{min_auc_area} aufgezeigt wurde, jeweils auf A und B angewandt und somit die Sortierung geschaffen, die für minimale AUC dennoch den vorgegebenen $PPV_k$ beibehält. 
Mithilfe der Formel aus Abschnitt \ref{min_auc_kap} kann also die minimale AUC berechnet werden, die ein Klassifikator erreichen muss, um eine Sortierung zu ermöglichen, die genau $\#PPV_{k}$ Elemente in den Bereich A befördern kann.

Zunächst werden nun die minimalen $AUC$ Werte für alle $\#PPV_{k}\in[1,k]$ berechnet und in einer Liste abgelegt. Jeder $\#PPV_{k}$ gehört zu genau einem der möglichen $PPV_k$ und diese Liste bildet exakt die Schwellenwerte ab, an denen durch einen Klassifikator ein nächsthöherer $PPV_k$ erreicht werden kann. Diese Eigenschaft resultiert direkt daraus, dass es, wie in Abschnitt \ref{min_auc_kap} gezeigt, keine Sortierung mit einer geringeren $AUC$ und derselben $PPV_k$ gibt.\\
Die resultierende Liste kann nun mit linearem Aufwand (vgl. Algorithmus \ref{algo_max_ppv}) auf die Stelle hin überprüft werden, an welcher der maximal zu erreichende $PPV_{k}$ für den gegebenen $AUC=b$ ist.
\lstset{language=Pascal,caption={Algorithmus zur Bestimmung des maximalen $PPV_{k_1}$ für feste $AUC$.},label=algo_max_ppv}
\begin{lstlisting}[frame=single,mathescape=true]
Liste_der_Grenzen =[]
for i:= 1 to k do
begin
	{Liste_der_Grenzen += AUC_min(i/k);}
end;
j=1;
while Liste_der_Grenzen[j]<b && j<k do
begin
	{j=j+1;}
end;
return Liste_der_Grenzen[j]
\end{lstlisting}

\subsubsection{Minimaler $PPV_k$ für feste $AUC$}
\begin{figure}[h!]
\centering
\includegraphics[width=\linewidth]{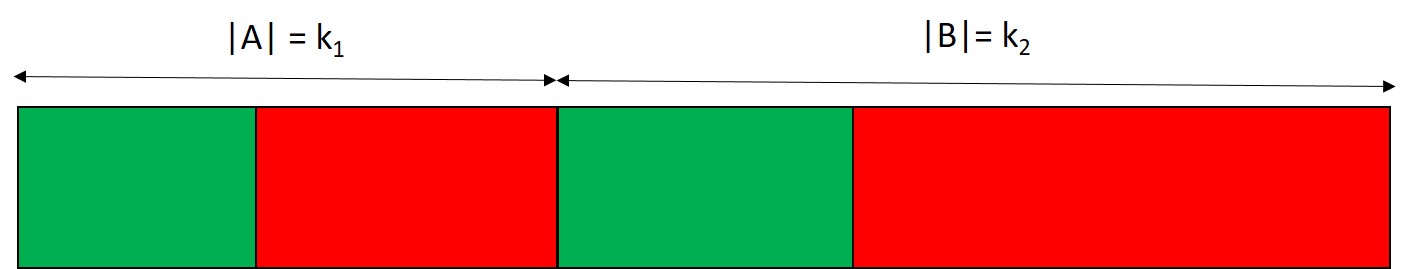}
\caption{\label{abb_3_5_1}Maximale $AUC$ in den Bereichen A und B}
\end{figure}

Der minimale $PPV_{k}$ lässt sich ähnlich zum maximalen bestimmen. Zunächst wird die Liste wieder, wie in Abbildung \ref{abb_ppv_aufteilung} skizziert, in die Bereiche A und B geteilt. Wie bereits erwähnt, beeinflusst die interne Sortierung der zwei Abschnitte die Berechnung des $PPV_k$ nicht (vgl. Definition $PPV_{k_1}$ in Kapitel \ref{def_ppv}) und es kann von der in Abbildung \ref{abb_3_5_1} gezeigten Sortierung ausgegangen werden, welche jeweils die $AUC$ innerhalb der Abschnitte von A und B maximiert (vgl. Beobachtung \ref{max_auc_area}). 
Somit kann man nun, mit der in Abschnitt \ref{max_auc_1} erläuterten Formel für alle $\#PPV_{k_1}\in[1,k_1]$ die maximale $AUC$ berechnen, die dann für den dazugehörigen $PPV_{k_1}$ als obere Schranke fungiert. 
Ähnlich zum maximalen $PPV_{k_1}$ muss jetzt nur noch die gegebene $AUC=b$ mit der Liste abgeglichen werden und man kann den minimalen $PPV_{k_1}$ ablesen (vgl. Algorithmus \ref{algo_min_ppv}), welcher mit der gegebenen AUC vereinbar ist.
\lstset{language=Pascal,caption={Algorithmus zur Bestimmung des minimalen $PPV_{k_1}$ für feste $AUC$.},label=algo_min_ppv}
\begin{lstlisting}[frame=single,mathescape=true]
Liste_der_Grenzen =[]
for i:= 1 to k  do
begin
	{Liste_der_Grenzen += AUC_max(i/k1);}
end;
j=1;
while Liste_der_Grenzen[j]<b && j<k do
begin
	{j=j+1;}
end;
return Liste_der_Grenzen[j]
\end{lstlisting}

\subsubsection{Vergleich maximaler und minimaler $PPV_k$ für feste $AUC$\label{verg_ppv}}
\begin{figure}[h!]
\begin{minipage}[t]{0.45	\textwidth}
\includegraphics[width=\linewidth]{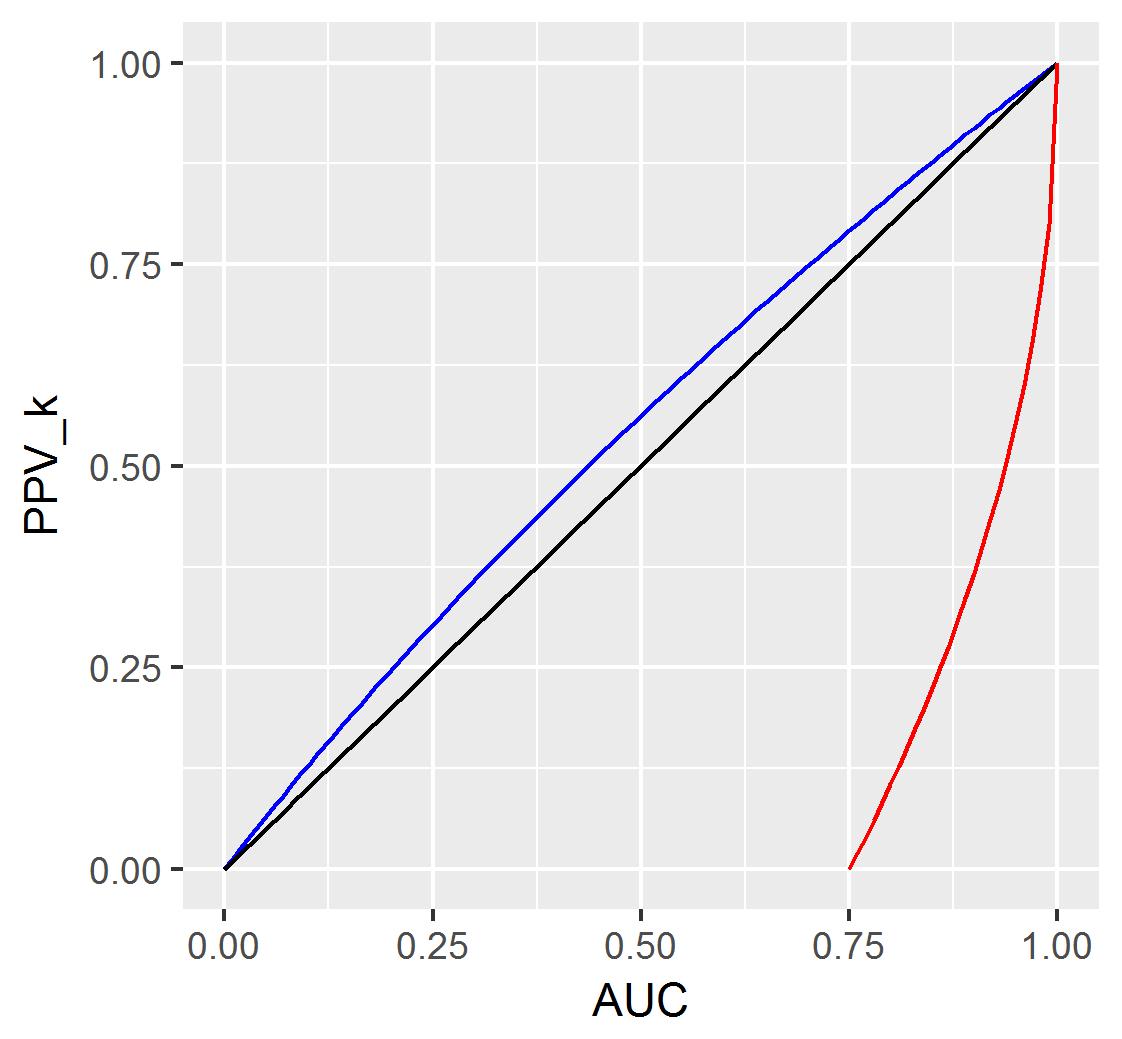}
\caption*{ $k_1: k_2$  ist $1:4$}

\end{minipage}
\hspace{\fill}
\begin{minipage}[t]{0.45\textwidth}
\includegraphics[width=\linewidth]{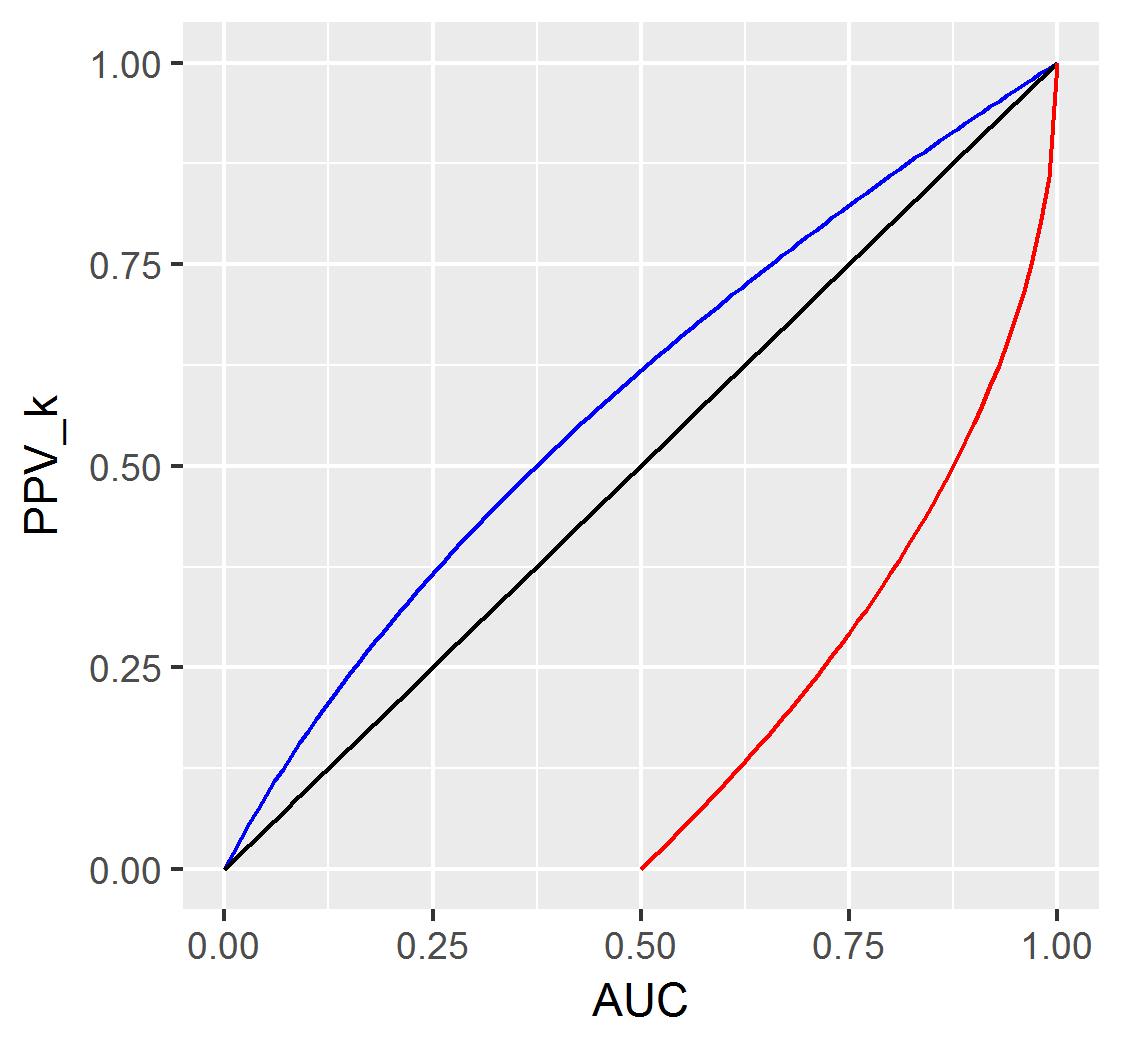}
\caption*{$k_1: k_2$  ist $2:4$}

\end{minipage}

\vspace*{0.5cm}
\begin{minipage}[t]{0.45\textwidth}
\includegraphics[width=\linewidth]{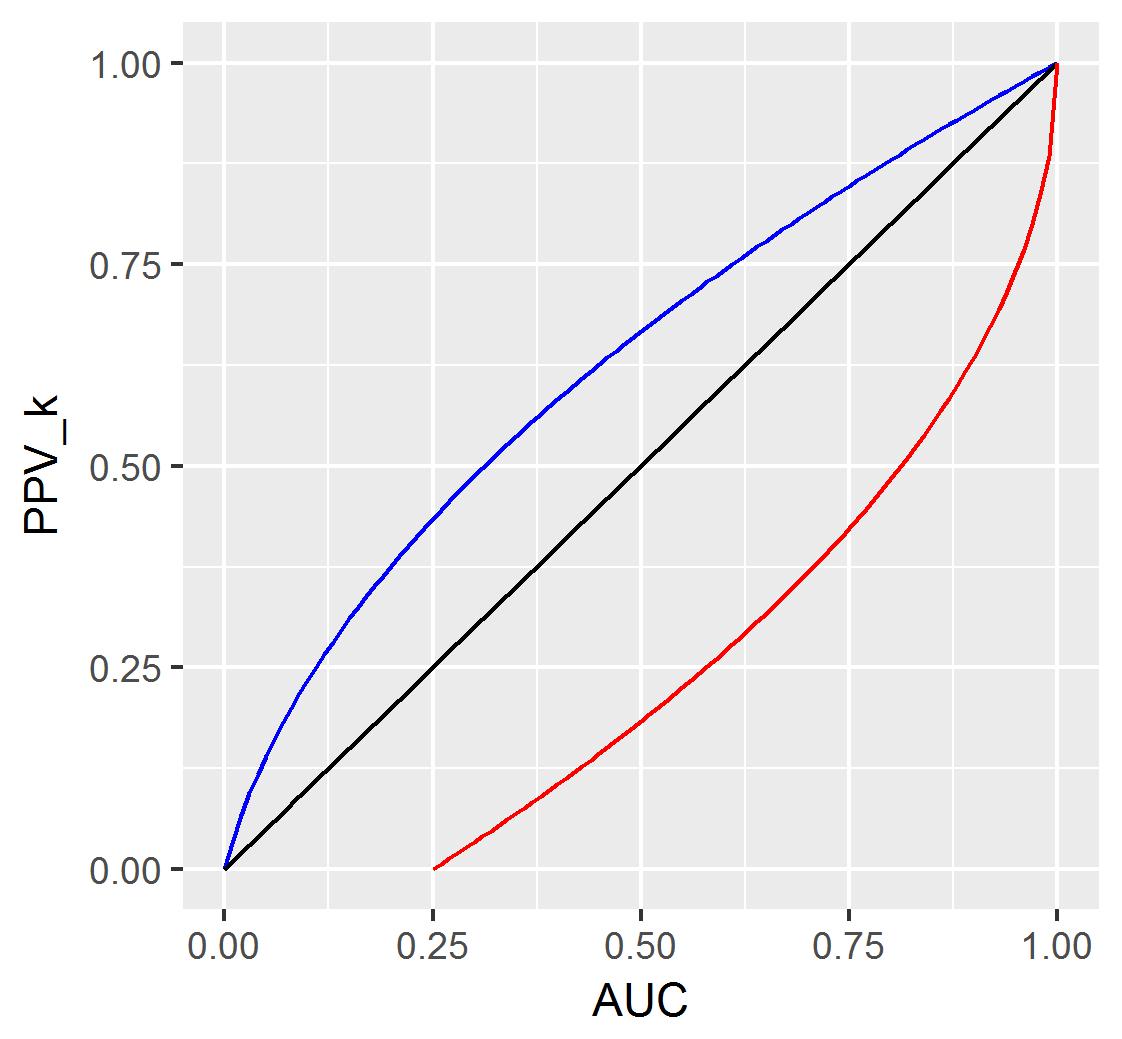}
\caption*{$k_1: k_2$  ist $3:4$}

\end{minipage}
\hspace{\fill}
\begin{minipage}[t]{0.45\textwidth}
\includegraphics[width=\linewidth]{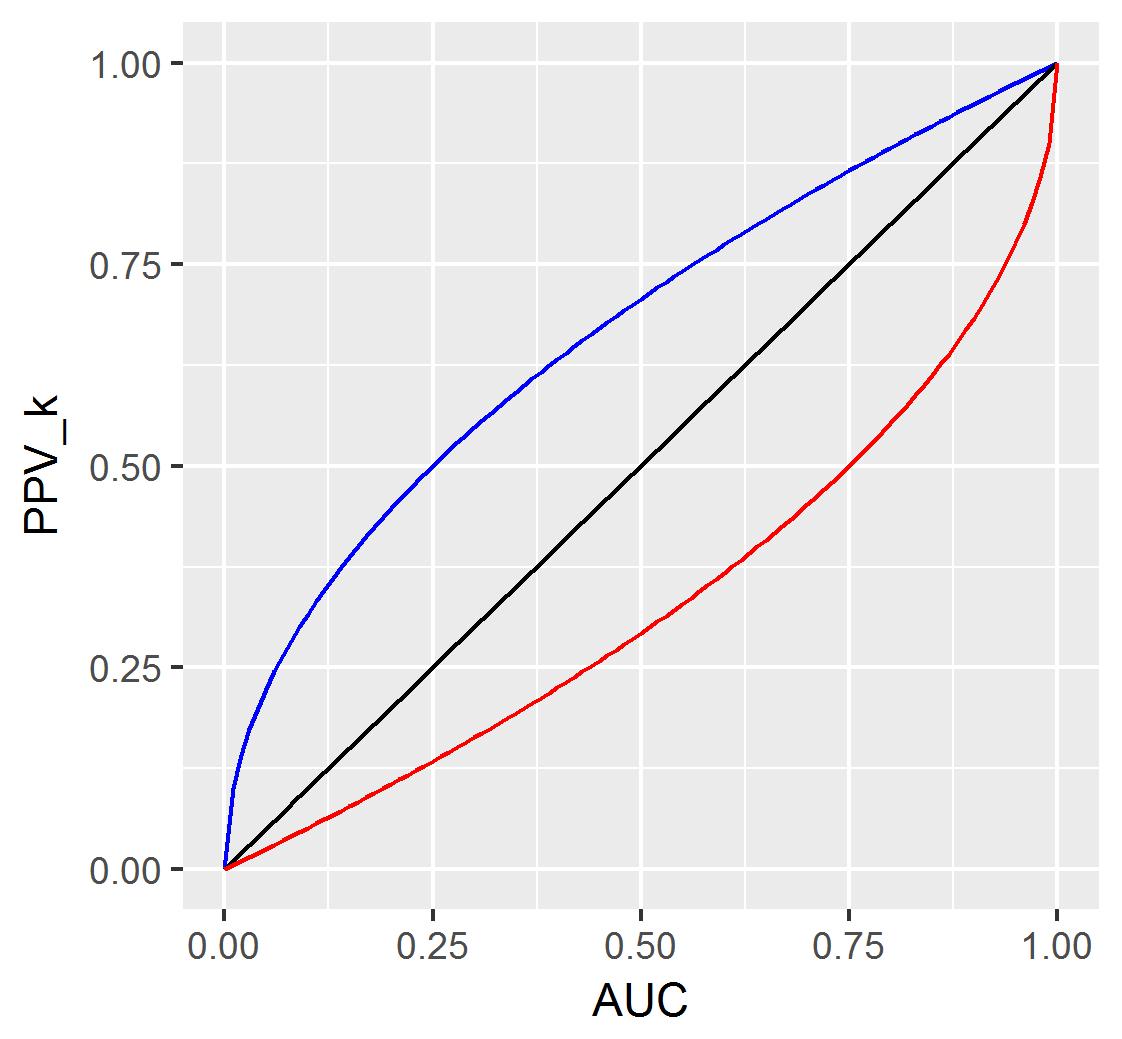}
\caption*{$k_1: k_2$  ist $1:1$}

\end{minipage}
\caption{Maximaler $PPV_k$ (blau) und minimaler $PPV_k$ (rot) für fixierte $AUC$ bei verschiedenen Klassengrößenverhältnissen}
\label{abb_ppv}
\end{figure}

Wie bereits eingangs erwähnt, wurde an dieser Stelle auf die im angehängten Paper erläuterte Formel zurückgegriffen, da sie als geschlossene Formel die Visualisierung der Ergebnisse deutlich vereinfacht hat.
Trägt man also sowohl den minimalen (rot) als auch den maximalen (blau) $PPV_k$ gegen die AUC auf, so erkennt man in Abbildung \ref{abb_ppv} das eklatante Ausmaß des möglichen $PPV_k$ Spielraums. Da gerade Rückfallrisiko-Klassifikatoren teilweise auf sehr unbalancierten Klassen arbeiten, könnte es sein, dass ein solcher mit einer AUC von 0.75 akzeptiert und eingesetzt wird (wie es üblich ist, vgl. Kapitel \ref{soa_auc}), im Extremfall lediglich einen $PPV_k$ von 0 aufweist. Das bedeutet, dass sich unter den k Personen, die laut Klassifikator die höchste Rückfallwahrscheinlichkeit aufweisen, keine Person befindet, die entsprechend der Ground-Truth rückfällig wird. Da sich die Justiz und die Gesellschaft aktuell noch lediglich auf die AUC als Bewertungsmaßstab für solche Prognose-Tools verlässt, ist der zu erkennende Spielraum mehr als alarmierend. Auch wenn nicht nicht unbedingt der Extremfall vorliegen muss, verbleibt die Frage warum die AUC dennoch immer noch als der beste Bewertungsmaßstab für solche Problemstellungen gilt~\cite[S.3]{Barnoski2007}.

\chapter{Überprüfung der Ergebnisse für das COMPAS Assessment Tool \label{ergebnisse}}
Die in Kapitel \ref{kap_bez_auc_ppv} erarbeiteten Ergebnisse geben zwar für eine binäre Klassifizierung Aufschlüsse über die Lage der jeweiligen maximal/minimal möglichen $PPV_k$ Werte bei fixierter $AUC$ und vice versa, jedoch können mittels dieser Formeln noch keine Rückschlüsse darüber getroffen werden, wie sich die Verteilung zwischen diesen Werten darstellt. Um die gesellschaftliche Brisanz der möglicherweise fehlenden Korrelation anhand von anwendungsbezogenen Daten zu erörtern, wurden die von ProPublica öffentlich bereitgestellten Datensätze\footnote{ \url{http://github.com/propublica/compas-analysis}} extrahiert und entsprechend analysiert. Diese enthalten alle 2013 und 2014 bewerteten Personen des Broward County Sheriff’s Office in Florida. Da hierbei das momentan in Wisconsin, USA in jeder Stufe des Justizprozesses angewandte~\cite{Wisconsin2017} \glq COMPAS Assessment Tool\grq~anhand echter Datensätze evaluiert werden kann, lässt sich überprüfen, ob im Anwendungsbezug die Korrelation von $AUC$ und $PPV_k$ gewahrt bleibt oder sie deutlich voneinander abweichen.
Es muss darauf hingewiesen werden, dass den folgenden Ergebnissen lediglich ein Algorithmus und ein lokal erfasster Datensatz zugrunde liegt. 
\newpage
\section{Definition und Erläuterung zum COMPAS Assessment Tool\label{compas}}
Die \glq \textbf{C}orrectional \textbf{O}ffender \textbf{M}anagement \textbf{P}rofile for \textbf{A}lternative \textbf{S}anctions\grq~, kurz COMPAS genannte Web-Applikation wurde vom \textit{Northpointe Institute for public manangement Inc}. als automatisierte Entscheidungsunterstützung zur Bewertung von Straffälligen entwickelt und vertrieben~\cite{Northpoint2017}. Das Tool wurde 1998  als "`Breitband"'-Bewertung konzipiert und kann anhand verschiedener Fragebögen in 22 verschiedenen Bedürfnis- und Risikobereichen Prognosen über Individuen erstellen~\cite{Northpoint2017}. So soll es den Hilfebedarf der Bewerteten erkennen und quantifizieren und bietet unter anderem die folgenden zwei Prognose-Möglichkeiten, die sich aus dem sogenannten \glq CORE Risk Assessment\grq~ - Fragebogen erstellen lassen, der aus 137 verschiedenen Fragen besteht~\cite{Angwin2016}: 
\begin{itemize}
	\item \glq \textit{General Recidivism Risk Scale}\grq :  Generelles Rückfallrisiko (GRRS)
	\item \glq \textit{Violent Recidivism Risk Scale}\grq :  Gewaltbasiertes Rückfallrisiko oder Rückfallrisko für Gewalttaten (VRRS)	
\end{itemize}
Im Folgenden wird vor allem auf diese beiden Modelle Bezug genommen, da sie, wie noch erläutert wird, in ihrer Anwendung einen binären Klassifikator (vgl. Kapitel \ref{def_klas}) abbilden. 
Für jeden Risikobereich ordnet das Prognose-Tool die Delinquenten in eine von zehn sogenannte Decile ein, die von Northpointe selbst geschaffen wurden.
Hierzu nahmen sie einen normativen Datensatz von 7.381 Personen (33.8\% mit kurzer Haftstrafe, meist kürzer als ein Jahr; 13,6\% mit längerer Haftstrafe und 52.6\% Personen mit Bewährungsstrafen)~\cite[S.11]{Northpointe2012} und beurteilten den Merkmalskatalog, welcher aus den Antworten der 137 Fragen resultiert, durch einen Bewertungsalgorithmus und erhielten jeweils einen \glq Raw Score\grq. Dieser soll die Prognose im jeweiligen Risikobereich widerspiegeln, jedoch fehlt zunächst der Maßstab, da er scheinbar willkürlich skaliert. Nach der Sortierung anhand des \glq Raw Score\grq~unterteilte Northpointe deshalb den Datensatz in  10 gleich große Gruppen, die sogenannten Decile~\cite[S.8]{Northpointe2012}. Ausschlaggebend für das Decile, in das eine Person kategorisiert wird, ist also der aus den Antworten errechnete Raw Score, wobei an dieser Stelle darauf verwiesen werden muss, dass die genaue Berechnung dieses Wertes als Firmengeheimnis geschützt wird und sich jeder demokratischen Kontrolle entzieht.\\
Northpointe sieht die entstehenden Decile wie folgt interpretiert~\cite[S.8]{Northpointe2012}:
\begin{itemize}
	\item 1-4: Niedriges relatives Risiko im Vergleich zur Normgruppe 
	\item 5-7: Mittleres relatives Risiko im Vergleich zur Normgruppe
	\item 8-10: Hohes relatives Risiko im Vergleich zur Normgruppe 	
\end{itemize}
Anhand dieser Kategorisierung wird deutlich, dass eine richterliche Nutzung, wie sie von Northpointe \cite{Northpointe2012} angedacht und auch unter anderem im Report "`Evaluation of the Pretrial Services Program"' \cite{Lukic2009} an das Broward Sheriff’s Office vom amerikanischen Justizwesen intern empfohlen wird, aus den drei Kategorien zwei schaffen wird. Die kontinuierliche Vorhersage der einzelnen Skalen wird im Justizwesen auf binäre Entscheidungen zu Fragen, ob der verurteilte Straftäter auf Bewährung das Gefängnis verlassen darf oder nicht, abgebildet. Es ist somit sehr wahrscheinlich, dass ein niedriges bis mittleres Rückfälligkeitsrisiko zusammengefasst und gegen die Gruppe der Personen mit einer hohen Rückfälligkeitswahrscheinlichkeit aufgeführt wird. Somit bietet sich in der Evaluation das Spektrum eines binären Klassifikators an, weshalb Northpointe selbst die hohe $AUC$ seines Algorithmus lobt ~\cite{Northpointe2012} und mit der $AUC$ in verschiedenen Studien wirbt, die den COMPAS GRRS mit den in Tabelle \ref{eval_tab} aufgeführten $AUC$-Werten beurteilen~\cite{Northpointe2012}.
 
\begin{table}[]
\centering
\small
\begin{tabular}{|l|l|l|l|l|}
\hline
\textbf{Quelle}                       & \textbf{Jahr} & \textbf{Stichprobengröße} & \textbf{\begin{tabular}[c]{@{}l@{}}Betrachtungszeitraum \\~~ der Rückfälligkeit\end{tabular}} & \textbf{$AUC$}  \\ \hline
\cite{Brennan2009a}\cite{Brennan2009} & 2009          & 2.328                     & 1 Jahr                                                                                      & \textbf{0.68} \\ \hline
\cite{Farabee2010}                    & 2010          & 25.009                    & 2 Jahre                                                                                     & \textbf{0.70} \\ \hline
\cite{Lansing2012}                    & 2012          & 11.289                    & 2 Jahre                                                                                     & \textbf{0.71} \\ \hline
\end{tabular}
\caption{Auszug der von Northpointe veröffentlichten Liste an Evaluationen des COMPAS Assessment Tool~\cite{Northpointe2012}}
\label{eval_tab}
\end{table}

\section{Auswertung der ProPublica Daten zum \mbox{COMPAS} Assessment Tool\label{propub_analyse}}
Dem vorliegenden ProPublica-Datensatz konnten für 11.777 Personen alle notwendigen Informationen\footnote{ eindeutige Personen ID, RAW Score für das GRRS/VRRS, wirkliche Rückfälligkeit der Person innerhalb von zwei Jahren} entnommen werden, um sowohl die $AUC$ als auch den erreichten $PPV_k$ zu bestimmen. Der Wert für die tatsächliche Rückfälligkeit eines Individuums bezieht sich in dem durch ProPublica offerierten Datensatz auf ein Zweijahresfenster, soweit es das Broward County Sheriff’s Office in Florida erfassen konnte.
Bezüglich der \glq General Recidivism Risk Scale\grq~(GRRS), also Northpointes Prognose für das generelle Rückfallrisiko, weisen die Daten eine Klassenverteilung von  $|\textrm{Rückfällige}| =  4.262$ und  $|\textrm{nicht~Rückfällige}| = 7.515$ auf. Somit enthält er mit 36,189\% eine Rückfallrate/Basisrate, welche den Basisraten in den von Northpointe selbst zitierten~\cite{Northpointe2012} Validationsstudien ähnelt. Lansing et al. erfassten so zum Beispiel auf ihrem 11 289 großen Datensatz eine generelle Rückfallrate von 34\% \cite{Lansing2012}. Auch konnten sie eine $AUC$ von 0.71 feststellen, welche die Akkuratheit des vorliegenden ProPublica-Datensatzes, die eine $AUC$ von 0.69  enthält, weiter untermauert. 
Legt man die dem Raw Score entspringende Sortierung zu Grunde, so schafft es das COMPAS Tool lediglich 2.260 tatsächlich rückfällige Straftäter unter die vorderen 36,189\% der Personen (4.262) zu bringen. Dies entspricht dem folgenden $PPV_k$:
$$PPV_k = \frac{2260}{4262} = 0.5302674800563116\approx 0.53$$

\begin{figure}
\begin{minipage}[t]{0.45\textwidth}
\includegraphics[width=\linewidth]{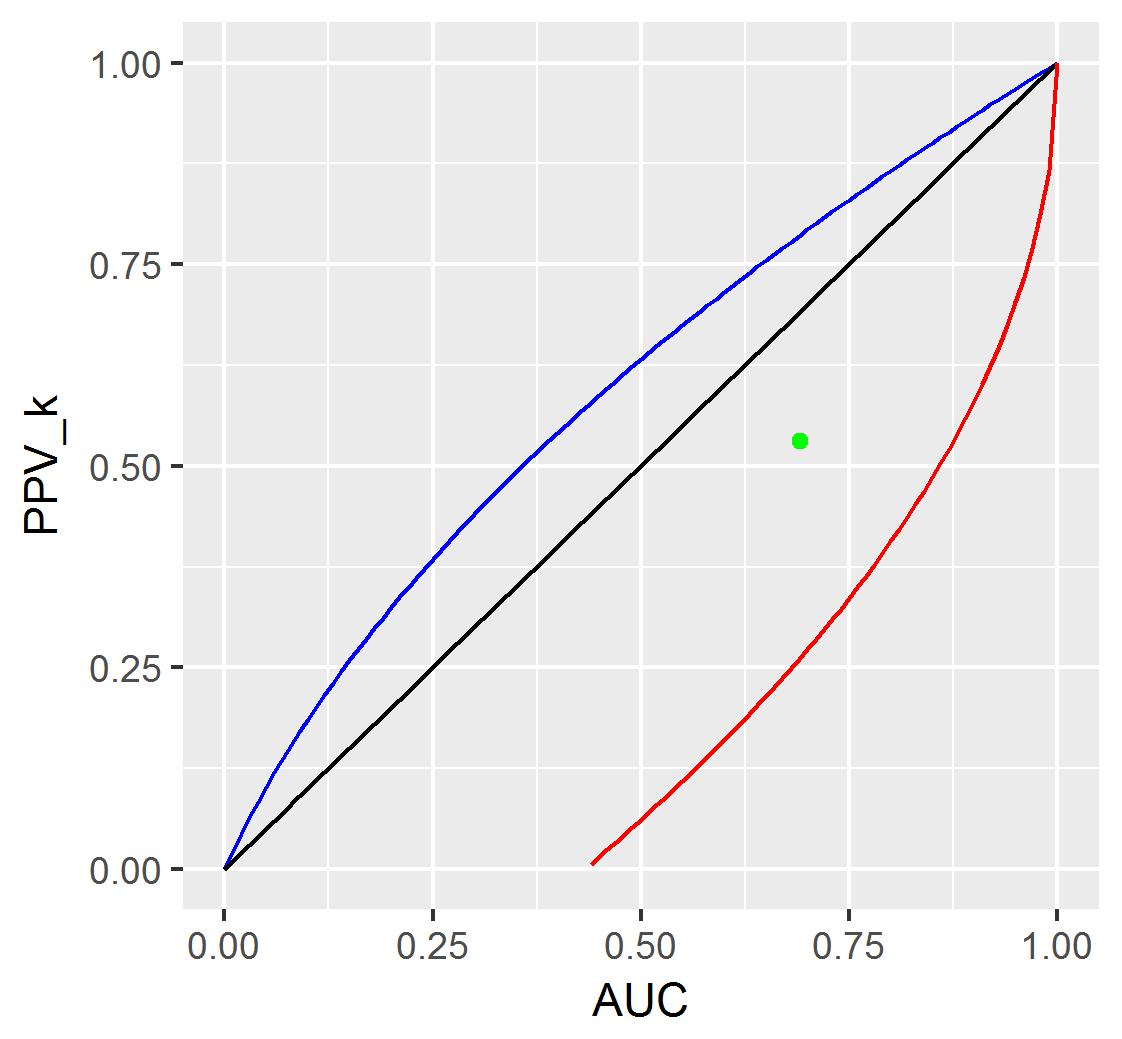}
\small
\caption{Auswertung des COMPAS Scores (GRRS) auf den ProPublica Daten (grüner Punkt) bei eingezeichnetem maximalen $PPV_k$ (blau) und minimalen $PPV_k$ (rot) für vorliegendes Klassenverhältnis:}
\label{abb_compas_recid}
\small
\begin{tabular}{lll}
$n$                        & = & 11.777              \\
$|\textrm{Rückfällige}|$            & = & 4.262               \\
$|\textrm{nicht~Rückfällige}|$ & = & 7.515               
\end{tabular}
\small
\begin{tabular}{lll}
$$AUC$$                      & = & 0.6909022561790231 $\approx$ 0.69 \\
$PPV_k$                  & = & 0.5302674800563116 $\approx$ 0.53
\end{tabular}

\end{minipage}
\hspace{\fill}
\begin{minipage}[t]{0.45\textwidth}
\includegraphics[width=\linewidth]{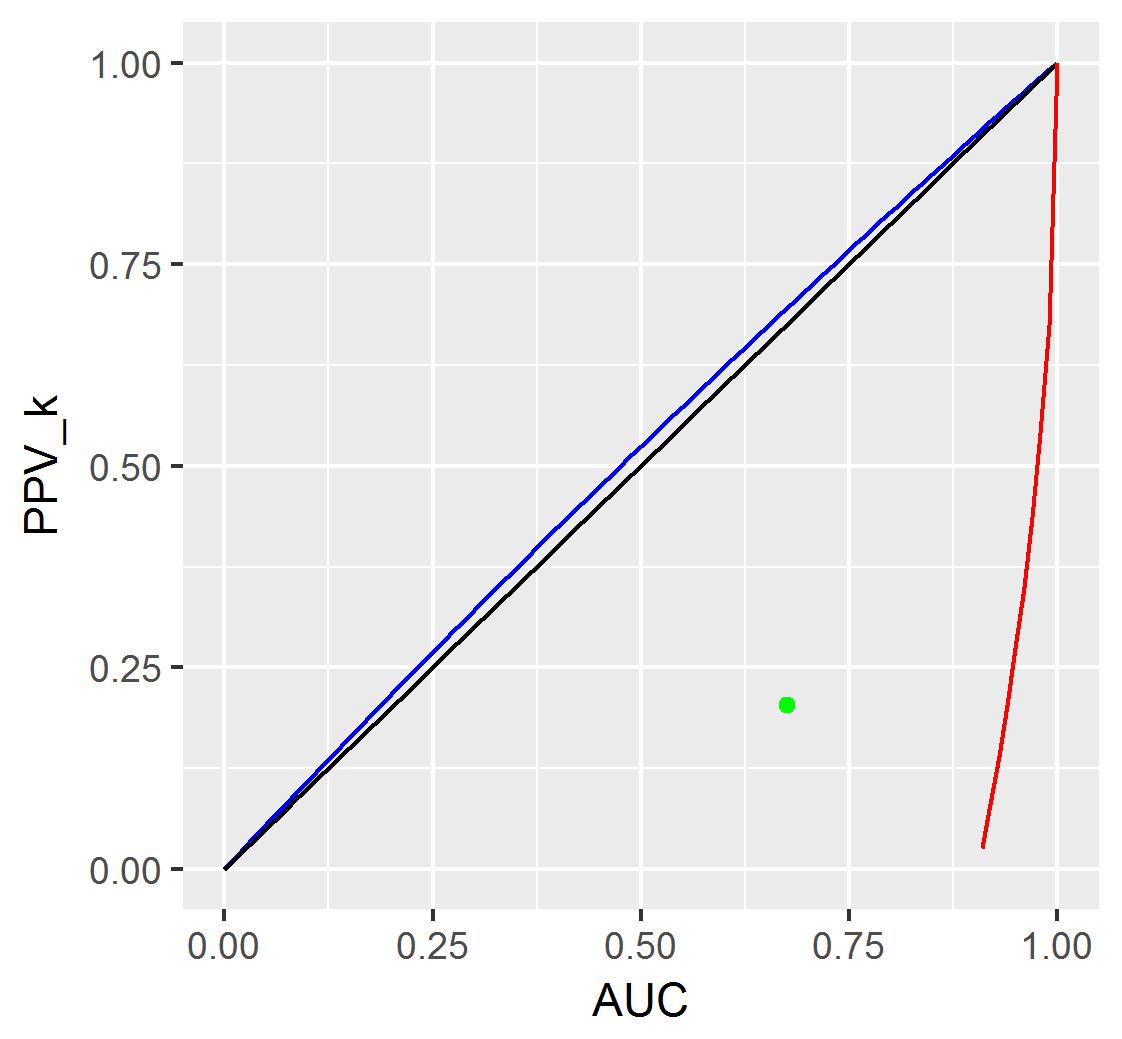}
\small
\caption{Auswertung des COMPAS Scores (VRRS) auf den ProPublica Daten (grüner Punkt) bei eingezeichnetem maximalen $PPV_k$ (blau) und minimalen $PPV_k$ (rot) für vorliegendes Klassenverhältnis:}
\small
\begin{tabular}{lll}
$n$                        & = & 12.526              \\
$|\textrm{gewaltbasiert~Rückfällige}$|& = & 1.085               \\
$|\textrm{nicht~gewaltbasiert~Rückfällige}|$& = & 11.441               
\end{tabular}
\small
\begin{tabular}{lll}
$$AUC$$                      & = & 0.6760433512426204~~$\approx$ 0.68 \\
$PPV_k$                  & = & 0.20276497695852536 $\approx$ 0.20
\end{tabular}
\label{abb_compas_violent}
\end{minipage}
\end{figure}

In Abbildung \ref{abb_compas_recid} ist grün visualisiert, dass der erreichte $PPV_k$ zwar im senkrechten Bereich des möglichen Wertebereiches (zwischen rot und blau), den dieser bei vorliegender $AUC$ (0.69) annehmen kann\footnote{ [0.2616143, 0.7862506]},  jedoch deutlich unter der $AUC$ liegt, die durch die schwarze Diagonale ersichtlich ist. Diese würde einer Identität der beiden Maße entsprechen. 

Wenn angenommen wird, dass ein Richter in seinem Distrikt den vorliegenden Datensatz zu bearbeiten hat, so hätte er bei der zufälligen Verurteilung eines Delinquenten entsprechend der im Datensatz vorherrschenden Basisrate eine Wahrscheinlichkeit von 36,189\%, dass dieser rückfällig wird. Nutzt er nun das vorgestellte COMPAS Tool und kann auf Grund seiner Erfahrung mit dem Distrikt die Basisrate korrekt abschätzen, verurteilt also nur die Straftäter, die in den oberen 36,189\% der Datenpunkte\footnote{ sortiert nach dem Raw Score} liegen, so erreicht er, der Interpretation des $PPV_k$ (siehe Kapitel \ref{def_ppv}) folgend, lediglich mit einer Wahrscheinlichkeit von ungefähr 53,023\% einen wirklich rückfällig Werdenden.
Es ist zu beachten, dass dieser Betrachtung die Annahme zugrunde liegt, welche der einfachheitshalber aufgestellt wurden, jedoch das Ergebnis für Northpointe noch fataler ausfallen lassen können. Zum einen muss den Richtern aufgrund ihrer Erfahrung eine deutlich höhere Trefferquote zugesprochen werden als in der erwähnten Monte-Carlo-Simulation, in der zufällig gezogen wird. Zum anderen wird kaum ein Richter die in seinem Distrikt vorherrschende Rückfallwahrscheinlichkeit explizit nennen können, da diese tagesaktuellen sowie individuellen Schwankungen unterliegt.   
Obwohl von diesen sehr idealisierten Annahmen ausgegangen wird, liegt der Anstieg in der Güte der Entscheidung lediglich bei 16,834 Prozentpunkten.

Viel drastischer stellt sich der Unterschied zwischen $AUC$ und $PPV_k$ für die \glq Violent Recidivism Risk Scale\grq~(VRRS) dar. Vorliegend schafft es das COMPAS Risk Assessment Tool in seiner Prognose für die gewaltbasierte Rückfälligkeit zwar auf einen $AUC$ von 0.68, jedoch kann es anhand der Raw Scores lediglich 220 Rückfällige in die ersten 1085 Straftäter sortieren, was sich in folgendem $PPV_k$ widerspiegelt: 
$$PPV_k = \frac{220}{1085}=0.20276497695852536\approx 0.20$$

In Abbildung \ref{abb_compas_violent} ist durch den grünen Punkt zu erkennen, dass es Northpointe mit diesem Klassifikator deutlich in das untere Drittel des möglichen Wertebereichs\footnote{ [0,0.6967742]} für den $PPV_k$ mit vorliegender $AUC$ von 0.69 schafft.
Es ist jedoch zu beachten, dass der Klassenunterschied mit geringer Basisrate von 8.662\% deutlich größer ist, sodass es wiederum schwieriger wird, eine Klassifizierung zu schaffen, welche die wenigen Rückfälligen von den vielen nicht Rückfälligen separiert. Wie oben betrachtet, würde ein Richter hier durch zufällige Urteile einen tatsächlich gewaltbasiert Rückfälligen lediglich mit einer der Basisrate entsprechenden Wahrscheinlichkeit von 8.662\% erfassen. Dieser Wert kann zwar durch die Anwendung des COMPAS Assessment Tool um 11.614 Prozentpunkte erhöht werden, jedoch liegt die Trefferwahrscheinlichkeit innerhalb der ersten 1.085 Straffälligen\footnote{ sortiert nach dem Raw Score} immer noch nur bei 20.276\% und gibt dem Anwender keine wirklich stichhaltige Prognose, auf der sein Urteil aufgebaut werden sollte.

Da es jedoch einem Richter wahrscheinlich schwerfallen wird, eine exakte, der aktuellen Situation und betrachteten Gegend entsprechende Prognose zur Basisrate der Rückfälligkeit aufzustellen und auch der zeitliche sowie mathematische Aufwand der Raw Score-Auswertung für einen amtierenden Richter unter Umständen zu aufwändig wäre, wird man gezwungen sein, die von Northpointe aufgestellten Decile und deren Interpretation zu nutzen. \\
Sollte sich der Richter allerdings vollständig auf die oben erörterte Decilen-Interpretation verlassen, also alle Straftäter mit einem hohen relativen Risiko im Vergleich zur Normgruppe (Decile 8-10) verurteilen, so wäre ein solcher Straftäter dem zugrundeliegenden Datensatz entsprechend für der GRRS nur zu 57.82\% \footnote{ 57.82060785767235\%} ein tatsächlich Rückfälliger. Obwohl in der GRRS eine Verbesserung\footnote{ 4.8 Prozentpunkte} zu verzeichnen ist, sinkt gerade im der bereits kritischen VRRS die letztendliche Trefferquote im Vergleich zum $PPV_k$ um ganze 2.515 \%, da hier lediglich 17.76\% \footnote{ 17.76061776061776\%} der Straftäter in den Decilen 8-10 eine Gewalttat begingen.

\subsection{Threats to validity --- Gefahren für die Gültigkeit der Ergebnisse}
Die erarbeiteten Ergebnisse weisen lediglich dann eine deutliche Diskrepanz zwischen $AUC$ und $PPV_k$ für das COMPAS Assessment Tool auf, sollte die von ProPublica erfassten Daten korrekt sein. Viele der erfassten Merkmale sind zwar kurz erläutert, die genauen Erfassungskriterien sind jedoch meist nicht näher geklärt. Zum Beispiel ist nicht ersichtlich, wie das Broward County Sheriff Rückfälligkeit exakt definiert. Selbiges gilt für den betrachteten Zweijahreszeitraum. Hier ist es fraglich, ob nur Straftäter aufgeführt werden, die tatsächlich zwei Jahre auf freiem Fuß waren, oder ob eine Reststrafe von einigen Monaten oder länger auch Betrachtung fand. Denn das Rückfallrisiko im Gefängnis könnte sich deutlich von dem außerhalb unterscheiden und der COMPAS ist für eine solche Prognose nicht ausgelegt. Auf diese und andere Gefahren bezüglich der Datenlage wird in der folgenden Abschlussbetrachtung näher eingegangen.
\chapter{Schlussfolgerung/Fazit \label{fazit}}

Die hohe Nutzungsquote der AUC bei binären Klassifikatoren für Rückfälligkeitsvorhersagen erscheint willkürlich und erklärt sich lediglich durch ihre verbreitete Anwendung in der Historie des maschinellen Lernens. Die daher in der vorliegenden Arbeit unternommene  Überprüfung der $AUC$ hinsichtlich ihrer Eignung als Bewertungsmaßstab in der ADM-gestützten Kriminalprognose blieb deutlich hinter den Erwartungen zurück.
Es wurde gezeigt, dass die Abweichung vom näher am richterlichen Entscheidungsprozess evaluierenden $PPV_k$ inakzeptabel hoch sein kann, sodass eine weitere zukünftige Heranziehung der $AUC$ als ausschließliches Qualitätsmaß kritisch zu hinterfragen ist. Bereits 2008 sprechen \textit{Urbanoik et al.} bei der ROC-Analyse von einer "`Dichotomomisierung des Rückfallrisikos"'~\cite{Urbaniok2008}.\\
Grundsätzlich muss für den Positive Predictive Value erwähnt werden, dass seine Verwendung als ausschließliches Qualitätsmerkmal in der Fachliteratur eher kritisch gesehen wird, worauf Daniel B. Kopans  bereits 1992 am Beispiel der Brustkrebsvorsorge (Mammographie)~\cite{Kopans1992} hinwies. Er zeigte unter anderem auf, dass zum Beispiel Sensitivität, Spezifität und Klassengrößen jeweils einen großen Einfluss auf die Güte eines Klassifikators haben, was gegen eine ausschließliche, unangepasste Nutzung dieser Maße spricht.\\
Sollte also die AUC weiterhin in der Justiz Verwendung finden, müssten Gutachter wie Richter die Aussagekraft eines hohen AUC-Wertes richtig zu interpretieren wissen und dürften die Fähigkeit der Instrumente, etwaige Rückfälle zu prognostizieren, aufgrund des Vorliegens hoher Validitätswerte nicht überschätzen~\cite{Eher2008}. Hier bedarf es der Aufklärung, da die Tendenz besteht, dass Menschen softwarebasierte Prognosen als verlässlicher, objektiver und aussagekräftiger als andere Informationen empfinden, was die Gefahr einer unkritischen Übernahme dieser Empfehlungen und Prognosen birgt. \\\\
Die in dieser Arbeit aufgezeigten Probleme bezüglich der Datenlagen haben weiterhin gezeigt, dass es von immenser Bedeutung ist, das Anwendungsgebiet und die dort vorherrschende Datenlage und Qualität genauestens mit der Lernumgebung der Algorithmen abzugleichen (vgl. \cite{Burnham2002}), denn "`\textit{in jedem Fall hängt der Klassifikationsvorgang eng mit der Beschaffung der Information zusammen; aufgrund dieser Interdependenz zwischen Klassifikation und Merkmals- bzw. Informationsgewinnung, die besonders bei der Behandlung realer Aufgabenstellungen wie z. B. modernen Bildverarbeitungssystemen oder Systemen zur automatischen Fehlerdiagnose zu Tage tritt, muss dem Gesamtsystem besonderer Stellenwert beigemessen werden. Grundsatz: der beste Klassifikationsalgorithmus ist gerade so gut wie die ihm vorliegende Information}"'~\cite{Hengen2004}.\\
Diese These wurde 2002 in der Medizin im speziellen Umfeld von Röteln und Herpes durch Burnham weiter untermauert, der den starken Einfluss des Abgleichs der Lernumgebung mit dem Datensatz, auf dem gearbeitet wird, auf das Klassifizierungsergebnis nachwies, indem er zeigte, dass der PPV durch die Einbeziehung der Krankengeschichte als weiteres Merkmal der Patienten sehr stark beeinflusst wird~\cite{Burnham2002}. \\\\

\textbf{Mögliche Konsequenzen und Forderungen:}\\\\
Sollte im deutschen Justizwesen die Einführung ADM-gesteuerter Prozesse zur Debatte stehen, wäre Deutschland in der glücklichen Lage von den Erfahrungen und Fehlern anderer Länder, wie den USA zu profitieren. Daher ist zu hoffen, dass dies ohne überstürzten politischen Aktionismus, sondern mit Bedacht nach einer ausführlichen, gesellschaftlichen Debatte erfolgt. Keinesfalls sollte es zu einer voreiligen Nutzung von Algorithmen kommen, wie es seinerzeit beim Jugendstrafrecht in den USA der Fall war~\cite[S.2]{Baird2009}.\\
Es muss betont werden, dass bei der Risikoprognose ein Mensch im Mittelpunkt algorithmischer Betrachtung steht, sodass Auswahl, Überprüfung und Nutzung von algorithmischen Entscheidungshilfen ein Höchstmaß an wissenschaftlicher Aufmerksamkeit geschenkt werden muss. Schließlich geht es hier nicht um die bloße Einordnung eines Menschen in die richtige Stufe einer KFZ-Versicherung, sondern um existenzielle Urteile für die Betroffenen. Fehlurteile könnten fatale Auswirkungen auf das Leben eines Betroffenen haben. Als Beispiel sei der \glq Feedback Loop\grq~genannt, nach dem ein \glq false positive\grq , also fälschlicherweise Verurteilter, tatsächlich -- im Sinne der \glq self fulfilling prophecy\grq\footnote{ selbsterfüllende Prophezeiung}~\cite{Azariadis1981}~-- kriminell wird.\\
Daher bedarf es im Vorfeld einer kritischen gesellschaftlichen Debatte, die sich aber ohne präzisere Einsicht in die algorithmische Entscheidungsfindung nicht führen lässt. Die Frage beispielsweise, ob ein ADM-Prozess ein adäquates Konzept von Fairness verwendet, kann ohne Kenntnis des spezifischen Modells und all seiner Annahmen nicht überprüft werden. Grundsätzlich sollten normative Entscheidungen, wie z. B. über Fairness-Kriterien, schon bei der Gestaltung eines ADM-Prozesses im Konsens mit der Gesellschaft gefällt werden, die natürlich dementsprechend informiert und geschult werden müsste. \\
Sinnvolle Handlungsempfehlungen für die gesellschaftliche Debatte gibt das\\ ADM-Manifest~\cite{AlgorithmWatch2016} in den Punkten 3 und 4:
\begin{enumerate}
	\setcounter{enumi}{2}
	\item \textit{ADM-Prozesse müssen nachvollziehbar sein, damit sie demokratischer Kontrolle unterworfen werden können.}
	\item \textit{Demokratische Gesellschaften haben die Pflicht, diese Nachvollziehbarkeit herzustellen: durch eine Kombination aus Technologien, Regulierung und geeigneten Aufsichtsinstitutionen }
\end{enumerate}
Zunächst sind hier die Vertreiber gefordert, die Algorithmen ihrer angebotenen Tools insoweit offenzulegen, "`dass Erklärbarkeit, Nachvollziehbarkeit, unabhängige Überprüfbarkeit und die Möglichkeiten zur forensischen Datenanalyse gegeben sind"'~\cite{Lischka2017}. Sollte dies nicht geschehen, läge es in der Verantwortung des Staates, durch geeignete Maßnahmen eine informierte Debatte zu ermöglichen. \\
Hierzu gehört auch, dass finanzielle Mittel zur Verfügung gestellt werden, damit in dieser Hinsicht eine weiterführende Forschung betrieben werden kann. In dem Zusammenhang sei darauf hingewiesen, dass es kein akzeptabler Status quo ist, wenn kritische Untersuchungen zu Risiken und gesellschaftlichen Folgen der ADM-Prozesse abhängig vom Interesse und verfügbaren Budget beliebiger Institutionen sind. So ist der Anstoß der Fairness Debatte um das COMPAS-Tool primär dem Engagement und den Studien des Recherchebüros ProPublica zu verdanken.\\\\
Abschließend lässt sich konstatieren, dass die Gesellschaft bei der Integration eines Risikovorhersage-Instruments die Auswahl des Bewertungsmaßstabs als eine der wichtigsten Modellierungsentscheidungen verstehen muss, denn:

\begin{center}

\textit{"`We recognize that creation of valid, reliable, and robust risk assessment instruments is both a science and an art."'\footnote{Wir stellen fest, dass die Erstellung eines fundierten, zuverlässigen und robusten Risikobeurteilugsinstruments sowohl eine Wissenschaft als auch eine Kunst ist.~\cite[S.10]{Baird2009}}
}
\end{center}

\RRLABbibliography{ml}

\include{appendix}



\begin{thebibliography}{}

\bibitem[Aggarwal 15]{Aggarwal2015}
\textsc{Aggarwal,C.~C.}: {\em Data mining: the textbook}. Springer, 2015

\bibitem[Albrecht 03]{Albrecht2003}
\textsc{Albrecht,G.}: Probleme der Prognose von Gewalt durch psychisch Kranke.
\newblock In: {\em Journal f{\"u}r Konflikt-und Gewaltforschung}. 5, (2003),
  Nr. 1, S.\ 97--126, 2003

\bibitem[Albrecht 04]{Albrecht2004}
\textsc{Albrecht,G.}: Sinn und Unsinn der Prognose von Gewalt. In
  \textsc{Heitmeyer,W.} (Hrsg.): {\em Gewalt Entwicklungen, Strukturen,
  Analyseprobleme}. Suhrkamp, 2004, S.\ 475--524

\bibitem[{Algorithm Watch} 16]{AlgorithmWatch2016}
\textsc{{Algorithm Watch}}.
\newblock {\em Das ADM-Manifest | The ADM Manifesto}.
\newblock \url{https://algorithmwatch.org/das-adm-manifest-the-adm-manifesto/}.
  2016, [Online; accessed 06-May-2017]

\bibitem[Altman 94]{Altman1994}
\textsc{Altman,D.~G.} ; \textsc{Bland,J.~M.}: Diagnostic tests. 1: Sensitivity
  and specificity.
\newblock In: {\em BMJ: British Medical Journal}. 308, (1994), Nr. 6943, S.\
  1552, 1994

\bibitem[{Amnesty International} 13]{AmnestyInternational2013}
\textsc{{Amnesty International}}.
\newblock {\em Amnesty Report - Vereinigte Staaten von Amerika}.
\newblock
  \url{https://www.amnesty.de/jahresbericht/2013/vereinigte-staaten-von-amerika}.
  2013, [Online; accessed 06-May-2017]

\bibitem[{Amnesty International} 15]{AmnestyInternational2015}
\textsc{{Amnesty International}}.
\newblock {\em Amnesty Report - Vereinigte Staaten von Amerika}.
\newblock
  https://www.amnesty.de/jahresbericht/2015/vereinigte-staaten-von-amerika.
  2015, [Online; accessed 06-May-2017]

\bibitem[Andrews 06]{Andrews2006}
\textsc{Andrews,D.~A.}: The Recent Past and Near Future of Risk and/or Need
  Assessment.
\newblock In: {\em Crime {\&} Delinquency}. 52, (2006), Nr. 1, S.\ 7--27, jan
  2006

\bibitem[Angwin 16]{Angwin2016}
\textsc{Angwin,J.} ; \textsc{Larson,J.} ; \textsc{Mattu,S.} ;
  \textsc{Kirchner,L.}: Machine Bias -There’s software used across the
  country to predict future criminals. And it’s biased against blacks.
\newblock In: {\em ProPublica}. 23.05.2016, (2016), 2016.
\newblock
  \url{https://www.propublica.org/article/machine-bias-risk-assessments-in-criminal-sentencing},
  . [Online; accessed 05-Mai-2017]

\bibitem[Azariadis 81]{Azariadis1981}
\textsc{Azariadis,C.}: Self-fulfilling prophecies.
\newblock In: {\em Journal of Economic Theory}. 25, (1981), Nr. 3, S.\
  380--396, dec 1981

\bibitem[Baird 09]{Baird2009}
\textsc{Baird,C.}: A question of evidence: A critique of risk assessment models
  used in the justice system.
\newblock In: {\em Madison, WI: National Council on Crime and Delinquency}.
  (2009), 2009

\bibitem[Barnoski 07]{Barnoski2007}
\textsc{Barnoski,R.} ; \textsc{Drake,E.}: Washington's Offender Accountability
  Act: Department of Corrections' Static Risk Instrument.
\newblock {T}echnischer {B}ericht, Washington State Institute for Public
  Policy, 2007

\bibitem[Beck 86]{Beck1986}
\textsc{Beck,J.~R.} ; \textsc{Shultz,E.~K.}: The use of relative operating
  characteristic (ROC) curves in test performance evaluation.
\newblock In: {\em Archives of pathology \& laboratory medicine}. 110, (1986),
  Nr. 1, S.\ 13--20, 1986

\bibitem[Beigi 11]{Beigi2011}
\textsc{Beigi,M.~M.} ; \textsc{Behjati,M.} ; \textsc{Mohabatkar,H.}: Prediction
  of metalloproteinase family based on the concept of Chou’s pseudo amino
  acid composition using a machine learning approach.
\newblock In: {\em Journal of Structural and Functional Genomics}. 12, (2011),
  Nr. 4, S.\ 191--197, dec 2011

\bibitem[Beizer 95]{Beizer1995}
\textsc{Beizer,B.}: {\em Black-box Testing: Techniques for Functional Testing
  of Software and Systems}. New York, NY, USA: John Wiley \& Sons, Inc., 1995

\bibitem[Bliesener 14]{Bliesener2014}
\textsc{Bliesener,T.} ; \textsc{L{\"o}sel,F.} ; \textsc{K{\"o}hnken,G.}: {\em
  Lehrbuch der Rechtspsychologie}. Hans Huber, 09 2014

\bibitem[Boetticher 07]{Boetticher2007}
\textsc{Boetticher,A.} ; \textsc{Kröber,H.-L.} ; \textsc{Müller-Isberner,R.}
  ; \textsc{Böhm,K.~M.} ; \textsc{Müller-Metz,R.} ; \textsc{Wolf,T.}:
  Mindestanforderungen für Prognosegutachten.
\newblock In: {\em Forensische Psychiatrie, Psychologie, Kriminologie}. 1,
  (2007), Nr. 2, S.\ 90--100, feb 2007

\bibitem[Bradley 97]{Bradley1997}
\textsc{Bradley,A.~P.}: The use of the area under the {ROC} curve in the
  evaluation of machine learning algorithms.
\newblock In: {\em Pattern Recognition}. 30, (1997), Nr. 7, S.\ 1145--1159, jul
  1997

\bibitem[Brennan 09a]{Brennan2009a}
\textsc{Brennan,T.} ; \textsc{Dieterich,B.} ; \textsc{Breitenbach,M.} ;
  \textsc{Mattson,B.}
\newblock {\em A Response to “Assessment of Evidence on the Quality of the
  Correctional Offender Management Profiling for Alternative Sanctions
  (COMPAS)”}.
\newblock
  \url{http://www.northpointeinc.com/files/whitepapers/Response_to_Skeem_Louden_Final_071509.pdf}.
  June 2009, [Online; accessed 28-March-2017]

\bibitem[Brennan 09b]{Brennan2009}
\textsc{Brennan,T.} ; \textsc{Dieterich,W.} ; \textsc{Ehret,B.}: Evaluating the
  Predictive Validity of the Compas Risk and Needs Assessment System.
\newblock In: {\em Criminal Justice and Behavior}. 36, (2009), Nr. 1, S.\
  21--40, jan 2009

\bibitem[Burgess 28]{Burgess1928}
\textsc{Burgess,E.~W.}: Factors determining success or failure on parole.
\newblock In: {\em The workings of the indeterminate sentence law and the
  parole system in Illinois}. (1928), S.\ 205--249, 1928

\bibitem[Burnham 02]{Burnham2002}
\textsc{Burnham,B.~R.}: Positive predictive value of a health history
  questionnaire.
\newblock In: {\em Military medicine}. 167, (2002), Nr. 8, S.\ 639, 2002

\bibitem[Burrell 96]{Burrell1996}
\textsc{Burrell,H.~C.} ; \textsc{Pinder,S.~E.} ; \textsc{Wilson,A. R.~M.} ;
  \textsc{Evans,A.~J.} ; \textsc{Yeoman,L.~J.} ; \textsc{Elston,C.~W.} ;
  \textsc{Ellis,I.~O.}: The positive predictive value of mammographic signs: A
  review of 425 non-palpable breast lesions.
\newblock In: {\em Clinical radiology}. 51, (1996), Nr. 4, S.\ 277--281, apr
  1996

\bibitem[Cesare 64]{Cesare1764}
\textsc{Cesare,B.}: Dei delitti e delle pene.
\newblock In: {\em Opera immortale, Vienna, Sam}. (1764), 1764

\bibitem[Chettiar 11]{Chettiar2011}
\textsc{Chettiar,I.} ; \textsc{Gupta,V.}: Smart Reform is Possible: States
  Reducing Incarceration Rates and Costs While Protecting Communities.
\newblock In: {\em {SSRN} Electronic Journal}. (2011), 2011

\bibitem[Dahle 06]{Dahle2006}
\textsc{Dahle,K.~P.}: Grundlagen und Methoden der Kriminalprognose. In: {\em
  Handbuch der Forensischen Psychiatrie: Band 3 Psychiatrische Kriminalprognose
  und Kriminaltherapie}. Steinkopff-Verlag, 2006, S.\ 1--67

\bibitem[Dahle 97]{Dahle1997}
\textsc{Dahle,K.-P.}: Kriminalprognosen im Strafrecht. Psychologische Aspekte
  individueller Verhaltenvorhersagen.
\newblock In: {\em Psychologie im Strafverfahren. Huber, Bern}. (1997), S.\
  119--140, 1997

\bibitem[Dieterich 16]{Dieterich2016}
\textsc{Dieterich,W.} ; \textsc{Mendoza,C.} ; \textsc{Brennan,T.}: COMPAS risk
  scales: Demonstrating accuracy equity and predictive parity.
\newblock {T}echnischer {B}ericht, Technical report, Northpointe, July 2016.
  http://www. northpointeinc. com/northpointe-analysis, 2016

\bibitem[Dittmann 03]{Dittmann2003}
\textsc{Dittmann,V.}: Was kann die Kriminalprognose heute leisten? In
  \textsc{H{\"a}{\ss}ler,F.} (Hrsg.): {\em Forensische Kinder-, Jugend-und
  Erwachsenenpsychiatrie: Aspekte der forensischen Begutachtung}. Schattauer
  Verlag, 2003, S.\ 173--187

\bibitem[D{\"o}bele 13]{Doebele2013}
\textsc{D{\"o}bele,A.-L.}: {\em Standardisierte Prognoseinstrumente zur
  Vorhersage des R{\"u}ckfallrisikos von Straft{\"a}tern: eine kritische
  Betrachtung des Einsatzes in der Strafrechtspflege aus juristischer Sicht}.
  Kovac, Dr. Verlag, 2013

\bibitem[Eher 08]{Eher2008}
\textsc{Eher,R.} ; \textsc{Rettenberger,M.} ; \textsc{Schilling,F.} ;
  \textsc{Pf{\"a}fflin,F.}: Validit{\"a}t oder praktischer Nutzen?
  R{\"u}ckfallvorhersagen mittels Static-99 und SORAG. Eine prospektive
  R{\"u}ckfallstudie an 275 Sexualstraft{\"a}tern.
\newblock In: {\em Recht \& Psychiatrie}. 26, (2008), S.\ 79--88, 2008

\bibitem[Endrass 08]{Endrass2008}
\textsc{Endrass,J.} ; \textsc{Urbaniok,F.} ; \textsc{Held,L.} ;
  \textsc{Vetter,S.} ; \textsc{Rossegger,A.}: Accuracy of the Static-99 in
  Predicting Recidivism in Switzerland.
\newblock In: {\em International Journal of Offender Therapy and Comparative
  Criminology}. 53, (2008), Nr. 4, S.\ 482--490, feb 2008

\bibitem[Farabee 10]{Farabee2010}
\textsc{Farabee,D.} ; \textsc{Zhang,S.} ; \textsc{Roberts,R.~E.} ;
  \textsc{Yang,J.}: COMPAS validation study: Final report.
\newblock In: {\em Semel Institute for Neuroscience and Human Behavior, UCLA,
  Los Angeles, CA}. (2010), 2010

\bibitem[Fergusson 77]{Fergusson1977}
\textsc{Fergusson,D.~M.} ; \textsc{Fifield,J.~K.} ; \textsc{Slater,S.~W.}:
  Signal Detectability Theory and the Evaluation of Prediction Tables.
\newblock In: {\em Journal of Research in Crime and Delinquency}. 14, (1977),
  Nr. 2, S.\ 237--246, jul 1977

\bibitem[{Focus Online} 09]{FocusOnline2009}
\textsc{{Focus Online}}.
\newblock {\em Häftlinge sollen ihren Gefängnisaufenthalt in den USA künftig
  bezahlen}.
\newblock
  \url{http://www.focus.de/politik/weitere-meldungen/usa-haeftlinge-sollen-ihren-gefaengnisaufenthalt-in-den-usa-kuenftig-bezahlen_aid_426102.html}.
  08 2009, [Online; accessed 06-May-2017]

\bibitem[Goutte 05]{Goutte2005}
\textsc{Goutte,C.} ; \textsc{Gaussier,E.}: A Probabilistic Interpretation of
  Precision, Recall and F-Score, with Implication for Evaluation.
\newblock In: {\em Lecture Notes in Computer Science}. Springer Berlin
  Heidelberg, 2005, S.\ 345--359

\bibitem[Guy 08]{Guy2008}
\textsc{Guy,L.~S.}: {\em Performance indicators of the structured professional
  judgment approach for assessing risk for violence to others: A meta-analytic
  survey}.
\newblock Dissertation, Simon Fraser University, 2008

\bibitem[Hanley 82]{Hanley1982}
\textsc{Hanley,J.~A.} ; \textsc{McNeil,B.~J.}: The meaning and use of the area
  under a receiver operating characteristic ({ROC}) curve.
\newblock In: {\em Radiology}. 143, (1982), Nr. 1, S.\ 29--36, apr 1982

\bibitem[Heiser 13]{Heiser2013}
\textsc{Heiser,J.}
\newblock {\em Zehntausende im Hungerstreik}.
\newblock \url{http://www.ag-friedensforschung.de/regionen/USA/haft.html}. 07
  2013, [Online; accessed 06-May-2017]

\bibitem[Hengen 04]{Hengen2004}
\textsc{Hengen,H.} ; \textsc{Feid,M.} ; \textsc{Pandit,M.}: {\"U}berwacht
  lernende Klassifikationsverfahren im {\"U}berblick, Teil 1 (Overview of
  Supervised learning Classification Methods, Part 1).
\newblock In: {\em at--Automatisierungstechnik/Methoden und Anwendungen der
  Steuerungs-, Regelungs-und Informationstechnik}. 52, (2004), Nr. 3/2004, S.\
  A1--A8, mar 2004

\bibitem[Honest 02]{Honest2002}
\textsc{Honest,H.} ; \textsc{Khan,K.~S.}: Reporting of measures of accuracy in
  systematic reviews of diagnostic literature.
\newblock In: {\em {BMC} Health Services Research}. 2, (2002), Nr. 1, S.\~4,
  2002

\bibitem[Horv{\'a}t 12]{Horvat2012}
\textsc{Horv{\'a}t,E.-{\'A}.} ; \textsc{Hanselmann,M.} ;
  \textsc{Hamprecht,F.~A.} ; \textsc{Zweig,K.~A.}: One plus one makes three
  (for social networks).
\newblock In: {\em PloS one}. 7, (2012), Nr. 4, S.\ e34740, 2012

\bibitem[Kerlikowske 93]{Kerlikowske1993}
\textsc{Kerlikowske,K.} ; \textsc{Grady,D.} ; \textsc{Barclay,J.} ;
  \textsc{Sickles,E.~A.} ; \textsc{Eaton,A.} ; \textsc{Ernster,V.}: Positive
  Predictive Value of Screening Mammography by Age and Family History of Breast
  Cancer.
\newblock In: {\em {JAMA}: The Journal of the American Medical Association}.
  270, (1993), Nr. 20, S.\ 2444--2450, nov 1993

\bibitem[Kopans 92]{Kopans1992}
\textsc{Kopans,D.~B.}: The positive predictive value of mammography.
\newblock In: {\em AJR. American journal of roentgenology}. 158, (1992), Nr. 3,
  S.\ 521--526, mar 1992

\bibitem[Kröber 06]{Kroeber2006}
\textsc{Kröber,H.~L.}: Kriminalprognostische Begutachtung. In: {\em Handbuch
  der Forensischen Psychiatrie}. Steinkopff-Verlag, 2006, S.\ 69--172

\bibitem[Kr{\"o}ber 99]{Kroeber1999}
\textsc{Kr{\"o}ber,H.-L.}: Gang und Gesichtspunkte der kriminalprognostischen
  psychiatrischen Begutachtung.
\newblock In: {\em Neue Zeitschrift f{\"u}r Strafrecht}. 12, (1999), S.\
  593--599, 1999

\bibitem[Lansing 12]{Lansing2012}
\textsc{Lansing,S.}: New York State COMPAS-probation risk and need assessment
  study: Examining the recidivism scale’s effectiveness and predictive
  accuracy.
\newblock In: {\em Retrieved March}. 1, (2012), S.\ 2013, 2012

\bibitem[Leushuis 09]{Leushuis2009}
\textsc{Leushuis,E.} ; \textsc{Steeg,J.~W.van~der,} ; \textsc{Steures,P.} ;
  \textsc{Bossuyt,P.~M.} ; \textsc{Eijkemans,M.~J.} ; \textsc{Veen,F.van~der,}
  ; \textsc{Mol,B.~W.} ; \textsc{Hompes,P.~G.}: Prediction models in
  reproductive medicine: a critical appraisal.
\newblock In: {\em Human reproduction update}. 15, (2009), Nr. 5, S.\ 537--552,
  2009

\bibitem[Liben-Nowell 03]{Liben-Nowell2003}
\textsc{Liben-Nowell,D.} ; \textsc{Kleinberg,J.}: The Link Prediction Problem
  for Social Networks.
\newblock In: {\em Proceedings of the Twelfth International Conference on
  Information and Knowledge Management}. New York, NY, USA: {ACM} Press, 2003,
  S.\ 556--559

\bibitem[Lischka 17]{Lischka2017}
\textsc{Lischka,K.} ; \textsc{Klingel,A.} ; \textsc{{Bertelsmann Stiftung}}.
\newblock {\em Wenn Maschinen Menschen bewerten}. 2017

\bibitem[Lukic 09]{Lukic2009}
\textsc{Lukic,E.~A.}: Evaluation of the Pretrial Services Program Administered
  by the Broward Sheriff’s Office, Report No. 09-07.
\newblock {T}echnischer {B}ericht, Broward Sheriff’s Office,
  \url{https://www.broward.org/Auditor/Documents/pretrial_final060909.pdf}, 05
  2009.
\newblock [Online; accessed 28-March-2017]

\bibitem[Manning 99]{Manning1999}
\textsc{Manning,C.~D.} ; \textsc{Sch{\"u}tze,H.} et al: {\em Foundations of
  statistical natural language processing}. MIT Press, 1999

\bibitem[Maschke 08]{Maschke2008}
\textsc{Maschke,W.}: Die Kriminalprognose im Einzelfall.
\newblock In: {\em Tagungsband der Jahrestagung 2008 - Deutsche Vereinigung
  für Jugendgerichte und Jugendgerichtshilfen e. V. Landesgruppe
  Baden-Württemberg}. Hochschule für Polizei Villingen-Schwenningen, 2008

\bibitem[McDermott 16]{McDermott2016}
\textsc{McDermott,R.} ; \textsc{Bar-Joseph,U.}: Pearl Harbor and Midway: the
  decisive influence of two men on the outcomes.
\newblock In: {\em Intelligence and National Security}. 31, (2016), Nr. 7, S.\
  949--962, mar 2016

\bibitem[Meehl 54]{Meehl1954}
\textsc{Meehl,P.~E.}: {\em Clinical versus statistical prediction: A
  theoretical analysis and a review of the evidence.} University of Minnesota
  Press, 1954

\bibitem[Meyer 59]{Meyer1959}
\textsc{Meyer,F.}: Der kriminologische Wert von Prognosetafeln.
\newblock In: {\em Mschr Krim}. 42, (1959), S.\ 214--245, 1959

\bibitem[Mohabatkar 13]{Mohabatkar2013}
\textsc{Mohabatkar,H.} ; \textsc{Beigi,M.~M.} ; \textsc{Abdolahi,K.} ;
  \textsc{Mohsenzadeh,S.}: Prediction of Allergenic Proteins by Means of the
  Concept of Chous Pseudo Amino Acid Composition and a Machine Learning
  Approach.
\newblock In: {\em Medicinal Chemistry}. 9, (2013), Nr. 1, S.\ 133--137, feb
  2013

\bibitem[Monahan 75]{Monahan1975}
\textsc{Monahan,J.}: The prediction of violence.
\newblock In: {\em Violence and criminal justice}. 15, (1975), S.\~32, 1975

\bibitem[Monahan 81]{Monahan1981}
\textsc{Monahan,J.} ; \textsc{Brodsky,S.~L.} ; \textsc{Shan,S.~A.}: {\em
  Predicting violent behavior: An assessment of clinical techniques}. Sage
  Publications Beverly Hills, CA, 1981

\bibitem[Monahan 96]{Monahan1996}
\textsc{Monahan,J.}: Violence prediction the past twenty and the next twenty
  years.
\newblock In: {\em Criminal Justice and Behavior}. 23, (1996), Nr. 1, S.\
  107--120, 1996

\bibitem[M{\"u}ller-Isberner 98]{Mueller-Isberner1998}
\textsc{M{\"u}ller-Isberner,R.} ; \textsc{J{\"o}ckel,D.} ;
  \textsc{Gonzalez~Cabeza,S.}: Die Vorhersage von Gewalttaten mit dem HCR-20.
\newblock In: {\em Institut f{\"u}r Forensische Psychiatrie, Haina}. (1998),
  1998

\bibitem[Murphy 87]{Murphy1987}
\textsc{Murphy,A.~H.} ; \textsc{Winkler,R.~L.}: A General Framework for
  Forecast Verification.
\newblock In: {\em Monthly Weather Review}. 115, (1987), Nr. 7, S.\ 1330--1338,
  jul 1987

\bibitem[{n-tv} 16]{NTV2016}
\textsc{{n-tv}}.
\newblock {\em US-Privatgefängnissen droht das Aus}.
\newblock
  \url{http://www.n-tv.de/politik/US-Privatgefaengnissen-droht-das-Aus-article18453536.html}.
  02 2016, [Online; accessed 06-May-2017]

\bibitem[Nedopil 05]{Nedopil2005}
\textsc{Nedopil,N.} ; \textsc{Gross,G.}: {\em Prognosen in der Forensischen
  Psychiatrie: ein Handbuch f{\"u}r die Praxis}. Pabst Lengerich, 2005

\bibitem[Nevin 69]{Nevin1969}
\textsc{Nevin,J.~A.}: Signal detection theory and operant {behaviorA} review of
  David M. Green and John A. Swets Signal detection theory and psychophysics.
\newblock In: {\em Journal of the Experimental Analysis of Behavior}. 12,
  (1969), Nr. 3, S.\ 475--480, may 1969

\bibitem[Northpointe 12a]{Northpoint2017}
\textsc{Northpointe}.
\newblock {\em COMPAS Risk \& Need Assessment System - Selected Questions Posed
  by Inquiring Agencies}.
\newblock \url{http://www.northpointeinc.com/files/downloads/FAQ_Document.pdf}.
  2012, [Online; accessed 27-March-2017]

\bibitem[Northpointe 12b]{Northpointe2012}
\textsc{Northpointe}.
\newblock {\em Practitioner’s Guide to COMPAS Core}.
\newblock
  \url{http://www.northpointeinc.com/downloads/compas/Practitioners-Guide-COMPAS-Core-_031915.pdf}.
  03 2012, [Online; accessed 27-March-2017]

\bibitem[Obergfell-Fuchs 11]{Obergfell-Fuchs}
\textsc{Obergfell-Fuchs,J.}: Gef{\"a}hrliche Straft{\"a}ter aus
  kriminologischer und psychologischer Sicht.
\newblock In: {\em Tagungsband „Sicherungsverwahrung und Führungsaufsicht.
  Wie gehen wir mit gefährlichen Straftätern um?“ Evangelische Akademie Bad
  Boll}. (2011), 2011

\bibitem[Peterson 54]{Peterson1954}
\textsc{Peterson,W.} ; \textsc{Birdsall,T.} ; \textsc{Fox,W.}: The theory of
  signal detectability.
\newblock In: {\em Transactions of the {IRE} Professional Group on Information
  Theory}. 4, (1954), Nr. 4, S.\ 171--212, sep 1954

\bibitem[Pogue 62]{Wohlstetter1962}
\textsc{Pogue,F.~C.} ; \textsc{Wohlstetter,R.}: {\em Pearl Harbor: Warning and
  Decision.} Stanford University Press, 1962

\bibitem[Potapov 12]{Potapov2012}
\textsc{Potapov,S.}: {\em Zur Verbesserung der Splitkriterien bei
  Klassifikationsb{\"a}umen und Ensemble-Methoden}.
\newblock Dissertation, m Institut für Medizininformatik, Biometrie und
  Epidemiologie der Friedrich-Alexander-Universität Erlangen-N{\"u}rnberg,
  2012

\bibitem[Provost 98]{Provost1998}
\textsc{Provost,F.~J.} ; \textsc{Fawcett,T.} ; \textsc{Kohavi,R.}: The case
  against accuracy estimation for comparing induction algorithms.
\newblock In: {\em ICML}. Bd. 98, 1998, S.\ 445--453

\bibitem[Rettenberger 13]{Rettenberger2013}
\textsc{Rettenberger,M.} ; \textsc{Franqu{\'e},F.von,}: {\em Handbuch
  kriminalprognostischer Verfahren}. Hogrefe Verlag, 2013

\bibitem[Rice 95]{Rice1995}
\textsc{Rice,M.~E.} ; \textsc{Harris,G.~T.}: Violent recidivism: assessing
  predictive validity.
\newblock In: {\em Journal of consulting and clinical psychology}. 63, (1995),
  Nr. 5, S.\ 737, 1995

\bibitem[R{\"o}ssner 00]{Roessner2000}
\textsc{R{\"o}ssner,D.} ; \textsc{Jehle,J.-M.}: {\em Beccaria als Wegbereiter
  der Kriminologie}. Forum Verlag Godesberg GmbH, 2000

\bibitem[Sack 78]{Sack1978}
\textsc{Sack,F.}: Probleme der Kriminalsoziologie.
\newblock In: {\em Handbuch der empirischen Sozialforschung}. 12, (1978), S.\
  192--492, 1978

\bibitem[Schiedt 36]{Schiedt1936}
\textsc{Schiedt,R.}
\newblock {\em Ein Beitrag zum Problem der Rückfallprognose}.
\newblock Münchener Zeitungsverl. 1936

\bibitem[Schneider 67]{Schneider1967}
\textsc{Schneider,H.~J.}: {\em Prognostische Beurteilung des Rechtsbrechers:
  Die ausl{\"a}ndische Forschung}. Verlag f{\"u}r Psychologie, 1967

\bibitem[Schwind 10]{Schwind2010}
\textsc{Schwind,H.-D.}: {\em Kriminologie: eine praxisorientierte
  Einf{\"u}hrung mit Beispielen}. CF M{\"u}ller GmbH, 2010

\bibitem[Singhal 01]{Singhal2001}
\textsc{Singhal,A.}: Modern information retrieval: A brief overview.
\newblock In: {\em IEEE Data Eng. Bull.} 24, (2001), Nr. 4, S.\ 35--43, 2001

\bibitem[statista 17a]{statista2017}
\textsc{statista}.
\newblock {\em Länder mit der größten Anzahl an Inhaftierten (Februar
  2017*)}.
\newblock
  \url{https://de.statista.com/statistik/daten/studie/3212/umfrage/laender-mit-den-meisten-gefangenen-im-jahr-2007/
  }. 2017, [Online; accessed 06-May-2017]

\bibitem[statista 17b]{statista2017a}
\textsc{statista}.
\newblock {\em USA: Gesamtbev{\"o}lkerung von 2007 bis 2017 (in Millionen
  Einwohner)}.
\newblock
  \url{https://de.statista.com/statistik/daten/studie/19320/umfrage/gesamtbevoelkerung-der-usa/
  }. 2017, [Online; accessed 10-May-2017]

\bibitem[Swets 64]{Swets1964}
\textsc{Swets,J.~A.}: {\em Signal Detection and Recognition in Human Observers:
  Contemporary Readings}. John Wiley and Sons, 1964

\bibitem[Swets 66]{Swets1966}
\textsc{Swets,J.~A.} ; \textsc{Green,D.~M.}: {\em Signal detection theory and
  psychophysics}. Wiley, 1966

\bibitem[Swets 73]{Swets1973}
\textsc{Swets,J.~A.}: The relative operating characteristic in psychology.
\newblock In: {\em Science}. 182, (1973), Nr. 4116, S.\ 990--1000, 1973

\bibitem[Swets 88]{Swets1988}
\textsc{Swets,J.~A.}: Measuring the accuracy of diagnostic systems.
\newblock In: {\em Science}. 240, (1988), Nr. 4857, S.\ 1285--1293, jun 1988

\bibitem[Urbaniok 08]{Urbaniok2008}
\textsc{Urbaniok,F.} ; \textsc{Rinne,T.} ; \textsc{Held,L.} ;
  \textsc{Rossegger,A.} ; \textsc{Endrass,J.}: Forensische Risikokalkulationen:
  Grundlegende methodische Aspekte zur Beurteilung der Anwendbarkeit und
  Validit{\"a}t verschiedener Verfahren.
\newblock In: {\em Fortschritte der Neurologie $\cdotp$ Psychiatrie}. 76,
  (2008), Nr. 08, S.\ 470--477, aug 2008

\bibitem[{US Supreme Court} 66]{USSC1966}
\textsc{{US Supreme Court}}: Baxstrom v. Herold, 383 U.S. 107 (1966).
\newblock In: {\em No. 219}. (1966), 1966.
\newblock \url{https://supreme.justia.com/cases/federal/us/383/107/case.html}

\bibitem[van Rijsbergen 79]{Rijsbergen1979}
\textsc{Rijsbergen,C.van,}.
\newblock {\em Information Retrieval. 1979}.
\newblock Dept. of Computer Science, University of Glasgow. 1979

\bibitem[Volckart 97]{Volckart1997}
\textsc{Volckart,B.}: {\em Praxis der Kriminalprognose: Methodologie und
  Rechtsanwendung}. Beck, 1997

\bibitem[Walmsley 14]{Walmsley2014}
\textsc{Walmsley,R.}: World Prison Population List.
\newblock In: {\em World Prison Brief}. (2014), 2014

\bibitem[Walpole 93]{Walpole1993}
\textsc{Walpole,R.~E.} ; \textsc{Myers,R.~H.} ; \textsc{Myers,S.~L.} ;
  \textsc{Ye,K.}: {\em Probability and statistics for engineers and
  scientists}. Macmillan New York, 1993

\bibitem[Webster 97]{Webster1997}
\textsc{Webster,C.~D.} et al: HCR-20: Assessing risk for violence.
\newblock In: {\em Mental Health, Law, and Policy Institute, Simon Fraser
  University, in cooperation with the British Columbia Forensic Psychiatric
  Services Commission}. (1997), 1997

\bibitem[Werner 16]{Werner2016}
\textsc{Werner,K.}
\newblock {\em Nudeln sind die neue W{"a}hrung in US-Gef{"a}ngnissen}.
\newblock S{\"u}ddeutsche Zeitung. 08 2016,
  \url{http://www.sueddeutsche.de/panorama/gefaengniswaehrung-zwei-suppen-fuer-ein-t-shirt-1.3132582}
  - [Online; accessed 06-May-2017]

\bibitem[{Wisconsin Department of Correction} 17]{Wisconsin2017}
\textsc{{Wisconsin Department of Correction}}.
\newblock {\em COMPAS Assessment Tool}.
\newblock
  \url{http://doc.wi.gov/about/doc-overview/office-of-the-secretary/office-of-reentry/compas-assessment-tool}.
  2017, [Online; accessed 27-March-2017]

\bibitem[Zweig 16]{Zweig2016}
\textsc{Zweig,K.~A.}
\newblock {\em 2. Arbeitspapier: Überprüfbarkeit von Algorithmen}.
\newblock Algorithm Watch. 07 2016,
  \url{http://algorithm-watch.org/zweites-arbeitspapier-ueberpruefbarkeit-algorithmen/}
  - [Online; accessed 06-May-2017]

\end{thebibliography}
\end{document}